%% file: Trans_info_theory.tex
\documentclass[journal,draftcls,onecolumn,twoside]{IEEEtran}
\normalsize
\input{./preamble}

\newif\ifproof
\prooftrue

\def\fig_path{./Figures}
\begin{document}
\title{Unsourced Random Access\\with Coded Compressed Sensing:\\Integrating AMP and Belief Propagation}

%
\author{\IEEEauthorblockN{\dag Vamsi K. Amalladinne, \emph{Student Member, IEEE},
\dag Asit Kumar Pradhan, \emph{Member, IEEE}, \\
\S Cynthia Rush, \emph{Member, IEEE},
\dag Jean-Francois Chamberland, \emph{Senior Member, IEEE}, \\
\dag Krishna R. Narayanan, \emph{Fellow, IEEE} \\
 \dag Department of Electrical and Computer Engineering, Texas A\&M University\\
 \S Department of Statistics, Columbia University
}
\thanks{
This material is based upon work supported, in part, by the National Science Foundation (NSF) under Grants CCF-1619085 \& CCF-1849883, and by Qualcomm Technologies, Inc., through their University Relations Program.
This work was presented in part at the International Symposium on Information Theory, 2020.
}
}

\maketitle

\begin{abstract}
Sparse regression codes with approximate message passing (AMP) decoding have gained much attention in recent times. 
The concepts underlying this coding scheme extend to unsourced random access with coded compressed sensing (CCS), as first demonstrated by Fengler, Jung, and Caire. 
Specifically, their approach employs a concatenated coding framework with an inner AMP decoder followed by an outer tree decoder. 
In their original implementation, these two components work independently of each other, with the tree decoder acting on the static output of the AMP decoder. 
This article introduces a novel framework where the inner AMP decoder and the outer tree decoder operate in tandem, dynamically passing information back and forth to take full advantage of the underlying CCS structure.
This scheme necessitates the redesign of the tree code as to enable belief propagation in a computationally tractable manner.
The enhanced architecture exhibits significant performance benefits over a range of system parameters.
The error performance of the proposed scheme can be accurately predicted through a set of equations, known as state evolution of AMP.
These findings are supported both analytically and through numerical methods.
\end{abstract}

\begin{IEEEkeywords}
Unsourced random access, sparse regression codes, approximate message passing, belief propagation, coded compressed sensing, concatenated coding.
\end{IEEEkeywords}

\section{Introduction and Background}
\label{section:Introduction}
Unsourced random access is a novel communication paradigm envisioned to accommodate the increasing traffic demands and heterogeneity of next generation wireless networks.
This framework garnered significant research interest owing to the emergence of Internet of Things (IoT) and machine-driven communications.
This model differs from the conventional multiple access paradigm in a number of ways.
Conventional multiple access schemes are designed primarily for human-centric communications with sustained connections wherein the cost of coordination can be amortized over a long time period.
However, this strategy may not be suitable for machine-centric communications because device transmissions are often sporadic with very short payloads.
This new reality invites the creation of protocols in which it is not mandatory for active devices to reveal their identities.
Rather, decoding is done only up to a permutation of the transmitted payloads, without regard for the identities of transmitting devices.
Active devices wishing to reveal themselves can embed such information in their payloads.
This approach enables all active devices to share a common codebook for their transmissions.

The unsourced random access channel was introduced in~\cite{polyanskiy2017perspective}, along with a random coding achievability bound for its capacity in the absence of complexity constraints.
Subsequently, several practical coding schemes that aim to perform close to this conceptual benchmark have been proposed in the literature \cite{ordentlich2017low,vem2019user,amalladinne2019coded,fengler2019sparcs,pradhan2019sparseidma,calderbank2018chirrup, amalladinne2020enhanced,marshakov2019polar,AKPolar}. 
These schemes can be broadly divided into two categories; schemes that are built on conventional channel codes like LDPC or polar codes \cite{vem2019user, pradhan2019sparseidma,  marshakov2019polar, AKPolar} and those that offer compressed sensing (CS) based solutions \cite{ amalladinne2019coded, fengler2019sparcs,  calderbank2018chirrup, amalladinne2020enhanced}.
Within this context, Amalladinne et al.~\cite{amalladinne2019coded} put forth a concatenated coding scheme that uses an inner CS code and an outer tree code.
In doing so, their scheme takes advantage of the connection between the equivalence of unsourced multiple access and support recovery in high dimensional compressed sensing.
Indeed, unsourced random access can be cast as a compressed sensing problem, albeit one whose excessive size precludes the straightforward application of existing solutions.
Accordingly, Amalladinne et al.\ leverage a divide-and-conquer approach to split the information messages of active users into several blocks, each amenable to standard CS solvers.
Redundancy is employed in the form of an outer tree code to bind the information blocks that correspond to one message together.
The performance of this scheme is extensively studied in \cite{amalladinne2019coded} and this revealed the natural trade-off between error performance and computational complexity afforded by the allocation of parity check bits across blocks.
This scheme, termed as coded compressed sensing (CCS), employs the parity-check bits added in the encoding phase solely for the purpose of stitching.
Yet, it turns out that the inner and outer decoders in CCS can be executed concurrently, and the redundancy employed during the transmission phase can be utilized to curtail the realm of possibilities for parity-check bits in subsequent stages~\cite{amalladinne2020enhanced}.
This algorithmic improvement developed for CCS offers significant benefits both in terms of error performance and computational complexity.
This enhancement also engenders new parameter allocation strategies beyond those observed in the CCS framework, some of which are studied in~\cite{amalladinne2020enhanced}.
We refer the reader to \cite{amalladinne2019coded,amalladinne2020enhanced} for more details regarding this line of work.
A motivation for our research is the hope that similar notions may apply to other schemes related to unsourced random access and beyond.

Approximate message passing (AMP) \cite{donoho2009message, montanari2012graphical, rangan2011generalized} refers to a broad class of iterative algorithms derived from message passing algorithms on dense factor graphs that have demonstrated good performance as decoders in the context of wireless communication.
An early instance of AMP applied to digital communication pertains to the efficient decoding of a sparse regression code (SPARC)~\cite{CIT-092, rush2017capacity, barbier2017approximate, barbier2014replica}. SPARCs are a coding scheme based on high-dimensional linear regression introduced for the single-user AWGN channel by Barron and Joseph \cite{joseph2013fast, joseph2012least}.
It was rigorously proved in \cite{rush2017capacity} that SPARCs with an AMP decoder and an appropriate power allocation achieves the capacity of a single-user AWGN channel asymptotically and, 
subsequently, there has been much effort to improve the finite block length performance of single user SPARC \cite{greig2017techniques, barbier2015approximate, rush2020capacity, hsieh2020modulated, rush2018error, rush2019spatially, hsieh2018spatially} by leveraging standard techniques like spatial coupling, concatenated coding, or 
higher order modulations.
The performance improvements demonstrated by these techniques provide a strong incentive to explore similar frameworks in the context of unsourced random access and we discuss one such approach in this article.

The first application of an AMP decoder to unsourced random access is due to Fengler, Jung, and Caire~\cite{fengler2019sparcs}.
Therein, the authors draw a connection between the structure of CCS and SPARC constructions.
They extend the CCS framework~\cite{ amalladinne2019coded} by using a design (measurement) matrix that does not assume a block diagonal structure, and they apply AMP as part of the message recovery process.
They rely on the tree decoder proposed in~\cite{ amalladinne2019coded} to accomplish the stitching process, once individual messages have been reconstructed by the AMP decoder.
Also, they show that this concatenated construction asymptotically achieves the symmetric sum rate Shannon capacity of the MAC channel, thereby extending the optimality results of SPARC in \cite{rush2017capacity} to scenarios with multiple users.
In this article, we build on the insights developed in \cite{amalladinne2020enhanced} to devise novel message passing rules that integrate the tree code and AMP.
The main contributions of this article are as follows.
\begin{enumerate}
\item A novel framework that facilitates dynamic interactions between inner and outer decoding components of the concatenated coding structure introduced in \cite{fengler2019sparcs} is proposed for the unsourced random access problem.
\item We show that the parity-check bits intrinsic to tree coding can be harnessed to assist the convergence of AMP.
\item We develop a modified tree code that allows passing meaningful messages back and forth between the inner AMP and the outer tree decoding components. This revised tree code enables us to employ fast Fourier transform (FFT) techniques to conduct belief propagation in a computationally efficient manner. The reader may notice a similarity between our framework and the way messages are passed in non-binary LDPC codes.
\item We provide finite-block length simulation results for this proposed scheme to demonstrate the performance gain it offers over previously proposed schemes for scenarios of interest~\cite{amalladinne2019coded, fengler2019sparcs}.
\item We develop a framework to track the state evolution and find that it accurately predicts the error performance of the proposed concatenated coding scheme. Also, within the proposed framework, we prove that the state evolution is accurate.
\end{enumerate}
Concatenated schemes that combine AMP with an outer code have been proposed in \cite{greig2017techniques, LiangCC} to improve the finite block length performance in the context of single user SPARC.
In \cite{ greig2017techniques}, a high rate LDPC code is used to protect sections of AMP with less allocated power.
The decoding operation involves three steps: the first step uses AMP decoding, the second step decodes the outer LDPC codes by employing the soft outputs produced by AMP in the first stage and the third step re-runs the AMP decoder after removing the contribution of successfully decoded sections from the second stage.
This approach results in a steep waterfall in error performance; a phenomenon that is not achieved by standalone AMP decoder for small block lengths.
However, this architecture does not allow dynamic interactions between AMP and LDPC decoder.
In other words, the LDPC decoder does not assist the convergence of AMP and it only works on soft outputs produced by AMP upon convergence.
Compressed-coding is proposed and analyzed in \cite{LiangCC}, where AMP is combined with an outer generic forward error correction (FEC) code.
Therein, the authors show that with a careful design of the underlying FEC code, the compressed-coding scheme can achieve the single user Gaussian capacity asymptotically.
Yet, their analysis hinges on the assumption that state evolution for AMP remains accurate even in the presence of an outer FEC decoder. 
In this paper, we do not rely on such assumptions; rather, we prove that state evolution for AMP is indeed accurate under certain conditions for the proposed architecture.
It is also worth noting that the schemes in \cite{greig2017techniques,LiangCC} do not admit straightforward extensions to the unsourced random access paradigm, which points to the novel and unique character of our contributions.

\subsection{Organization}

The remainder of this article is organized as follows.
In Section~\ref{section:Motivation}, we describe the system model and the broad CCS-AMP architecture.
We also introduce the reader to the relevant notation used throughout this paper.
In Section~\ref{section:TreeCodeRevisited}, we provide a detailed description of the revised tree code and outline the key distinguishing factors between the revised tree code and the one developed in~\cite{amalladinne2019coded}.
We then introduce a framework that allows soft decoding of the outer tree code, which can be employed in tandem with the decoding of the inner AMP code.
A fast implementation of the above algorithm, which leverages FFT techniques, is also detailed in this part.
Section~\ref{section:AMP} focuses on the inner code and the AMP decoder.
Therein, we explain how the structure of underlying tree code can be harnessed to assist the convergence of AMP through a novel denoising function.
Furthermore, we provide guiding principles to design good tree codes which serve the purpose of assisting AMP convergence, as well as stitching individual messages together once AMP has converged.
Finally, for the proposed scheme, we describe a framework to track state evolution, a tool used to predict the error performance of AMP decoder.
Simulation results are reported in Section~\ref{section:SimulationResults}, and conclusions are drawn in Section~\ref{section:Conclusion}.

\section{System Model and Conceptual Framework}
\label{section:Motivation}

Consider a situation where $\Ka$ active devices out of a population of $\Ktot$ devices ($\Ka \ll \Ktot$) each wish to send a message to an access point.
Without loss of generality, we label active devices using integers from one to $\Ka$.
The transmission process takes place over a multiple access channel, with a time duration of $n$ channel uses (real degrees of freedom).
We denote the message of device~$i$ by $\wv_i$, and we use $\xv_i$ to represent the corresponding transmitted waveform.
The signal received at the destination takes the form
\begin{equation} \label{equation:ChannelModel}
\yv = \sum_{i=1}^{\Ka} \xv_i + \zv
\end{equation}
where $\xv_i \in \mathbb{R}^n$. 
The additive noise component $\zv$ is composed of an independent sequence of Gaussian elements, each with distribution $\mathcal{N}(0,1)$.
All the devices share a common codebook $\mathcal{C}$, as is customary in unsourced random access.
Consequently, $\xv_i$ is a function of payload $\wv_i$, but not of the identity of device~$i$ itself.
For ease of exposition, we assume that a frame synchronization beacon or an alternate mechanism enables coherent transmission, although we note that this framework can be extended to more realistic settings~\cite{amalladinne2019asynchronous}.

The access point is tasked with producing an unordered list $\widehat{W}(\yv)$ of candidate messages based on the received signal $\yv$.
The size of the output list is constrained and cannot exceed $\Ka$, i.e., $\big| \widehat{W}(\yv) \big| \leq \Ka$.
The performance of this communication scheme is assessed according to the per-user probability of error (PUPE), which is the predominant evaluation criterion for unsourced random access~\cite{polyanskiy2017perspective}.
Mathematically, we have
\begin{equation} \label{equation:PUPE}
\Pe = \frac{1}{\Ka} \sum_{i = 1}^{\Ka}
\Pr \left( \wv_i \notin \widehat{W}(\yv) \right)
\end{equation}
where $\wv_i$ is the payload of active device~$i$.
Our prime design goal consists of creating a pragmatic, low-complexity scheme that enables the communication of $\Ka$ messages to the access point under the unsourced random access paradigm with a probability of failure $\Pe \leq \epsilon$.
Furthermore, we wish to do so utilizing a small number of channel uses ($n \approx 30,000$), and with devices expending as little energy as possible.

At this stage, we are ready to initiate our treatment of CCS-AMP.
We begin with a brief overview of the proposed architecture.
We then discuss the specifics of an outer tree code tailored to its integration with the AMP framework.
We finish the treatment of our new scheme with the description and analysis of the enhanced AMP algorithm.

\subsection{CCS-AMP Architecture}
\label{subsection:Architecture}

The selection of a codeword by an active device follows the general structure obtained by combining the tree code of Amalladinne et al.~\cite{amalladinne2018couple} and the SPARC-like encoding of Fengler et al.~\cite{fengler2019sparcs}.
Specifically, consider a payload $\wv \in \{0, 1\}^w$.
Redundancy is first added to this message in the form of a tree code.
Explicitly, this $w$-bit binary message $\wv$ is partitioned into $L$ blocks, where the $\ell$th fragment contains $w_{\ell}$ information bits and $\sum_{\ell=1}^{L} w_{\ell} = w$.
In our treatment of tree codes, we frequently represent message $\wv$ as a concatenation of fragments,
\begin{equation} \label{equation:ConcatenatedFragments}
\wv = \wv(1) \wv(2) \cdots \wv(L) .
\end{equation}
The tree encoder appends $p_{\ell}$ parity check bits to fragment~$\wv(\ell)$, bringing the total length of block $\vv(\ell) = \wv(\ell) \pv(\ell)$ to $v_{\ell} = w_{\ell} + p_{\ell}$.
The structure of the message produced by the outer code assumes the form
\begin{equation}
\vv = \underbrace{\wv(1)}_{\vv(1)} \; \underbrace{\wv(2) \pv(2)}_{\vv(2)} \cdots \underbrace{\wv(L) \pv(L)}_{\vv(L)} . \label{vv}
\end{equation}
Equivalently, vector $\vv$ can be viewed as containing $L$ disjoint blocks: $\wv(1), \wv(2) \pv(2), \ldots, \wv(L) \pv(L)$.
For the sake of uniformity, we regard $\pv(1)$ as a parity segment of length zero, i.e., $p_1 = 0$.

The original tree code found in \cite{amalladinne2018couple,fengler2019sparcs} admits blocks containing a combination of information and parity-check bits akin to \eqref{vv}.
In our revised implementation, we focus on homogeneous blocks: every coded block features either information bits or parity-check bits, but not a combination of both.
Thus, we have either $w_{\ell}=0$ or $p_{\ell}=0$ for all ${\ell} \in [1:L]$.
As we will see shortly, this structure eliminates certain dependencies among blocks and it enables us to exploit a circular convolution structure propitious to the application of FFT methods during decoding.
Such a design naturally leads to a partition of $[1:L]$ into the set of information blocks $\mathcal{W}$ and the collection of parity blocks $\mathcal{P} = [1:L] \setminus \mathcal{W}$.
The parity-check bits contained in $\pv(\ell)$, $\ell \in \mathcal{P}$, act as constraints on the bits coming from other blocks.

As part of the next encoding step, every block is turned into a message index of length $m_{\ell} = 2^{v_{\ell}}$.
This action is emblematic of CCS and, therefore, it is worth going over carefully.
Conceptually, the message index is
\begin{equation} \label{equation:IndexFunction}
\begin{split}
\mv(\ell) &= f_{\mathbb{F}_2^{v_{\ell}} \rightarrow \{ 0, 1 \}^{m_{\ell}}} (\vv(\ell)) ,
\end{split}
\end{equation}
where the function $f_{\mathbb{F}_2^{v_{\ell}} \rightarrow \{ 0, 1 \}^{m_{\ell}}} (\vv (\ell))$ can be grasped by regarding argument $\vv(\ell)$ as an integer in binary form.
The output is a length-$2^{v_{\ell}}$ real vector with zeros everywhere, except for a one at location $[\vv(\ell)]_2$.
The shorthand notation $[\cdot]_2$ designates an integer expressed using a radix of 2 (possibly with leading zeros), and the location indexing of entries in $\mv(\ell)$ ranges from zero to $2^{v_{\ell}}-1$.
For example, if $\vv(\ell)=101$ then $v_{\ell} = 3$, $m_{\ell} = 8$, and $\mv(\ell) = 00000100$ because $[\vv(\ell)]_2 = 5$.
A message $\mv$ is subsequently created by concatenating individual sections,
\begin{equation} \label{equation:SPARCs}
\begin{split}
&\mv = \mv(1) \cdots \mv(L) \\
&= f_{\mathbb{F}_2^{v_1} \rightarrow \{ 0, 1 \}^{m_1}} (\vv (1)) \cdots
f_{\mathbb{F}_2^{v_L} \rightarrow \{ 0, 1 \}^{m_L}} (\vv (L)) .
\end{split}
\end{equation}
with every block being one-sparse.
The induced vector is reminiscent of a SPARC codeword~\cite{joseph2013fast, joseph2012least,CIT-092}.
The structure in \eqref{equation:SPARCs} is slightly more general than the form in~\cite{fengler2019sparcs} because it admits the possibility of having sections of different sizes.
The resemblance is nevertheless manifest.

Next, we describe how signal $\xv$ is generated from $\mv$.
Let $\Am$ be an $n \times m$ matrix over the real numbers, where $m = \sum_{\ell=1}^L m_{\ell}$, and let $\Dm$ be diagonal with non-negative entries.
Transmitted signals in $\mathcal{C}$ are created via the product $\xv = \Am \Dm \mv$ over the field of real numbers.
Matrix $\Dm$ accounts for the power allocated to every section and, accordingly, diagonal entries are constant within each block.
The amplitude of the signal for section~$\ell$ is denoted by $d_{\ell}$.
Given that all active devices utilize a same codebook, this process yields a received vector $\yv$ of the form
\begin{equation} \label{equation:SPARClike}
\yv = \sum_{i=1}^{\Ka} \mathbf{A} \Dm \mv_i + \zv
= \Am \Dm \underbrace{\left( \sum_{i=1}^{\Ka} \mv_i \right)}_{\sv} + \zv
= \Am\Dm \sv + \zv
\end{equation}
where $\sv$ is such that all its sections, $\sv(\ell) = \sum_{i=1}^{\Ka} \mv_i (\ell)$ with $\ell \in [1:L]$, are $\Ka$ sparse.\footnote{Every section is at most $\Ka$ sparse; improbable index collisions may occasionally reduce the number of non-zero entries.}
Matrix $\Am$ is normalized in that the 2-norm of every column is one.
The interpretation of \eqref{equation:SPARClike} as a SPARC-like model, originally put forth in~\cite{fengler2019sparcs}, immediately extends to the present case. 
Moreover, as in \cite{fengler2019sparcs}, the resulting multiple access channel can be viewed as the combination of a point-to-point channel $\sv \rightarrow \Am \Dm \sv + \zv$ and an outer binary adder MAC $\sv = \sum_{i=1}^{\Ka} \mv_i$.
Therein, the authors refer to these components as the \emph{inner} and \emph{outer channels}, respectively.
They also draw a distinction between the \emph{inner} and \emph{outer encoder/decoder} pairs.

Our article embraces the aforementioned categorization for the channel components, \emph{inner} and \emph{outer channels}.
However, we do not subscribe to the latter dissociation between the decoders.
Rather, we seek to exploit the fact that decoding can be improved if information is allowed to flow dynamically between the inner and outer channel components while iterative decoding takes place.
As mentioned above, the impetus behind our approach stems from a potential algorithmic enhancement that was first noticed in the context of coded compressed sensing~\cite{amalladinne2020enhanced}.
In this simpler scenario, the authors show how decoded blocks in earlier stages can inform the CS recovery process at subsequent stages through tracking sets of admissible parity patterns.
Indeed, the collection of active paths in the tree decoder at a particular stage determines the set of possible parity patterns at the subsequent stage.
This information can be leveraged to reduce the difficulty of the support recovery task at the later stages and concurrently improve performance.
Additional details regarding this algorithmic improvement for CCS can be found in~\cite{amalladinne2020enhanced}.

The relationship between parity bits in the outer tree code and support recovery in the inner channel is more subtle in the present context, which we call CCS-AMP.
There are two architectural aspects that can help guide the convergence of our iterative decoding process.
The inherent block sparsity contained in $\sv$ provides a foundation for the AMP denoiser~\cite{fengler2019sparcs}.
Moreover, there is an embedded factor graph structure in the tree code that, when designed carefully, is amenable to belief propagation.
Thus, within each iteration of the AMP algorithm, the state estimates can be updated via message passing on the factor graph induced by the tree code.
Under this novel approach, as the ambiguity on some sections diminishes, the uncertainty on the belief states of their (graph) neighbors also decreases.
Incorporating these two modalities within the iterative decoding process offers significant performance benefits, beyond the concentration afforded by the AMP algorithm alone.
We will see shortly how these pieces of information can be integrated into a consolidated iterative scheme.
Admittedly, in the actual decoding process, the interactions between the tree decoder and the AMP algorithm are more complex than described above.
Nevertheless, this simplified characterization strongly hints at an opportunity to improve performance by running AMP and the tree decoder in tandem, dynamically passing information back and forth between these two components.
The specifics of our implementation and how it facilitates such dynamic exchanges of information are contained in the upcoming sections.

\subsection{Notation}

Notation in this article is heavy despite our best effort to keep it to a minimum.
We therefore offer a brief overview of the notation we adopt, along with a table summary, for the sake of readability.
Table~\ref{table:Representations} is useful because we frequently transition between equivalent representations to explain concepts.
A tree codeword is a sequence of bits of the form $\vv = \vv(1) \cdots \vv(L)$.
The number of bits in $\vv(\ell)$ is $v_{\ell}$.
The value $k_{\ell}$ of binary block $\vv(\ell)$ is obtained by regarding the payload as an integer expressed in radix-2, which we write as $k_{\ell} = \left[ \vv(\ell) \right]_2$.
The block index for $\vv(\ell)$ is a standard basis element with a one at location~$k_{\ell}$ and zeros everywhere else, i.e., the entries of $\mv(\ell)$ are given by $\mv(\ell,k) = \delta_{k_{\ell}}(k)$ where $\delta_{k_{\ell}}(k) = 1$ if $k= k_{\ell}$ and $\delta_{k_{\ell}}(k) = 0$ otherwise.
Sparse vector $\mv$ is obtained by concatenating the index messages, $\mv = \mv(1) \cdots \mv(L)$.
The integer vector representation of this same message is $\kv = \left( k_1, \ldots, k_L \right)$.
The message produced by device~$i$ is annotated with either a subscript, $\vv_i$ or $\mv_i$, or a superscript $\kv^{(i)}$.
The aggregate signal is equal to $\sv = \sum_{i = 1}^{\Ka} \mv_i$.
There is no equivalent representation for $\sv$ in compact form or integer form.
Consequently, we resort to collections wherever needed.
We define sections of $\sv$ through the sum $\sv(\ell) = \sum_{i = 1}^{\Ka} \mv_i(\ell)$ and we employ $\sv(\ell,k)$ to refer to the $k$th entry of its $\ell$th section.
We extend this convention to all the vectors that admit a fragmented representation.

\begin{table}[tbh]
    \centering
    \begin{tabular}{|c||c|c|c|}
    \hline
    & \multicolumn{3}{c|}{\textbf{Representations}} \tabularnewline
    \hline
    \textbf{Objects} & Compact & Integer & Sparse Vector \tabularnewline
    \hline
    Block & $\vv(\ell) \in \{0,1\}^{v_{\ell}}$ & $k_{\ell} = \left[ \vv(\ell) \right]_2$ & $\mv(\ell) : \mv(\ell,k) = \delta_{k_{\ell}}(k)$ \tabularnewline
    \hline
    Message & $\vv = \vv(1) \cdots \vv(L)$ & $\kv = \left( k_1, \ldots, k_L \right)$ & $\mv = \mv(1) \cdots \mv(L)$ \tabularnewline
    \hline
    Signal & $\left\{ \vv_i : i \in [\Ka] \right\}$ & $\left\{ \kv^{(i)} : i \in [\Ka] \right\}$ & $\sv = \sum_i \mv_i$ \tabularnewline
    \hline
    \end{tabular}
    \caption{Summary of various representations.}
    \label{table:Representations}
\end{table}

\section{Tree Code Revisited}
\label{section:TreeCodeRevisited}

This section focuses on the \emph{outer code}, which takes the form of a modified tree code.
An important distinction between the original CCS framework and CCS-AMP from a tree code perspective stems from the fact that the former applies consistency checks sequentially to short lists on the order of $\Ka$ items, whereas CCS-AMP produces large belief vectors where complete blocks are processed to update estimates.
To accommodate this new reality, the \emph{outer code} employed in this article differs from the original tree code introduced in~\cite{amalladinne2018couple,amalladinne2019coded}.
Below, we describe the revised encoding process for our system and how it deviates from the original tree code incarnation.
We also introduce a soft decoder for our tree code suitable for dynamic interactions with the \emph{inner code} and AMP.

\subsection{Alternate Tree Encoding}
\label{subsection:AlternateTreeEncoding}
In a manner akin to the original CCS scheme, the construction of our tree code starts by partitioning information bits into fragments.
Parity patterns are then added to blocks in a causal fashion, leading to vector $\vv$.
To this extent, the alternate tree code subscribes to the same structure as the original one found in~\cite{amalladinne2018couple}.
However, in our revised construction, parity bits are created differently.
For $\ell \in \mathcal{P}$, we use $\mathcal{W}_{\ell}$ to denote the collection of blocks on which parity block $\vv(\ell)$ operates.
Every parity block is obtained using the following three-step sequence.
We first take a random linear combination of the bits in each fragment of $\mathcal{W}_{\ell}$; specifically, $\vv(j) \mathbf{G}_{j,\ell}$ for $j \in \mathcal{W}_{\ell}$, where $\mathbf{G}_{j,\ell}$ is selected at random from $\{0,1\}^{v_j \times v_{\ell}}$.
Within this step, vector operations are taken over Galois field $\mathbb{F}_2$.
We transpose these combinations from the space of length-$v_{\ell}$ binary vectors over $\mathbb{F}_2$ to the ring of integers modulo-$2^{v_{\ell}}$, which we denote by $\mathbb{Z}/2^{v_{\ell}}\mathbb{Z}$.
We then add the resulting elements of $\mathbb{Z}/2^{v_{\ell}}\mathbb{Z}$, which we call parity precursors, using modulo-$2^{v_{\ell}}$ arithmetic.
Finally, we convert the ensuing sum back to a vector in $\mathbb{F}_2^{v_{\ell}}$.
The sequence of bits obtained through this process determines parity block $\vv(\ell)$.
Mathematically, these operations can be expressed as
\begin{equation} \label{equation:ParityGeneration}
\vv(\ell)
\equiv \sum_{j \in \mathcal{W}_{\ell}}
\left[ \vv(j) \mathbf{G}_{j,\ell} \right]_{\mathbb{Z}/2^{v_{\ell}}\mathbb{Z}} .
\end{equation}
The notation $[ \cdot ]_{\mathbb{Z}/2^{v_{\ell}}\mathbb{Z}}$ emphasizes that the argument is interpreted as an element of the quotient ring $\mathbb{Z}/2^{v_{\ell}}\mathbb{Z}$, and the equivalence relation `$\equiv$' denotes equality in $\mathbb{Z}/2^{v_{\ell}}\mathbb{Z}$.
Conceptually, $\vv(\ell)$ contains non-linear constraints on the information bits from fragments in $\mathcal{W}_{\ell}$.
It is worth mentioning that the parity encoding process in \eqref{equation:ParityGeneration} is more contrived than the random linear combinations utilized in the original tree code~\cite{amalladinne2018couple}.
The value of this intricate encoding process is that it
induces a cyclic structure on parity precursors $\left( \left[ \vv(j) \mathbf{G}_{j,\ell} \right]_{\mathbb{Z}/2^{v_{\ell}}\mathbb{Z}}  : j \in \mathcal{W}_{\ell} \right)$ conducive to circular convolution, which is befitting to the eventual application of FFT techniques.\footnote{This strategy allows us to employ fast transform techniques to dynamically pass information between outer and inner decoders.
 The reader may notice that these binary vectors can be mapped to elements of the finite field $\mathbf{GF}({2^{v_\ell}})$ instead of elements of the quotient ring $\mathbb{Z}/2^{v_{\ell}}\mathbb{Z}$.
 These two representations exhibit similar structural properties under the proposed framework, and we choose to adhere to the ring notation throughout this article.}
The advantages of this construction will become clear shortly.

We recall that the tree code construction admits a factor graph representation.
Our upcoming discussion relies heavily on this abstraction and, as such, we elaborate on this analogy.
We denote the variable nodes by $\left\{ s_{\ell} : \ell \in [L] \right\}$, and we label the factors using $\left\{ a_{p} : p \in \mathcal{P} \right\}$.
By definition, an edge necessarily exists between $a_{p}$ and $s_{p}$.
In addition, an edge is placed between $s_j$ and $a_p$ whenever $j \in \mathcal{W}_p$.
The factor associated with $a_p$, where $p \in \mathcal{P}$, is derived from the computation of the corresponding parity block and local observations.
Given $p \in \mathcal{P}$, we emphasize that candidate sections $\left( \hat{\vv}(j) : j \in \mathcal{W}_p \cup p \right)$ can only come from the same message if they are collectively parity consistent.
We introduce indicator function $\mathcal{G}_{a_p} (\cdot)$ to assess local consistency,
\begin{equation} \label{equation:ParityConsistency}
\mathcal{G}_{a_p} \left( \hat{\vv}(j) : j \in \mathcal{W}_p \cup p \right)
= \mathbf{1} \left( \sum_{j \in \mathcal{W}_p}
\left[ \hat{\vv}(j) \mathbf{G}_{j,p} \right]_{\mathbb{Z}/2^{v_p}\mathbb{Z}}
\equiv \hat{\vv}(p) \right) .
\end{equation}
This function returns one when the block indices in its argument are parity consistent, and it outputs zero otherwise.
As a side note, we adopt an overloaded notation for factor $\mathcal{G}_{a_p} (\cdot)$ in that we employ the same notation for all the message representations listed in Table~\ref{table:Representations} although, technically, these functions have different domains.
Since the overloaded functions have equivalent meanings, this should not lead to confusion yet it simplifies exposition.

\begin{figure}[tbh]
  \centering
  \input{Figures/graph1}
  \caption{
  In the factor graph interpretation of the tree code, every block yields a variable node, $\left\{ s_{\ell} : \ell \in [L] \right\}$, and every parity equation produces a factor, $\left\{ a_{p} : p \in \mathcal{P} \right\}$.}
  \label{figure:subvectors1}
\end{figure}
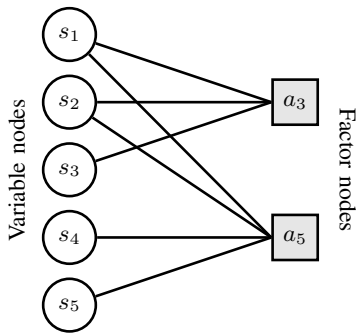

Using factor graph notation, the neighbors of factor $a_{p}$, written $N ( a_{p} )$, are the elements of $\mathcal{W}_{p} \cup \{ p \}$.
Moreover, the graph neighbors of variable $s_{\ell}$, written $N ( s_{\ell} )$, are the factors associated with the parity equations where $\vv(\ell)$ appears either as a summand or as the parity. 
We offer an elementary example to help cement these notions.
Consider a tree code where $\vv(1), \vv(2), \vv(4)$ are information blocks, and parity blocks are given by
\begin{align}
\vv(3) &\equiv \big[ \vv(1) \mathbf{G}_{1,3} \big] + \big[ \vv(2) \mathbf{G}_{2,3} \big] \\
\vv(5) &\equiv \big[ \vv(1) \mathbf{G}_{1,5} \big] + \big[ \vv(2) \mathbf{G}_{2,5} \big]
+ \big[ \vv(4) \mathbf{G}_{4,5} \big]
\end{align}
Then, $N(s_1) = N(s_2) = \{ a_3, a_5 \}$, $N(s_3) = \{ a_3 \}$, and $N(s_4) = N(s_5) = \{ a_5 \}$;
likewise, $N(a_3) = \{ s_1, s_2, s_3 \}$ and $N(a_5) = \{ s_1, s_2, s_4, s_5 \}$.
The ensuing factor graph, which captures parity consistencies among blocks, appears in Fig.~\ref{figure:subvectors1}.

\subsection{Validation of Candidate Codewords}
\label{subsection:InvalidCodewords}

As mentioned above, all active devices use a common encoding scheme and, based on their own messages, they collectively produce vectors $\vv_1, \ldots, \vv_{\Ka}$.
The bijection between $\vv$ and $\mv$ defined in \eqref{equation:SPARCs} naturally introduces a correspondence between $\sv (\ell)$ and subset $\left\{ \vv_1 (\ell), \ldots, \vv_{\Ka} (\ell) \right\}$.
Specifically, $\vv(\ell)$ is contained in the latter set whenever $\sv (\ell)$ has a value of one (or greater) at index location $\left[ \vv (\ell) \right]_2$.
We recall that the original tree decoder (with hard decisions) operates on $\sv$ and it is tasked with recovering the collection $\left\{ \vv_1, \dots, \vv_{\Ka} \right\}$.
That is, it must stitch the segments of valid codewords together.
At the onset of the process, the decoder extracts lists of blocks from $\sv$, one for every level~$\ell$.
Tentative codewords are formed by concatenating one fragment from each list, abiding by the natural progression from level~$1$ to level~$L$.
Given one such candidate codeword, written as
\begin{equation}
\vv_{\mathrm{c}} = \wv_{i_1}(1) \wv_{i_2}(2) \pv_{i_2}(2) \cdots \wv_{i_L}(L) \pv_{i_L}(L) ,
\end{equation}
the decoder attempts to validate its structure by recreating its parity patterns.
For instance, it begins with information fragment $\wv_{i_1}(1)$ and determines parity pattern $\pv(2)$ using \eqref{equation:ParityGeneration}.
If the resulting parity pattern matches $\pv_{i_2}(2)$, the decoder proceeds forward to the next block; else candidate codeword $\vv_{\mathrm{c}}$ is marked as invalid and it is discarded immediately.
If $\vv_{\mathrm{c}}$ gets through stage~$(\ell-1)$, then the tree decoder continues by constructing $\pv(\ell)$ using fragments $\wv_{i_1}(1), \ldots, \wv_{i_{\ell-1}}(\ell-1)$ and \eqref{equation:ParityGeneration}.
Once again, the candidate codeword is retained when $\pv(\ell)$ matches $\pv_{i_{\ell}}(\ell)$; else it is dropped.
Altogether, $\vv_{\mathrm{c}}$ remains a codeword candidate for as long as parity blocks are consistent with the information fragments that precede them.
To justify this procedure, it is relevant to recognize that $\pv(\ell)$ is computed using fragments $\wv_{i_1}(1), \ldots, \wv_{i_{\ell-1}}(\ell-1)$; whereas $\pv_{i_{\ell}}(\ell)$, coming from a valid codeword, was created by applying $\eqref{equation:ParityGeneration}$ to $\wv_{i_{\ell}}(1), \ldots, \wv_{i_{\ell}}(\ell-1)$.
Thus, the ability of the tree decoder to detect invalid codewords is predicated on the probability that parity patterns generated through mismatched fragments $\wv_{i_1}(1), \wv_{i_2}(2), \ldots  \wv_{i_L}(L)$ correspond to the embedded parity patterns $\pv_{i_2}(2), \ldots, \pv_{i_{L}}(L)$.
We stress that valid codewords whose fragments are on the lists are never discarded through tree decoding because their parity patterns are necessarily self-consistent.

A detailed characterization of tree decoding (with hard decisions) and its expected performance can be found in~\cite{amalladinne2019coded} when parity bits are generated through random linear combinations, with the entries in $\mathbf{G}_{j,\ell} \in \{0,1\}^{w_j \times p_{\ell}}$ being independent Rademacher trials (Bernoulli trials with parameter half).
The analysis therein relies on three key properties, which we enumerate below.
Consider erroneous pseudo-codeword
\begin{equation}
\vv_{\mathrm{e}} = \wv_{i_1}(1) \wv_{i_2}(2) \pv_{i_2}(2) \cdots \wv_{i_L}(L) \pv_{i_L}(L) ,
\end{equation}
where indices $i_1, i_2, \ldots, i_L$ are not all from a same encoded message.
Given that $\vv_{\mathrm{e}}$ is invalid, the tree decoder is tasked with detecting and discarding it.

\begin{remark} \label{remark:AlternateTreeCode}
In examining the probability that a tree decoder is successful in this singular endeavor and to assess average computational complexity, the following facts come into play.
\begin{enumerate}
    \item The collection $\pv_{\mathrm{e}}(\ell)$ of parity bits is either statistically discriminating or, as a block, it is uninformative.
    In the former case, the probability that $\pv_{i_{\ell}}(\ell)$ is consistent with pseudo-codeword $\vv_{\mathrm{e}}$ is equal to $2^{-p_{\ell}}$.
    \item The list of statistically discriminating parity blocks within $\pv_{i_2}(2), \ldots, \pv_{i_L}(L)$ depends on the index sequence $i_1, \ldots, i_L$, with a re-entry to a previously visited level reducing the probability that the corresponding parity block remains statistically discriminating.
    For example, if $i_3 = i_1$, then we say that the sequence has re-entered level $i_1$ during stage~3.
    \item The conditional distribution on the collection of statistically discriminating blocks, given $i_1, \ldots, i_L$, is permutation invariant.
    That is, the precise state labeling $i_1, \ldots, i_L$ is unimportant; only the order in which previously visited states are re-entered matters (over the random ensemble of generating matrices).
    Mathematically, if $\pi(\cdot)$ is a permutation function on admissible indices, the probability that the tree decoder tags $\vv_{\mathrm{e}}$ as invalid is the same as the probability that the decoder recognizes
    \begin{equation}
    \vv_{\mathrm{e}_{\pi}} = \wv_{\pi(i_1)}(1) \wv_{\pi(i_2)}(2) \pv_{\pi(i_2)}(2) \cdots \wv_{\pi(i_L)}(L) \pv_{\pi(i_L)}(L)
    \end{equation}
    as an erroneous codeword.
\end{enumerate}
\end{remark}


An important aspect of the novel encoding scheme introduced in \eqref{equation:ParityGeneration} is the fact that it preserves the three properties listed above whenever the entries in $\mathbf{G}_{j,\ell} \in \{0,1\}^{v_j \times v_{\ell}}$ are independent Rademacher trials.
Because the performance characterization of tree codes is based on expected behavior, the fact that the conditional probability of an erroneous codeword being detected remains the same, conditioned on index sequence $i_1, \ldots, i_L$, is enough to ensure that findings derived for the original tree code extend to the new encoding process.
In other words, the performance results presented in our previous work on tree codes are also valid in the current context.
The third item enumerated above guarantees that the complexity reduction exposed in~\cite{amalladinne2019coded}, and related to the Bell numbers, is present under the new encoding as well.
This connection is pertinent because it offers performance guarantees for detection applied to the revised tree code, under certain conditions, thereby pointing to the practicality of our alternate approach.
Also, the last step of the overall CCS-AMP decoding process proposed in this work is essentially (hard) tree decoding.
We formalize these findings below.

\begin{theorem} \label{theorem:AlternateTreeCode}
Under hard decision decoding, the expected performance of the revised tree code based on \eqref{equation:ParityGeneration} is equal to that of the original tree code found in \cite{amalladinne2019coded}, provided that the elements of $\{ \mathbf{G}_{j,\ell} \}$ are i.i.d.\ Rademacher trials.
The complexity of the decoding process in both settings is analogous, with the respective tree decoders validating candidate codewords by checking the consistency of parity patterns sequentially, starting from root fragments.
\end{theorem}
\begin{IEEEproof}
To relate the expected performance of the alternate tree code to that of the original tree code, it suffices to check properties~1 and~2 in Remark~\ref{remark:AlternateTreeCode}.
The complexity reduction statement relies on the third property therein.
These three properties and, hence, Theorem~\ref{theorem:AlternateTreeCode} are established in Appendix~\ref{Section:PerformanceTreeCodeAppendix}.
\end{IEEEproof}

\subsection{Soft Tree Decoding}
\label{subsection:TreeDecoding}

This section focuses on the soft decoding of the \emph{outer tree code}, which can be employed in tandem with the iterative decoding of the \emph{inner code}.
The encoding process and, specifically, the parity generation defined in \eqref{equation:ParityGeneration} induces a generalized Markov structure on the tree code.
The AMP composite iterative algorithm, as we will see, utilizes two steps: the computation of a residual, and an update of the state estimate based on an effective observation.
In its original form~\cite{fengler2019sparcs}, the denoising step leverages solely the sparse structure of $\sv$.
Yet, the tree code imposes parity consistency conditions, beyond sparsity; the corresponding graphical structure can too inform the denoising step.
To describe this algorithmic opportunity, it is useful to think of the soft tree decoder as getting an observation $\rv$ of the form
\begin{equation} \label{equation:BP-EffectiveObservation}
\rv = \mathbf{D} \sv + \tau \zetav
\end{equation}
where $\zetav$ is an i.i.d.\ $\mathcal{N}(0,1)$ random vector and $\tau$ is a scaling parameter that captures the standard deviation.
Based on the underlying factor graph inherited from the tree code, this sub-component of the denoiser seeks to produce estimates for the elements of $\sv$ using iterative message passing.
Naively, the construction of $\sv$ in \eqref{equation:SPARClike} imposes a sparsity constraint within each block, whereas the generalized Markov structure described above captures dependencies between information and parity bits across blocks.
In our proposed framework, sparsity is dictated primarily through the AMP iteration; whereas parity factors are leveraged within the denoiser via belief propagation.
As a side note, we emphasize that the complexity of implementing an optimal state estimator for $\sv$ based on effective observation $\rv$ is often cost prohibitive for realistic designs.
Nevertheless, it is possible to focus on local aspects of the factor graph associated with the outer code, computing beliefs using suitable message passing rules, extrinsic information, and applying factor functions derived from the parity generation mechanism of \eqref{equation:ParityGeneration}.

The message passing rules we utilize for the outer factor graph are presented below.
We expound on the rationale behind them in Appendix~\ref{appendix:MessagePassingRulesGraph}.
The factor function associated with check nodes admits a product decomposition and it is given by
\begin{equation} \label{equation:BP-FactorFunctions}
\begin{split}
\mathcal{G} \left( \kv \right)
= \prod_{a_p \in \mathcal{P}}
\mathcal{G}_{a_p} \left( \kv_{a_p} \right)
\end{split}
\end{equation}
where $\kv = \left( k_{\ell} : \ell \in [L] \right)$, $\kv_{a} = \left( k_{\ell} : \ell \in N(a) \right)$, and $k_{\ell} = \left[ \hat{\vv}(\ell) \right]_2$ with the function $\mathcal{G}_{a_p}$ defined in \eqref{equation:ParityConsistency}.
Equivalently, $k_{\ell}$ can be interpreted as the index of the one in $\hat{\mv}(\ell)$.
In words, function $\mathcal{G} \left( \kv \right)$ verifies the parity consistency of its vector argument $\kv$ under the tree code parity structure, as defined in \eqref{equation:ParityGeneration}.

\begin{figure}[tbh]
  \centering
  \input{Figures/graph2}
  \caption{
  This illustration shows the augmented factor graph for the tree code with variable nodes, parity check constraints, and extra factors associated with local observations.
  Local factor messages $\left\{ \lambdav_{\ell} \right\}$ appear for mathematical convenience.
  They do not change during the belief propagation process when $\rv$ is fixed.}
  \label{figure:FactorGraph2}
\end{figure}
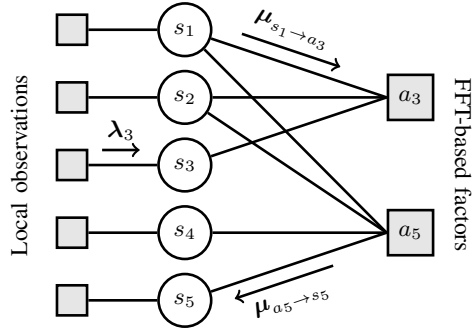

Having specified factor functions in \eqref{equation:BP-FactorFunctions}, we can write standard expressions for the message passing rules.
A message passed from check node $a_p$ to variable node $s \in N(a_p)$ has the form
\begin{equation} \label{equation:BP-Check2Variable}
\muv_{a_p \to s} (k)
= \sum_{\kv_{a_p}: k_p = k} \mathcal{G}_{a_p} \left( \kv_{a_p} \right)
\prod_{s_j \in N(a_p) \setminus s} \muv_{s_j \to a_p} (k_j) .
\end{equation}
Similarly, a message going from variable node $s_{\ell}$ to check node $a \in N(s)$ abides by
\begin{equation} \label{equation:BP-Variable2Check}
\muv_{s_{\ell} \rightarrow a} (k)
\propto \lambdav_{\ell} (k) \prod_{a_p \in N(s_{\ell}) \setminus a} \muv_{a_p \to s_{\ell}} (k) .
\end{equation}
The `$\propto$' symbol indicates that the measure is normalized before being sent out as a message.
Block vector $\lambdav_{\ell}$ in \eqref{equation:BP-Variable2Check} can be interpreted as a collection of local estimates, where entry $\lambdav_{\ell}(k)$ represents the estimate that a designated device has sent integer fragment~$k$ within section~$\ell$ as part of its message.
We have some flexibility in selecting a suitable estimator for each such component, but this value should be calculated solely based on the intrinsic information afforded by $\rv(\ell)$, as is customary in the derivation of message passing rules.
We delay the treatment of $\lambdav_{\ell}$ until Section~\ref{section:AMP}.
Still, it may be helpful to note that its components are closely linked to the likelihood ratio for binary classification in additive Gaussian noise, namely
\begin{equation} \label{equation:BP-LikelihoodRatio}
\mathcal{L}_{\ell} (k)
= \exp \left({- \frac{\left( \rv(\ell, k) - d_{\ell} \right)^2 - \rv(\ell, k)^2}{2 \tau^2}}\right)
= \exp \left({\frac{2 d_{\ell} \rv(\ell, k) - d_{\ell}^2}{2 \tau^2}}\right) .
\end{equation}
Parameter $\tau$ in \eqref{equation:BP-LikelihoodRatio} corresponds to the standard deviation of the noise component in the effective observation of \eqref{equation:BP-EffectiveObservation}, where $\rv(\ell) = d_{\ell} \sv(\ell) + \tau \zetav(\ell)$ and recall that $d_{\ell}$ is the amplitude of the signal for section $\ell$.
Following established notation, $\rv(\ell)$ denotes the $\ell$th section of $\rv$ and element $\rv(\ell,k)$ refers to the $k$th entry of the $\ell$th section of $\rv$.
The message passing rules are depicted in Fig.~\ref{figure:FactorGraph2}.
All the dynamic messages are initialized with $\muv_{s \to a} = \onev$ and $\muv_{a \to s} = \onev$.
The parallel sum-product algorithm then iterates between \eqref{equation:BP-Check2Variable} and \eqref{equation:BP-Variable2Check}.


At any stage of this iterative process, the belief vector on section~$\ell$ based on extrinsic information is proportional to the product of the messages from adjoining parity factors.
That is, it is proportional to
\begin{equation} \label{equation:EquivalentPriors}
\muv_{s_{\ell}}(k) = \prod_{a \in N(s_{\ell})} \muv_{a \to s_{\ell}} (k) .
\end{equation}
Likewise, the estimated marginal distribution of a specific device having transmitted index~$k$ at variable node~$s_{\ell}$ is proportional to the product of the current messages from all adjoining factors, including the intrinsic information,
\begin{equation} \label{equation:BP-TildeM}
\begin{split}
p_{s_{\ell}} (k) &\propto \lambdav_{\ell} (k) \prod_{a \in N(s_{\ell})} \muv_{a \to s_{\ell}} (k)
= \lambdav_{\ell} (k) \muv_{s_{\ell}} (k) .
\end{split}
\end{equation}
That is, a normalized version of $\lambdav_{\ell} (k) \muv_{s_{\ell}} (k)$ can be viewed as an estimate for the event $\left\{ \mv_i(\ell,k) = 1 \right\}$, where $i$ is fixed.
The expected value state vector $\sv$ is given component-wise by
\begin{equation} \label{equation:BP-TildeS}
\begin{split}
\hat{\sv}(\ell,k) &= \mathbb{E} \left[ \sv(\ell,k) \middle| \rv \right]
= \mathbb{E} \left[ \sum_{i \in [\Ka]} \mv_i(\ell,k) \middle| \rv \right] \\
&= \sum_{i \in [\Ka]} \mathbb{E} \left[ \mv_i(\ell,k) \middle| \rv \right]
\approx \Ka p_{s_{\ell}} (k) .
\end{split}
\end{equation}
This is precisely the information needed to provide estimates for the support of $\sv$ to the AMP denoiser.
In general, such an iterative procedure is guaranteed to converge for acyclic graphical models, but not for arbitrary graphs~\cite{kschischang2001factor}.
Nevertheless, it is known to perform well in many cases where factor graphs have cycles.
Having said that, it is intuitively appealing to construct outer codes whose factor graphs do not contain short cycles.
In the framework we envision, the denoiser performs only one (or a select few) composite steps of the belief propagation algorithm before returning the updated state estimate $\hat{\sv}$ back to the AMP algorithm, which subsequently seeks to improve the effective observation $\rv$.
We will come back to this particular point in Section~\ref{section:AMP}, where the reason for halting belief propagation early can be explained adequately.


\subsection{Design Considerations for Fast Execution}

A naive implementation of \eqref{equation:BP-Check2Variable} would have the denoiser parse through $\prod_{j : s_j \in N(a)} v_j$ distinct paths to compute message $\muv_{a \to s}$, an approach which is intractable for the parameters of interest.
Our goal, then, is to exploit the structure of the revised tree code to get a low-complexity solution.
This is where the cyclic aspect of the parity precursors in \eqref{equation:ParityGeneration} comes into play.
Let $k = \left[\hat{\vv}(\ell)\right]_2$ and define $g = [\hat{\vv}(\ell) \mathbf{G}_{\ell,p}]_{\mathbb{Z}/2^{v_p}\mathbb{Z}}$, then we can write
\begin{equation} \label{equation:BP-CircConv1}
\begin{split}
\muv_{a_p \to s_{\ell}} (k)
&\propto \sum_{\kv_{a_p}: k_p = k} \mathcal{G}_{a_p} \left( \kv_{a_p} \right)
\prod_{s_j \in N(a_p) \setminus s_{\ell}} \muv_{s_j \to a_p} (k_j) \\
&= \underbrace{\sum_{\substack{\gv_{a_p} : g_{\ell} = g \\
\sum g_j \equiv 0}} \prod_{s_j \in N(a_p) \setminus s_{\ell}}
\left( \sum_{\left[ \hat{\vv}(j) \mathbf{G}_{j,p} \right]_{\mathbb{Z}/2^{v_p}\mathbb{Z}} \equiv g_j}
\muv_{s_j \to a_p} \left( \left[\hat{\vv}(j)\right]_2 \right) \right)}_{\text{circular convolution structure}} ,
\end{split}
\end{equation}
where we have implicitly introduced matrix $\mathbf{G}_{p,p} = - \mathbf{I}$ for the sake of exposition.
This definition transforms parity check equation \eqref{equation:ParityGeneration} into the symmetric form
\begin{equation*}
\sum_{j : s_j \in N(a_p)} \left[ \vv(j) \mathbf{G}_{j,p} \right]_{\mathbb{Z}/2^{v_{p}}\mathbb{Z}} \equiv 0 .
\end{equation*}
The circular discrete convolution structure identified above suggests the application of the discrete Fourier transform and related techniques.
In addition, since the underlying period is $2^{v_p}$ (a factor of two), these operations can be performed with the FFT algorithm.
Thus, through the structure of the parity patterns produced by \eqref{equation:ParityGeneration}, the computation of messages becomes manageable under \eqref{equation:BP-CircConv1}, even for large values of $2^{v_p}$.
This fact alone forms the impetus behind the development of an alternate tree code and the adoption of its more intricate parity generation process in Section~\ref{subsection:AlternateTreeEncoding}.

Pragmatically, message $\muv_{a_p \to s_{\ell}}$, where $p \in \mathcal{P}$, can be computed as follows.
For every $j$ such that $s_j \in N(a_p)$, a collection of static binary vectors $\big\{ \gv_{j,p}^{(g)} \in \{0, 1\}^{m_j} \big\}$ is maintained, one vector for every $g \in \mathbb{Z}/2^{v_p}\mathbb{Z}$.
Vector $\gv_{j,p}^{(g)}$ features a one at every index location where $\hat{\vv}(j)$ is such that $\left[ \hat{\vv}(j) \mathbf{G}_{j,p} \right]_{\mathbb{Z}/2^{v_p}\mathbb{Z}} \equiv g$, and zeros everywhere else.
Given factor $p \in \mathcal{P}$, this vector marks all the locations associated with parity precursor $g$ at level~$j$.
Through this partitioning, the aggregate weight of this group becomes
\begin{equation} \label{equation:SvComponents}
\begin{split}
\Lv_{j,p} (g) &=
\sum_{ \left[ \hat{\vv}(j) \mathbf{G}_{j.p} \right]_{\mathbb{Z}/2^{v_p}\mathbb{Z}} \equiv g}
\muv_{s_j \to a_p} \left( \left[ \hat{\vv}(j) \right]_2 \right) \\
&= \left\langle \muv_{s_j \to a_p}, \gv_{j, p}^{(g)} \right\rangle .
\end{split}
\end{equation}
When ordered and stacked, the values in \eqref{equation:SvComponents} yield a vector $\Lv_{j,p} \in \mathbb{R}^{2^{v_p}}$.
Given the circular convolution structure identified above, message $\muv_{a_p \to s_{\ell}}$ can be computed at once as
\begin{equation} \label{equation:BlockFFT}
\muv_{a_p \to s_{\ell}} \left( \left[ \hat{\vv}(\ell) \right]_2 \right)
\propto
\frac{1}{\left\| \gv_{\ell, p}^{(g)} \right\|_0}
\left( \operatorname{FFT}^{-1} \left( \prod_{s_j \in N(a_p) \setminus s_{\ell}} \operatorname{FFT} \left( \Lv_{j,p} \right) \right) \right) (g)
\end{equation}
where $\left[ \hat{\vv}(\ell) \mathbf{G}_{\ell,p} \right]_{\mathbb{Z}/2^{v_p}\mathbb{Z}} \equiv g$.
The leading coefficient in \eqref{equation:BlockFFT} accounts for the fact that the weight of a certain parity precursor $g$ within section~$\ell$, based on the information afforded by an adjoining factor, is evenly distributed among entries that map to $g$.
This highlights the need to avoid situations where several entries in $p_{s_{\ell}}$ map to a same $g \in \mathbb{Z}/2^{v_p}\mathbb{Z}$.
Again, the `$\propto$' symbol accounts for the normalization of the belief vector.

\subsubsection{Decoding with Extended Lists}

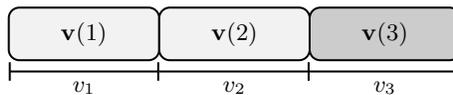
\begin{figure}[tbh]
  \centering
  \input{Figures/subvectors2_new}
  \caption{An information and parity allocation that is conducive to the application of FFT-based techniques appears above.
  Blocks are homogeneously composed of information or parity bits, but not both.
  The shaded component denotes a parity block.}
  \label{figure:subvectors2}
\end{figure}

In the hard version of tree decoding, lists at various stages contain $\Ka$ entries (or slightly more), and paths moving forward are pruned aggressively.
The tools described herein and, specifically, \eqref{equation:BlockFFT} enable the propagation of soft estimates over much larger collections.
This is key in creating a soft denoising that accounts for the underlying tree code.
Also, we see how having both information and parity bits within a same block is not conducive to the effective application of \eqref{equation:BlockFFT} because of weight splitting.
Such a situation would reduce the cardinality of $\mathbb{Z}/2^{v_p}\mathbb{Z}$, thereby reducing the number of masks $\big\{ \gv_{j, p}^{(g)} : g \in \mathbb{Z}/2^{v_p}\mathbb{Z} \big\}$ and, coincidentally, lowering resolution.
This justifies the approach we adopted in Section~\ref{subsection:Architecture}: a codeword $\vv$ contains homogeneous blocks of information bits, with sections of parity bits interspersed in-between.
In view of these insights, we examine systems where sections of information bits are followed by a section of parity bits, as depicted in Fig~\ref{figure:subvectors2}.

\subsubsection{Tree Pruning and Block Stitching}

Our discussion so far has focused on how the tree code structure can be integrated into the composite AMP iterate.
Once the AMP has converged and returned a reliable estimate for $\sv$, the tree decoder must perform the stitching process that binds consistent blocks together.
After stitching, the tree decoder outputs $\left\{ \hat{\mv}_1, \ldots, \hat{\mv}_{\Ka} \right\}$ or, equivalently, $\widehat{W}(\yv)$, which concludes the decoding of $\yv$.
The stitching process for a tree code is discussed at length in~\cite{amalladinne2019coded} and, as such, we do not reproduce this treatment in the present article.
Still, it is pertinent to mention a couple implementation tricks that can be applied after AMP has converged to fuse blocks together.
Note that, once AMP has converged and the iteration process is terminated, the effective observation remains fixed.
One can keep iterating on the factor graph of the tree code, and progressively apply thresholding to the components of $\left( \lambdav_{\ell} : \ell \in [L] \right)$, setting small entries to zero.
Effectively, this turns very unlikely entries into impossible locations, which propagates through the factor graph through message passing in a natural manner.

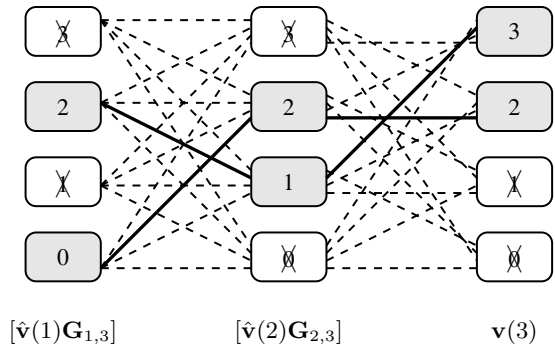
\begin{figure}[tbh]
  \centering
  \input{Figures/subvectors4_new}
  \caption{In this example with $v_1 = v_2 = v_3 = 2$, the 64 parity consistent paths get pruned to two groupings (super-sections) after the belief vector of each list has concentrated on two elements.}
  \label{figure:subvectors4}
\end{figure}

A second interesting trick arises as a consequence of our homogeneous tree code design: every block features information bits or parity bits, but not both.
For such tree codes, binding can be performed on local neighborhoods of the form $\left( \sv(j) : j \in N(a) \right)$.
Suppose that the lists in $N(a)$ have already been pruned with high statistical confidence as described above, leaving a few candidate blocks per section.
Then, parity constraints can be enforced on aggregates of the surviving members (per section), resulting in a reduced list of most likely super-sections.
This procedure is portrayed in Fig.~\ref{figure:subvectors4}, where 64 possible paths are distilled into two super-sections after halving individual lists and checking parity constraints.
Extensions to local binding, beyond members of $N(a)$, are conceptually straightforward.
For instance, repeated applications of these concepts can be scaffolded in a cascading, hierarchical, or mixed fashion.

\subsection{Synopsis of Revised Tree Code}

In summary, the revised tree code is designed to facilitate belief propagation on its factor graph at scale, while providing pertinent statistical information to the inner decoder at every step.
This is accomplished by using homogeneous blocks composed of either information bits or parity bits, but not both.
The parity bits are redefined to establish a circular convolution structure conducive to the application of FFT techniques.
Yet, the revised design preserves the foundation on which the analysis of \cite{amalladinne2019coded} is built and, as such, performance guarantees can be obtained when the generator matrices are random binary matrices whose entries are independent Rademacher trials.

\section{Inner Code and AMP Decoding}
\label{section:AMP}

We are ready to initiate our description of the enhanced CCS-AMP algorithm.
As mentioned in Section~\ref{section:Motivation}, the inner code introduced in \eqref{equation:SPARClike} operates on a received signal of the form
\begin{equation} \label{equation:ReceivedSignal}
\yv = \mathbf{A} \mathbf{D} \sv + \zv
\end{equation}
where $\sv = \sv(1) \cdots \sv(L)$ is fragmented into $L$ sections.
Recall that every section in $\sv$ is $\Ka$-sparse because $\sv(\ell) = \sum_{i=1}^{\Ka} \mv_i(\ell)$, a structure which suggests that AMP can perform well in this setting.
Matrix $\Am$ is normalized in that the 2-norm of every column is one.
Matrix $\Dm$ is diagonal with equal, non-negative diagonal entries within each section.
It accounts for the transmit power allocated to each block; the amplitudes are labeled $d_{\ell} = \sqrt{n P_{\ell}}$.
In the spirit of AMP for sparse regression codes~\cite{joseph2013fast,CIT-092,rush2017capacity,greig2017techniques}, the challenge in designing the AMP decoder is to create a composite iterative process to recover sparse vector $\sv$.
The composite algorithm iterates through two equations:
\begin{align}
\zv^{(t)} &= \yv - \Am \Dm \sv^{(t)} + \frac{\zv^{(t-1)}}{n} \operatorname{div} \Dm \etav_{t-1} \left( \rv^{(t-1)} \right) \label{equation:FinalAMP00} \\
\sv^{(t+1)} &= \etav_t \Big( \underbrace{\Am^{\mathrm{T}} \zv^{(t)} \big. + \Dm \sv^{(t)} }_{\rv^{(t)}} \Big)
 \label{equation:FinalAMP01}
\end{align}
with initial conditions $\sv^{(0)} = \zerov$ and $\zv^{(0)} = \yv$.
The first equation can be interpreted as a computation of the \emph{residual} enhanced with an Onsager correction~\cite{bayati2011dynamics,donoho2013information}.
The second equation updates the state estimate through denoising.
The collection of functions $\left( \etav_t (\cdot) \right)_{t \geq 0}$ will be defined shortly.
For the time being, it suffices to say that $\etav_t (\cdot)$ seeks to leverage the structure embedded in $\sv$ while computing a state update.
The argument of the denoiser in \eqref{equation:FinalAMP01}, termed the \emph{effective observation} $\rv^{(t)}$, also plays an important role in the upcoming discussion.
We emphasize that our specific AMP characterization falls within the extended framework for non-separable functions characterized by Berthier, Montanari, and Nguyen~\cite{berthier2017state}.

The first application of AMP to the unsourced MAC with a CCS tree outer code is due to Fengler, Jung, and Caire~\cite{fengler2019sparcs}.
The authors therein demonstrate that AMP can be adapted to the application scenario at hand, although this is only possible after addressing several technical challenges rooted in the sparse structure of the problem.
We briefly revisit some of these advances below for the sake of completeness.
We also describe our contributions and point out places where our envisioned framework differs from the scheme put forth by Fengler et al.

\subsection{Prior Art}
\label{subsection:TestStatistics}

To gain a better understanding of the proposed AMP decoder, we begin by looking at \eqref{equation:FinalAMP01}, where $\rv^{(t)}$ acts as a test statistic.
A remarkable fact about AMP is that, under certain conditions on the denoising functions, the effective observation $\rv^{(t)}$ is asymptotically distributed as $\mathbf{D} \sv + \tau_t \zetav_t$ where $\zetav_t$ is an i.i.d.\ $\mathcal{N}(0,1)$ random vector and $\tau_t$ is a deterministic quantity.
This property typically hinges on the presence of an Onsager correction term in the iterative algorithm and it assumes that the denoiser is sufficiently smooth~\cite{berthier2017state}.
In the original AMP for SPARCs implementation~\cite{rush2017capacity, barbier2017approximate, barbier2014replica}, which is designed for a single user, the denoiser is applied independently to every section.
Specifically, the denoising function they adopt therein is an instance of a minimum mean square error (MMSE) estimator that accounts for a one-sparse structure per block (single user).
Given test statistic $\rv(\ell)$, their updated block estimate takes the form
\begin{equation} \label{equation:SPARCsDenoiser}
\begin{split}
&\mathbb{E} \left[ \sv(\ell) \middle\vert d_{\ell} \sv(\ell) + \tau_t \zetav_t(\ell) = \rv(\ell) \right] \\
&\propto \sum_{\mv(\ell)} \mv(\ell) \exp \left( - \frac{\left\| \rv(\ell) - d_{\ell} \mv(\ell) \right\|^2}{2 \tau_t^2} \right)
\end{split}
\end{equation}
where `$\propto$' accounts for normalization and the sum is over the $m_{\ell}$ possible one-sparse blocks.
While this approach works adequately for a per block sparsity of one and reasonably small sections, it does not extend easily to the unsourced MAC problem.
Indeed, in the latter scenario, an optimal Bayesian denoiser must take into account the $\Ka$-sparse constraint on the support of $\sv(\ell)$.
That is, instead of the $m_{\ell}$ distinct possibilities in the original SPARC setting, an optimal solution must parse through $\binom{m_{\ell}}{\Ka}$ arrangements for the support of $\sv(\ell)$, a computationally expensive task.

A suitable approximation to this solution, for select parameters, is the marginal posterior mean estimate (PME) of Fengler et al.~\cite{fengler2019sparcs}.
To introduce the concept, we begin with an elementary building block.
Consider scalar binary signal $s \in \{ 0, 1 \}$ embedded in Gaussian noise.
Suppose the prior distribution for this signal is $q = \Pr (s = 1) = 1 - \Pr (s = 0)$, and the observation model is $r = d s + \tau \zeta$.
The PME, which is also the conditional probability that $s$ is equal to one given observation $r$, is characterized in Lemma~\ref{lemma:PME}.

\begin{lemma}[PME~\cite{fengler2019sparcs}] \label{lemma:PME}
The posterior mean estimator (PME) for binary signal $s$, conditioned on observation $r = d_{\ell} s + \tau \zeta$ where $\zeta \sim \mathcal{N}(0,1)$, takes the form
\begin{equation} \label{equation:OriginalPME}
\hat{s}_{\ell} \left( q, r, \tau \right)
= \frac{q \exp \left( - \frac{ \left( r - d_{\ell} \right)^2}{2 \tau^2} \right)}
{ q \exp \left( - \frac{ \left( r -  d_{\ell} \right)^2}{2 \tau^2} \right)
+ (1-q) \exp \left( -\frac{r^2}{2 \tau^2} \right)} .
\end{equation}
where $q$ is the prior probability of entry $s$ being equal to one, and $(1-q)$ is the probability of it being zero.
\end{lemma}
\begin{IEEEproof}
For a random variable $s$ taking value one with probability $q$ and zero otherwise, the conditional expectation can be written as
\[\hat{s}_{\ell} \left( q, r, \tau \right)
= \mathbb{E} \left[ s \middle\vert d_{\ell} s + \tau \zeta = r \right],\]
where $\zeta$ is a standard Gaussian random variable independent of $s$.
Therefore, we get the explicit formula
\begin{equation*}
\begin{split}
&\hat{s}_{\ell} \left( q, r, \tau \right)
= \frac{0 \cdot (1-q) f \left( \frac{r}{\tau} \right) + 1 \cdot q f \left( \frac{r - d_{\ell}}{\tau} \right)}
{(1-q) f \left( \frac{r}{\tau} \right) + q f \left( \frac{r - d_{\ell}}{\tau} \right)} \\
&= \frac{q \exp \left( - \frac{ \left( r - d_{\ell} \right)^2}{2 \tau^2} \right)}
{(1-q) \exp \left( -\frac{r^2}{2 \tau^2} \right)
+ q \exp \left( - \frac{ \left( r - d_{\ell} \right)^2}{2 \tau^2} \right)},
\end{split}
\end{equation*}
where $f(\cdot)$ is the probability density function of a normal random variable.
\end{IEEEproof}

Constant $d_{\ell} = \sqrt{n P_{\ell}}$ in~\eqref{equation:OriginalPME} comes from the value along the diagonal of $\mathbf{D}$ in the $\ell$th section; it captures the amplitude of the transmitted symbol.
The purpose of this rudimentary lemma is to lay a foundation for the upcoming denoising functions.
Another quantity that we will need shortly is the partial derivative of the marginal PME, which we present in Lemma~\ref{lemma:PartialSestimate}.
It is instructive to highlight the close resemblance between $\hat{s}_{\ell}(q, r, \tau)$ and the logistic function before stating the lemma.
In view of this connection, the form of the derivative should not come as a surprise.

\begin{lemma} \label{lemma:PartialSestimate}
The partial derivative with respect to $r$ of the posterior mean estimator (PME) defined in Lemma~\ref{lemma:PME} is
\begin{equation}
\frac{\partial\hat{s}_{\ell}(q, r, \tau)}{\partial r}
= \frac{d_{\ell}}{\tau^2}\hat{s}_{\ell}(q, r, \tau)
\left( 1 - \hat{s}_{\ell}(q, r, \tau) \right) .
\end{equation}
\end{lemma}
\begin{IEEEproof}
First, we note that the PME can be rewritten as
\begin{equation}
\begin{split}
\hat{s}_{\ell}(q, r, \tau)
&= \frac{q}{q + (1-q) \exp \left( \frac{d_{\ell}^2 - 2 r d_{\ell}}{2 \tau^2} \right)} .
\end{split}
\end{equation}
Then, by the chain rule of differentiation, we get
\begin{equation*}
\begin{split}
\frac{\partial \hat{s}_{\ell}(q, r, \tau)}{\partial r}
&= \frac{d_{\ell}}{\tau^2} \frac{q (1-q) \exp \left( \frac{d_{\ell}^2 - 2 r d_{\ell}}{2 \tau^2} \right)}
{\left( q + (1-q) \exp \left( \frac{d_{\ell}^2 - 2 r d_{\ell}}{2 \tau^2} \right) \right)^2} \\
&= \frac{d_{\ell}}{\tau^2} \hat{s}_{\ell}(q, r, \tau)
\left( 1 - \hat{s}_{\ell}(q, r, \tau) \right),
\end{split}
\end{equation*}
as stated.
\end{IEEEproof}

The denoiser introduced by Fengler et al.~\cite{fengler2019sparcs} can be described using the PME of Lemma~\ref{lemma:PartialSestimate} as follows.
First, recall that $q$ act as a proxy for the probability that the signal is non-zero.
If there are $m_{\ell}$ locations in $\sv(\ell)$ and $\Ka$ active devices, each picking a location independently, then the probability that any particular entry is non-zero can be expressed as $q_{\ell} = 1 - \left( 1-\frac{1}{m_{\ell}} \right)^{\Ka} \approx \frac{\Ka}{m_{\ell}}$.
Estimates for the entries in $\sv(\ell)$, given observation $\rv(\ell)$, can be obtained via
\begin{equation} \label{equation:OriginalDenoiserBlock}
\hat{\sv}_{\ell}^{\mathrm{OR}} \left( \rv(\ell), \tau \right)
= \left( \hat{s}_{\ell} \left( q_{\ell}, \rv(\ell, k), \tau \right) : k \in 0, \ldots, m_{\ell} - 1 \right)
\end{equation}
where $\rv(\ell, k)$ denotes the $k$th entry of the $\ell$th section of $\rv$, and $q_\ell$ is the fixed constant mentioned above.
Their overall vector estimate is obtained by concatenating $L$ blocks,
\begin{equation} \label{equation:AggrageOriginalDenoiser}
\hat{\sv}^{\mathrm{OR}} \left( \rv, \tau \right) =
\hat{\sv}_{1}^{\mathrm{OR}} \left( \rv(1), \tau \right) \cdots
\hat{\sv}_{L}^{\mathrm{OR}} \left( \rv(L), \tau \right) .
\end{equation}
This low-complexity strategy offers very good empirical performance when combined with AMP, as reported in~\cite{fengler2019sparcs}.
When integrated within the AMP composite algorithm, \eqref{equation:FinalAMP00}--\eqref{equation:FinalAMP01}, the denoiser defined in \eqref{equation:AggrageOriginalDenoiser} yields section estimate
\begin{equation} \label{equation:AggrageOriginalDenoiserBlock}
\hat{\sv}_{\ell}^{\mathrm{OR}} \left( \Am^{\mathrm{T}} \zv^{(t)}(\ell) + \Dm \sv^{(t)}(\ell), \tau_t \right) .
\end{equation}
The state vector is then updated according to \eqref{equation:AggrageOriginalDenoiserBlock} utilizing the aggregate form of \eqref{equation:AggrageOriginalDenoiser}.
This is a well-behaved denoiser, with desirable properties.

\begin{lemma} \label{lemma:PME-LipschitzContinuity}
The denoiser $\hat{\sv}^{\mathrm{OR}} \left( \rv, \tau \right)$ is Lipschitz continuous.
\end{lemma}
\begin{IEEEproof}
From Lemma~\ref{lemma:PartialSestimate} and the fact that $ \hat{s}_{\ell}(q, r, \tau) \in [0, 1]$, we gather that
\begin{equation}
\left| \frac{\partial \hat{s}_{\ell}(q, r, \tau)}{\partial r} \right| \leq \frac{d_{\ell}}{\tau^2} .
\end{equation}
Moreover, we have
$\frac{\partial \hat{s}_{\ell} \left( q_{\ell}, \rv(\ell, k_{\ell}), \tau \right)}{\partial \rv (j, k)} = 0$.
whenever $j \neq \ell$ or $k \neq k_{\ell}$.
It follows that, by the mean value theorem,
\begin{equation*}
\begin{split}
\left\| \hat{\sv}^{\mathrm{OR}} \left( \rv, \tau \right) - \hat{\sv}^{\mathrm{OR}} \left( \rv', \tau \right) \right\|^2
&= \sum_{\ell \in [L]} \sum_{k_{\ell} = 0}^{m_{\ell} - 1}
\left( \hat{s}_{\ell} \left( q_{\ell}, \rv(\ell, k_{\ell}), \tau \right)
- \hat{s}_{\ell} \left( q_{\ell}, \rv'(\ell, k_{\ell}), \tau \right) \right)^2 \\
&\leq \sum_{\ell \in [L]} \sum_{k_{\ell} = 0}^{m_{\ell} - 1} 
\left( \frac{d_{\ell}}{\tau^2} \right)^2 \left( \rv(\ell, k_{\ell}) - \rv'(\ell, k_{\ell}) \right)^2 \\
&\leq \left( \frac{d_{\max}}{\tau^2} \right)^2 \left\| \rv - \rv' \right\|^2
\end{split}
\end{equation*}
where $d_{\max} = \max_{\ell} d_{\ell}$.
This establishes the Lipschitz continuity of this denoiser.
\end{IEEEproof}

A possible explanation for this denoiser is that it seeks to compute
\begin{equation} \label{equation:ExplanationDenoiser}
\mathbb{E} \left[ \sv(\ell,k)  \middle\vert \sqrt{n P_{\ell}} \sv(\ell,k)
+ \tau_t \zetav_t(\ell,k)  =  \rv^{(t)}(\ell,k)  \right] ,
\end{equation}
This is a loose interpretation because $\hat{s}_{\ell} \left( q_{\ell}, \rv(\ell, k), \tau \right)$ disregards the fact that, with low probability, $\sv(\ell,k)$ can take integer values other than zero or one.
Nonethless, it is very close to \eqref{equation:SPARCsDenoiser} in spirit.
A key insight behind this approach is that, instead of pursuing the computationally challenging task of estimating signal $\sv$ optimally from the effective observation, it suffices to infer its support using marginal distributions.

\subsection{Dynamic PME Denoising}

The alternate viewpoint we propose in this article stems from the realization that, given the presence of an outer code, estimates of the elements of $\sv(\ell)$ can be improved by taking advantage of the underlying code construction, as described in Section~\ref{section:TreeCodeRevisited}.
Adopting constant $q_{\ell}$ in \eqref{equation:OriginalDenoiserBlock} disregards the factor graph structure of the outer code altogether.
It is equivalent to using uninformative prior marginal distributions within the PME at every stage.
This situation reveals an opportunity to enhance CCS-AMP by accounting for connections between neighboring blocks through local factors.
This alternative approach forms a marked departure from AMP for SPARCs in the context of the unsourced MAC~\cite{fengler2019sparcs}, and it has the potential to accelerate convergence and improve performance significantly.
An incentive to explore this research direction is that, although $\vv$ is a vector of length $2^v$, the tree encoding process essentially forces it to lie in a much smaller space that contains only $2^w$ elements.
Likewise, the tree code confines every section of $\sv$ to take value in a potentially much smaller subset when conditioned on graph neighboring sections through local factors.
The factor graph structure of the tree code, paired with message passing, gives rise to beliefs for the components of $\sv$ based on extrinsic information, as highlighted in \eqref{equation:EquivalentPriors}.
These can play the role of prior probability vector $\qv(\ell)$ in evaluating \eqref{equation:ExplanationDenoiser}.
Altogether, exploiting the graphical structure of the tree code is attainable via FFT techniques, and it can help improve the inference process.
Thus, our next goal is to define a denoiser in a manner analogous to \eqref{equation:OriginalDenoiserBlock}, but to incorporate the extrinsic information embedded in $\left\{ \rv^{(t)}(j) : j \in [L] \setminus \ell \right\}$ (or a good approximation thereof) in calculating estimates for the elements of $\sv^{(t+1)}(\ell)$.

Formally, we propose to create denoiser function $\etav^{\mathrm{PME}}_t \left( \rv^{(t)} \right)$ as follows.
We initiate local estimate vector $\lambdav_{j}^{\mathrm{PME}}(k)$ using \eqref{equation:OriginalDenoiserBlock}, with
\begin{equation} \label{eqruation:InitialLambdaPME}
\lambdav_{j}^{\mathrm{PME}}(k)
= \hat{s}_{j} \left( q_{j}, \rv(j, k), \tau_t \right)
\end{equation}
and $q_{j} = 1 - \left( 1-{1}/{m_{j}} \right)^{\Ka} \approx {\Ka}/{m_{j}}$.
We then run $b$ rounds of belief propagation using message passing rules according to \eqref{equation:BP-Check2Variable} and \eqref{equation:BP-Variable2Check}.
We aggregate the messages coming to variable node $s_{\ell}$ from adjoining parity factors and compute belief vector
\begin{equation} \label{equation:DynamicDenoiserPriors}
\qv(\ell,k) = 1 - \left( 1-\frac{\muv_{s_{\ell}}(k)}{\| \muv_{s_{\ell}} \|_1} \right)^{\Ka}
\end{equation}
based on extrinsic information by applying \eqref{equation:DynamicDenoiserPriors}.
As a final denoising step, we compute section estimate using \eqref{equation:OriginalPME} in Lemma~\ref{lemma:PME}.
The culmination of these steps leads to the following denoiser.

\begin{definition}[Dynamic PME Denoiser] \label{definition:BP-PME-Denoiser}
The functions in \eqref{equation:FinalAMP01} for the dynamic PME denoiser are given by
\begin{equation}
\etav_t^{\mathrm{PME}}(\rv) = \hat{\sv}_{1}^{\mathrm{PME}} \left( \rv, \tau_t \right) \cdots \hat{\sv}_{L}^{\mathrm{PME}} \left( \rv, \tau_t \right) .
\end{equation}
where individual components are equal to
\begin{equation} \label{equation:PME-b}
\hat{\sv}_{\ell}^{\mathrm{PME}} \left( \rv, \tau_t \right)
= \big( \hat{s}_{\ell} \left( \qv(\ell, k), \rv(\ell, k), \tau_t \right) : k \in 0, \ldots, m_{\ell} - 1 \big) .
\end{equation}
The effective observation $\rv = \Am^{\mathrm{T}} \zv + \Dm \sv$ is provided by the AMP algorithm,
whereas the belief vector $\qv(\ell)$ is derived from extrinsic information using \eqref{equation:EquivalentPriors} and \eqref{equation:DynamicDenoiserPriors} with initial conditions $\lambdav_{j}^{\mathrm{PME}}$.
Constants $\tau_t^2$ in \eqref{equation:PME-b} can be obtained deterministically through the state evolution, which we discuss later.
Alternatively, it can be approximated as $\tau_t^2 \approx \left\| \zv^{(t)} \right\|^2/n$ for $t \ge 0$ \cite{SchniterAMP}.
\end{definition}

It is worth mentioning that the denoiser in Definition~\ref{definition:BP-PME-Denoiser} reduces to the PME with uninformative prior probabilities introduced in~\cite{fengler2019sparcs} when the number of composite steps performed on the factor graph of the tree code is zero.
Next, we turn to the divergence of $\Dm \etav_t^{\mathrm{PME}} \left( \rv \right)$, which gives rise to the Onsager correction term found in \eqref{equation:FinalAMP00}.
Although $\etav_t^{\mathrm{PME}} (\cdot)$ is a non-separable function, its divergence admits an elegant and tractable structure, provided that $\qv(\ell)$ is computed based exclusively on extrinsic information.
This condition is ensured by the requirement that the number of BP iterations on the factor graph be strictly less than its shortest cycle.
In fact, this is the main motivation behind introducing this constraint on the number of BP iterations.
When this condition applies, $\qv(\ell)$ depends exclusively on $\left\{ \rv(j) : j \in [L] \setminus \ell \right\}$; we will soon see how this is also highly desirable for the computation of the Onsager correction.
We begin with a preliminary result that focuses on individual components.

\begin{lemma} \label{lemma:BP-PME-PartialSestimate}
If the number of message passing iterations on the tree graph is strictly less than the length of its shortest cycle, then the partial derivative of $\hat{s}_{\ell}^{\mathrm{PME}} \left( k_{\ell}, \rv, \tau \right)$ with respect to $\rv \left( \ell, k_{\ell} \right)$ is given by
\begin{equation} \label{equation:BP-PME-PartialSestimate}
\frac{\partial
\hat{s}_{\ell}^{\mathrm{PME}} \left( k_{\ell}, \rv, \tau \right)}{\partial \rv \left( \ell, k_{\ell} \right)}
= \frac{d_{\ell}}{\tau^2} \hat{s}_{\ell}^{\mathrm{PME}} \left( k_{\ell}, \rv, \tau \right)
\left( 1 - \hat{s}_{\ell}^{\mathrm{PME}} \left( k_{\ell}, \rv, \tau \right) \right) .
\end{equation}
\end{lemma}
\begin{IEEEproof}
We note that, predicated on the number of BP iterations being strictly less than the shortest loop in the factor graph, we are guaranteed to have
\begin{equation*}
\frac{\partial \qv(\ell, k_{\ell})}{\partial \rv \left( \ell, k_{\ell} \right)} = 0
\end{equation*}
because $\qv(\ell)$ only depends on extrinsic information or, equivalently, it is computed based on $\left\{ \rv(j) : j \in [L] \setminus \ell \right\}$.
Also, recall that $\tau$ is a deterministic constant.
Consequently, using the chain rule of differentiation, we get
\begin{equation*}
\begin{split}
&\frac{\partial
\hat{s}_{\ell}^{\mathrm{PME}} \left( k_{\ell}, \rv, \tau \right)}{\partial \rv \left( \ell, k_{\ell} \right)}
= \frac{\partial \hat{s}_{\ell} \left( \qv(\ell, k_{\ell}), \rv \left( \ell, k_{\ell} \right), \tau \right)}
{\partial \rv \left( \ell, k_{\ell} \right)} \\
&= \frac{d_{\ell}}{\tau^2} \hat{s}_{\ell} \left( \qv(\ell, k_{\ell}), \rv \left( \ell, k_{\ell} \right), \tau \right)
\left( 1 - \hat{s}_{\ell} \left( \qv (\ell, k_{\ell}), \rv \left( \ell, k_{\ell} \right), \tau \right) \right) \\
&= \frac{d_{\ell}}{\tau^2} \hat{s}_{\ell}^{\mathrm{PME}} \left( k_{\ell}, \rv, \tau \right)
\left( 1 - \hat{s}_{\ell}^{\mathrm{PME}} \left( k_{\ell}, \rv, \tau \right) \right)
\end{split}
\end{equation*}
The partial derivative of $\hat{s}_{\ell} \left( q, r, \tau \right)$ with respect to $r$ comes from Lemma~\ref{lemma:PartialSestimate}.
\end{IEEEproof}

Interestingly, the derivative in \eqref{equation:BP-PME-PartialSestimate} does not depend on the number of BP iterations computed on the factor graph, provided that the conditions of Lemma~\ref{lemma:BP-PME-PartialSestimate} are met.
The divergence associated with the denoiser is obtained below.

\begin{proposition} \label{proposition:DivComputationBP}
The divergence of $\Dm \etav_t^{\mathrm{PME}} \left( \rv \right)$ with respect to $\rv$ is equal to
\begin{equation} \label{equation:BP-PME-OnsagerCorrection}
\operatorname{div} \Dm \etav_t^{\mathrm{PME}} \left( \rv \right)
= \frac{1}{\tau_t^2} \left( \left\|\Dm^2 \etav_t^{\mathrm{PME}} \left( \rv \right) \right\|_1 - \left\| \Dm \etav_t^{\mathrm{PME}} \left( \rv \right) \right\|^2 \right) .
\end{equation}
\end{proposition}
\begin{IEEEproof}
First, we expand the $\operatorname{div}$ operator as
\begin{equation} \label{equation:PME-Finalbt}
\begin{split}
&\operatorname{div} \Dm \etav_t^{\mathrm{PME}} \left( \rv \right)
= \sum_{\ell=1}^L d_{\ell} \operatorname{div} \hat{\sv}_{\ell}^{\mathrm{PME}} \left( \rv, \tau_t \right) \\
&= \sum_{\ell=1}^L d_{\ell} \sum_{k=0}^{m_{\ell}-1} \frac{\partial
\hat{s}_{\ell}^{\mathrm{PME}} \left( k, \rv, \tau_t \right)}{\partial \rv(\ell, k)} \\
&= \sum_{\ell=1}^L d_{\ell} \sum_{k=0}^{m_{\ell}-1} \frac{d_{\ell}}{\tau_t^2} \hat{s}_{\ell}^{\mathrm{PME}} \left( k, \rv, \tau_t \right)
\left( 1 - \hat{s}_{\ell}^{\mathrm{PME}} \left( k, \rv, \tau_t \right) \right) .
\end{split}
\end{equation}
The last equality follows from substituting the expression for the partial derivative of $\hat{s}_{\ell}^{\mathrm{PME}} \left( k_{\ell}, \rv, \tau_t \right)$ obtained in Lemma~\ref{lemma:BP-PME-PartialSestimate}.
Again, we emphasize that $\qv(\ell, k)$, being derived from extrinsic information, does not depend on the elements of block $\rv (\ell)$, which implies $\frac{\partial \qv(\ell,k)}{\partial \rv(\ell,k)} = 0$.
This property greatly limits the difficulty in computing the summands of \eqref{equation:PME-Finalbt}.
Recognizing $\left\{ \hat{s}_{\ell}^{\mathrm{PME}} \left( k, \rv, \tau_t \right) \right\}$ as the components of $\etav_t^{\mathrm{PME}} \left( \rv \right)$, the equation above can be interpreted as an inner product.
Accounting for the signal amplitudes through diagonal matrix $\Dm$ yields
\begin{equation}
\begin{split}
\operatorname{div} \Dm \etav_t^{\mathrm{PME}} \left( \rv \right)
&= \frac{1}{\tau_t^2} \left\langle \Dm \etav_t^{\mathrm{PME}} \left( \rv \right),
\Dm \onev - \Dm \etav_t^{\mathrm{PME}} \left( \rv \right) \right\rangle \\
&= \frac{1}{\tau_t^2} \left( \left\| \Dm^2 \etav_t^{\mathrm{PME}} \left( \rv \right) \right\|_1
- \left\| \Dm \etav_t^{\mathrm{PME}} \left( \rv \right) \right\|^2 \right)
\end{split}
\end{equation}
where $\onev$ is a vector whose entries are all ones.
This is precisely the format of the Onsager correction in \eqref{equation:BP-PME-OnsagerCorrection}.
\end{IEEEproof}

Proposition~\ref{proposition:DivComputationBP}, together with the facts that $\Am$ is a normalized matrix and $\sv^{(t)} = \etav_{t-1}^{\mathrm{PME}} \left( \rv^{(t-1)} \right)$, yields correction coefficient
\begin{equation} \label{equation:PME-OnsagerCorrection}
\frac{1}{n} \operatorname{div} \Dm \etav_{t-1}^{\mathrm{PME}} \left( \rv^{(t-1)} \right)
= \frac{1}{n \tau_{t-1}^2} \left( \left\| \Dm^2 \sv^{(t)} \right\|_1 - \left\| \Dm \sv^{(t)} \right\|^2 \right).
\end{equation}
Overall, this produces an AMP algorithm that takes advantage of the structure of the underlying tree code throughout the decoding of the inner code.
In our implementation, we update $\qv(\ell)$ dynamically at every AMP iteration, essentially by running BP on a truncated factor graph.
Still, it is possible to estimate $\operatorname{div} \Dm \etav_t^{\mathrm{PME}} \left( \rv \right)$ in practice, without updating $\qv(\ell)$ at every iteration, although such a variant would deviate from a strict AMP definition.
The dynamical PME denoising is an appealing solution because it meshes nicely with prior art.
It offers a conceptual bridge between using uninformative prior probabilities and performing multiple rounds of message passing on the factor graph of the tree code.

The dynamic PME denoiser is not as smooth as the original PME denoiser.
This can be seen through the fact that collections of impossible entries in neighboring sections may lead to vanishing elements in $\qv(\ell)$.
From a decoding perspective, this is desirable because it can rapidly prune down the space of possibilities.
However, it is much more difficult to obtain Lipschitz conditions under such circumstances.
In Appendix~\ref{appendix:LipschitzPMEDenoiser}, we show that the dynamic PME denoiser with one round of message passing on the factor graph of the tree code is Lipschitz continuous.
This property brings credibility to our upcoming study of the state evolution.

\subsection{Good Tree Code Structures}
\label{subsection:TreeCodeStructure}

The state evolution for CCS-AMP is intimately linked to the structure of the underlying tree code whenever the algorithm dynamically updates the state estimate as part of every AMP composite iteration.
To boost the benefits of incorporating the outer tree code within the AMP framework, a careful redesign of its graph structure is necessary.
The parity allocation of the original tree code aims at limiting the growth of active paths during sequential tree decoding, while also maintaining high performance~\cite{amalladinne2019coded}.
This leads to a parity assignment where information bits and parity bits coexist in several blocks.
Yet, as mentioned earlier, such a strategy is not conducive to fast transform methods whereby all possible paths are assessed concurrently.
Rather, it becomes desirable within CCS-AMP to have homogeneous blocks composed of either information bits or parity bits, but not both.
Naively, the discriminating power of block~$\ell$ is $2^{p_{\ell}}$, and it is greatest when $p_{\ell} = v_{\ell}$.
Moreover, heterogeneous blocks introduce dependencies that complicate the straightforward implementation of message passing with fast transform methods.
This leads us to restrict our attention to tree code designs with homogeneous blocks.

Another guiding principle for the redesign of the tree code stems from mixing considerations and the addition of parity precursors.
When sections of parity bits are interspersed in-between information blocks, the number of likely patterns associated with a parity section is equal to the product of the number of likely parity precursors in every precursory section.
Thus, for a parity section to remain informative while running AMP iterations, it should not be created by combining too many information sections.
We elucidate this phenomenon through a crude, back-of-the-envelope motivating example below.

\begin{example} \label{example:TreeRedesign}
Consider a tree code with three sections, $\ell = 1, 2, 3$, as in Fig~\ref{figure:subvectors2}.
Suppose that every section has common size $m_{\ell} = 2^{16}$ and that they are homogeneous, containing either information bits or parity bits, but not both.
Let $\Ka$ be the number of active devices.
Also, for illustrative purposes, suppose that the probability distribution associated with each section has concentrated over $H$ entries of $\hat{\sv}(\ell)$, including all the $\Ka$ legitimate message indices.
We wish to identify the expected number of indices in each section that retain a high probability after the parity constraints have been enforced.
We note that, by design, all the legitimate indices are validated and maintained.
Consider an erroneous parity pattern, i.e., one that does not arise from genuine parity precursors in previous sections.
For this pattern to survive the validation step, it must pair up with probable indices in the preceding sections in a parity consistent manner.
When parities are created using two information sections, there are (at most) $H^2$ admissible parity consistent patterns based on likely neighbors, out of which $H^2 - \Ka$ are erroneous.
The probability that the selected pattern falls within this set is approximately
\begin{equation*}
\frac{H^2 - \Ka}{2^{16} - \Ka} \quad \left( H < 2^8 \right) .
\end{equation*}
Given that there are $H-\Ka$ likely erroneous parity patterns before these constraints are checked, the expected number of surviving erroneous indices after applying the local factor inherited from the tree graph becomes approximately
\begin{equation*}
\frac{(H-\Ka)(H^2-\Ka)}{2^{16} - \Ka} \quad \left( H < 2^8 \right) .
\end{equation*}
Consequently, as distributions start to concentrate over $H = \sqrt{2^{16}}=256$ indices, the parity constraints of the tree code begin to softly discount erroneous blocks.
This accelerates the AMP convergence process and reduces the probability of erroneous indices surviving the decoding of the inner code.
In comparison, when the parity patterns act on three information sections, a similar pruning effect only starts to take place once the distributions have concentrated on approximately $H = \sqrt[3]{2^{16}} \approx 40$ values.
That is, the probability that an erroneous block is parity consistent with other sections remains high for much longer.
While it is possible to calculate $\left\{ \muv_{s_{\ell}} \right\}$ in such cases, the statistics obtained remain uninformative for many more AMP cycles, which nullifies the potential benefits of the enhanced CCS-AMP algorithm with its dynamic denoiser.
This becomes progressively worse as the number of information sections over which parity patterns are computed increases.
\end{example}

A third consideration pertaining to the design of our tree code is the fact that we should avoid short cycles in the factor graph representation.
Empirically, factor graphs that only contain long cycles are amenable to belief propagation, whereas message passing over factor graphs with short loops may be problematic.
Of course, the fact that Lemma~\ref{lemma:BP-PME-PartialSestimate} only applies when the number of message passing iterations on the tree graph is strictly less than the length of its shortest cycle is also an incentive to design tree codes with no short cycles.

\begin{figure}[tbh]
  \centering
  \input{Figures/architecture}
  \caption{This graph shows the type of triadic connections between precursory information sections and parity blocks utilized in the design of the revised tree code.
  Shaded blocks denote parity sections, whereas light blocks correspond to information fragments.}
  \label{figure:architecture}
\end{figure}
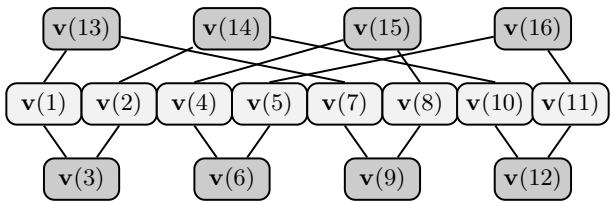

Given these observations, combined with the evidence afforded by numerical simulations, we focus on tree structures based on triadic designs whereby parity sections are determined based on two of the preceding blocks.
Although Example~\ref{example:TreeRedesign} overlooks issues such as correlations and feedback, the results obtained by applying these guidelines are excellent for the operating parameters we are interested in.
Moreover, alternate graphical constructions for the tree code where parity sections are formed using more than two precursory sections do not perform as well empirically.
For these reasons, our analysis of CCS-AMP moving forward assumes a triadic tree structure such as the construction depicted in Fig.~\ref{figure:architecture}.
That is, every parity section is attached to two other blocks, and short loops are avoided altogether in the factor graph.

We point out, again, that the proposed approach warrants making a decision about the structure of the outer code before being able to tackle the state evolution in AMP.
This differs from the algorithm presented in~\cite{fengler2019sparcs}, whereby the state evolution is completely detached from the outer code.
With a candidate design class in mind, we proceed to evaluating the progression of the AMP algorithm.

\subsection{State Evolution}
\label{subSection:SE}

The state evolution is a standard object in the treatment of AMP~\cite{bayati2011dynamics}.
It offers a blueprint to determine the sequence $\{ \tau_t^2 \}_{t \geq 0}$ and, concurrently, it provides a predictor of algorithmic performance.
The computations for the state evolution rest on the fact that, asymptotically in the number of dimensions, the effective observations become Gaussian.
Specifically, the analysis hinges on $\rv^{(t)}$ being distributed according to
\begin{equation} \label{equation:StateEvolutionObservation}
\rv^{(t)} \sim \mathbf{D} \sv + \tau_t \zetav_t
\end{equation}
where $\zetav_t$ is an i.i.d.\ $\mathcal{N}(0,1)$ random vector, as mentioned at the beginning of Section~\ref{subsection:TestStatistics}.
This broad setting is common to most AMP analyses.
The standard deviation parameter $\tau_t$ is computed iteratively.

As mentioned above, we delayed the treatment of the state evolution until now because it is predicated on the structure of the outer tree code for the proposed CCS-AMP framework.
Indeed, this is a byproduct of the fact that the tree code appears within the denoising functions and, consequently, its structure must be specified explicitly.
Likewise, defining the outer code is needed to examine the performance benefit associated with our more intricate scheme when compared to the performance of the original, tree-agnostic version in~\cite{fengler2019sparcs}.
Below, we study designs wherein every information section is involved in exactly two disjoint triadic parity connections.
The rationale for this condition can be found in Section~\ref{subsection:TreeCodeStructure}.
This structure is general enough to accommodate the scenarios we are interested in, and to provide insight on the suitability of competing designs.
Finally, we assume that there are no collisions at the section level;
that is, $\| \sv(\ell) \|_0 = \Ka$ for $\ell \in [L]$.
This is typical for sparse settings with a finite number of levels.

The state evolution captures the progression of the AMP algorithm, as a function of the iteration count~$t$.
We parallel the development found in \cite{berthier2017state} to characterize the evolution of this system, and we specialize their results to the application at hand.
Define
\begin{align*}
\hat{\zv}^{(t)} &= \sigma_t \Xiv_t \\
\hat{\rv}^{(t)} &= \Dm \sv + \tau_t \zetav_t
\end{align*}
where $\Xiv_t$ and $\zetav_t$ are i.i.d.\ $\mathcal{N}(0,1)$ random vectors.
The values of parameters $\sigma_t$ and $\tau_t$ can be obtained through a composite iteration process.
The first equation for the variance parameters is simply
\begin{equation} \label{equation:TauEvolution}
\begin{split}
\tau_t^2
= \sigma^2 + \lim_{n \rightarrow \infty} \frac{1}{n} \mathbb{E} \left[ \left\langle \hat{\zv}^{(t)}, \hat{\zv}^{(t)} \right\rangle \right]
= \sigma^2 + \lim_{n \rightarrow \infty} \frac{1}{n} \mathbb{E} \left[ \left\| \sigma_t \Xiv_t \right\|^2 \right]
= \sigma^2 + \sigma_t^2 ,
\end{split}
\end{equation}
where $\sigma^2$ is the variance of observation noise in \eqref{equation:ChannelModel}.
The second equation is more contrived with
\begin{equation} \label{equation:SigmaEvolution}
\sigma_{t+1}^2
= \lim_{n \rightarrow \infty} \frac{1}{n} \mathbb{E} \left[ \left\| \Dm \left( \etav_{t} \left( \Dm \sv + \tau_{t} \zetav_{t} \right) - \sv \right) \right\|^2 \right] .
\end{equation}
The initial conditions are $\tau_0^2 = \sigma^2 + \sigma_0^2 = \lim_{n \rightarrow \infty} \| \yv \|^2 / n$.
The foundation of the state evolution is rooted in the following proposition.
\begin{proposition}[Berthier, Montanari, and Nguyen]
Let the AMP iteration $\left\{ \zv^{(t)}, \rv^{(t)} \right\}_{t \geq 1}$ be generated via \eqref{equation:FinalAMP00} and \eqref{equation:FinalAMP01} with initial conditions $\zv^{(0)} = \yv$, $\sv^{(0)} = \zerov$ and assuming that $\tau_{t}$ is taken from \eqref{equation:TauEvolution}.
Consider the state evolution $\left\{ \hat{\zv}^{(t)}, \hat{\rv}^{(t)} \right\}_{t \geq 1}$ where the variance parameters are also those defined in \eqref{equation:TauEvolution} with initial condition $\tau_0 = \lim_{n \rightarrow \infty} \| \yv \|/\sqrt{n}$.
Then, under some regularity conditions, $\zv^{(t)} \rightarrow \hat{\zv}^{(t)}$ and $\rv^{(t)} \rightarrow \hat{\rv}^{(t)}$ in probability for all $t \geq 0$.
\end{proposition}
\begin{IEEEproof}
This proposition is a restriction of Theorem~1 and Corollary~2 in \cite{berthier2017state}.
In applying this result, it is pertinent to mention that its regularity conditions are fulfilled.
In particular, $\Am$ is an i.i.d.\ Gaussian matrix with normalized columns.
Also, from Lemma \ref{lemma:PME-LipschitzContinuity}, the original PME denoiser is (uniformly) Lipschitz.
For the triadic designs considered in this paper, we show in Appendix~\ref{appendix:LipschitzPMEDenoiser} that the dynamic PME denoiser is (uniformly) Lipschitz when one composite BP step is performed on the outer factor graph per AMP iteration.
This implies that the state evolution is also accurate in the presence of this dynamic PME denoiser.
\end{IEEEproof}

To leverage the state evolution to compare the two denoisers discussed thus far, we must compute (or approximate) the right-hand-side of \eqref{equation:SigmaEvolution} for these various cases.
We begin by writing
\begin{equation} \label{equation:StateEvolution}
\mathbb{E} \left[ \left\| \Dm \left( \etav_{t} \left( \Dm \sv + \tau_{t} \zetav_{t} \right) - \sv \right) \right\|^2 \right]
= \sum_{\ell \in [L]} d_{\ell}^2 \mathbb{E} \left[ \left\| \hat{\sv}_{\ell} \left( d_{\ell} \sv (\ell) + \tau_{t} \zetav_{t} (\ell), \tau_{t} \right) - \sv(\ell) \right\|^2 \right] .
\end{equation}
We note that this equation has the same form irrespective of the number of BP iterations in the denoiser.
The challenge in applying \eqref{equation:StateEvolution} comes from the fact that a closed-form expression is not available and a numerical evaluation of this expectation involves a very large number of random components.

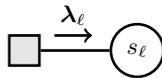
\begin{figure}[tbh]
  \centering
  \input{Figures/graphBP0}
  \caption{
  This illustration shows the local neighborhood of $s_{\ell}$ under the original PME denoiser.
  }
  \label{figure:LocalTreeBP0}
\end{figure}

\paragraph{Original PME Denoiser}
We explore the specifics of state evolution by first considering the simpler case: the original posterior mean estimate without message passing.
There are exactly $\Ka$ locations in vector $\sv(\ell)$ where the entry is one, and the remaining entries are equal to zero.
We treat these two cases separately.
Along these lines, we introduce the convenient notation $\mathcal{S}_{\ell}^1 = \left\{ k : \sv(\ell, k) = 1 \right\}$ and $\mathcal{S}_{\ell}^0 = \left\{ k : \sv(\ell, k) = 0 \right\}$.
This partition delineates how we approach the expectation and, ultimately, what quantities need to be evaluated through numerical methods.
Recall that the state estimate for each PME element has the form
\begin{equation}
\hat{s}_{\ell} \left( q, r, \tau \right)
= \frac{q \exp \left( - \frac{ \left( r - d_{\ell} \right)^2}{2 \tau^2} \right)}
{ q \exp \left( - \frac{ \left( r -  d_{\ell} \right)^2}{2 \tau^2} \right)
+ (1-q) \exp \left( -\frac{r^2}{2 \tau^2} \right)} ,
\end{equation}
where $q = 1 - \left( 1-{1}/{m_{\ell}} \right)^{\Ka} \approx {\Ka}/{m_{\ell}}$.
For a specific section, we can then write
\begin{equation*}
\begin{split}
&\mathbb{E} \left[ \left\| \hat{\sv}_{\ell} \left( d_{\ell} \sv (\ell) + \tau_{t} \zetav_{t} (\ell), \tau_{t} \right) - \sv(\ell) \right\|^2 \right] \\
&= \sum_{k \in \mathcal{S}_{\ell}^1} \mathbb{E} \left[ \left( \hat{s}_{\ell} \left( q, d_{\ell} + \tau_{t} \zetav_{t} (\ell, k), \tau_{t} \right) - 1 \right)^2 \right]
+ \sum_{k \in \mathcal{S}_{\ell}^0} \mathbb{E} \left[ \left( \hat{s}_{\ell} \left( q, \tau_{t} \zetav_{t} (\ell, k), \tau_{t} \right) - 0 \right)^2 \right] \\
&= \Ka \mathbb{E} \left[ \left( \frac{q \exp \left( - \frac{ \left( \tau_{t} \zeta \right)^2}{2 \tau_{t}^2} \right)}
{ q \exp \left( - \frac{ \left( \tau_{t} \zeta \right)^2}{2 \tau_{t}^2} \right)
+ (1-q) \exp \left( -\frac{ \left( d_{\ell} + \tau_{t} \zeta \right)^2}{2 \tau_{t}^2} \right)} - 1 \right)^2 \right] \\
&+ (m_{\ell} - \Ka) \mathbb{E} \left[ \left( \frac{q \exp \left( - \frac{ \left( \tau_{t} \zeta - d_{\ell} \right)^2}{2 \tau_{t}^2} \right)}
{ q \exp \left( - \frac{ \left( \tau_{t} \zeta -  d_{\ell} \right)^2}{2 \tau_{t}^2} \right)
+ (1-q) \exp \left( -\frac{ \left( \tau_{t} \zeta \right)^2}{2 \tau_{t}^2} \right)} \right)^2 \right] \\
&= \Ka \mathbb{E} \left[ \left( \frac{(1-q) \exp \left( \frac{ d_{\ell} \zeta}{\tau_{t}} \right)}
{ q \exp \left( \frac{d_{\ell}^2}{2 \tau_{t}^2} \right) + (1-q) \exp \left( \frac{d_{\ell} \zeta }{\tau_{t}} \right)} \right)^2 \right] \\
&+ (m_{\ell} - \Ka) \mathbb{E} \left[ \left( \frac{q \exp \left( \frac{d_{\ell} \zeta}{\tau_{t}} \right)}
{ q \exp \left( \frac{ d_{\ell} \zeta}{\tau_{t}} \right)
+ (1-q) \exp \left( \frac{d_{\ell}^2}{2 \tau_{t}^2} \right)} \right)^2 \right] .
\end{split}
\end{equation*}
Both expectations in this sum can be evaluated via Monte Carlo simulation or through numerical integration techniques.
The resulting values can subsequently be employed in conjunction with \eqref{equation:StateEvolution} to get numerical approximations for the state evolution in \eqref{equation:TauEvolution} and \eqref{equation:SigmaEvolution}.
The simplicity of this case stems from the fact that only local information is used in the denoiser, as depicted in Fig.~\ref{figure:LocalTreeBP0}.
This notional diagram can be compared with the situation associated with the dynamic PME denoiser, which we turn to next.

\begin{figure}[tbh]
  \centering
  \input{Figures/graphBP1}
  \caption{
  For the triadic tree code design considered in this section, there are two types of local neighborhoods.
  The diagram on the left represent the truncated tree associated with a parity section, whereas the one of the right represents the truncated tree of an information block.
  }
  \label{figure:LocalTreeBP1}
\end{figure}
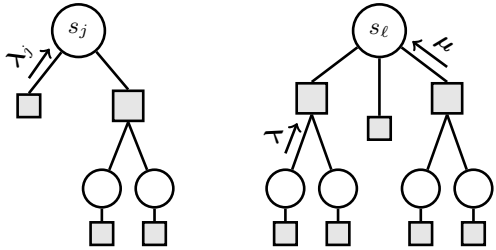

\paragraph{Dynamic PME Denoiser}
In spirit, the state evolution equation for the dynamic PME denoiser is similar to the scenario above, except for vector $\qv$ which is obtained from extrinsic information through belief propagation on the factor graph of the tree code.
Partly motivated by the fact that the dynamic PME denoiser is Lipschitz when the number of composite iterations performed on the factor graph of tree code is one (see Appendix~\ref{appendix:LipschitzPMEDenoiser}), we restrict our treatment of the state evolution to this specific case;
extending this analysis to accommodate multiple iterations on the outer code is not trivial.
For the scenario of interest, the governing section equation can be written as
\begin{equation} \label{equation:PME-Section-Evolution}
\begin{split}
&\mathbb{E} \left[ \left\| \hat{\sv}_{\ell}^{\mathrm{PME}} \left( \Dm \sv + \tau_{t} \zetav, \tau_t \right) - \sv(\ell) \right\|^2 \right] \\
&= \sum_{k \in \mathcal{S}_{\ell}^1} \mathbb{E} \left[ \left( \hat{s}_{\ell} \left( \qv (\ell, k), d_{\ell} + \tau_{t} \zetav_{t} (\ell, k), \tau_{t} \right) - 1 \right)^2 \right] \\
&+ \sum_{k \in \mathcal{S}_{\ell}^0} \mathbb{E} \left[ \left( \hat{s}_{\ell} \left( \qv (\ell, k), \tau_{t} \zetav_{t} (\ell, k), \tau_{t} \right) - 0 \right)^2 \right] \\
&= \sum_{k \in \mathcal{S}_{\ell}^1} \mathbb{E} \left[ \left( \frac{\left( 1-\qv (\ell, k) \right) \exp \left( \frac{ d_{\ell} \zetav_{t} (\ell, k)}{\tau_{t}} \right)}
{ \qv (\ell, k) \exp \left( \frac{d_{\ell}^2}{2 \tau_{t}^2} \right) + \left( 1-\qv (\ell, k) \right) \exp \left( \frac{d_{\ell} \zetav_{t} (\ell, k) }{\tau_{t}} \right)} \right)^2 \right] \\
&+ \sum_{k \in \mathcal{S}_{\ell}^0} \mathbb{E} \left[ \left( \frac{\qv (\ell, k) \exp \left( \frac{d_{\ell} \zetav_{t} (\ell, k)}{\tau_{t}} \right)}
{ \qv (\ell, k) \exp \left( \frac{ d_{\ell} \zetav_{t} (\ell, k)}{\tau_{t}} \right)
+ \left( 1-\qv (\ell, k) \right) \exp \left( \frac{d_{\ell}^2}{2 \tau_{t}^2} \right)} \right)^2 \right] .
\end{split}
\end{equation}
Recall that $\qv(\ell,k) = 1 - \left( 1-\frac{\muv_{s_{\ell}}(k)}{\| \muv_{s_{\ell}} \|_1} \right)^{\Ka}$, where $\muv_{s_{\ell}}(k) = \prod_{a \in N(s_{\ell})} \muv_{a \to s_{\ell}} (k)$ first appears in \eqref{equation:EquivalentPriors}.
In computing $\muv_{s_{\ell}}(k)$, the connectivity of the factor graph matters.
For the class of triadic designs described in Section~\ref{subsection:TreeCodeStructure}, information blocks and parity blocks must be treated separately because the structure of their local neighborhoods differs.
These local neighborhoods are depicted in Fig.~\ref{figure:LocalTreeBP1}.
When $\ell$ is a parity section, $N(s_{\ell})$ is a singleton because parity sections are connected to only one check node.
On the other hand, when $\ell$ is an information section, then $N(s_{\ell})$ has a cardinality of two since every information fragment is attached to two check nodes.
For triadic designs, by \eqref{equation:EquivalentPriors}, we have
\begin{equation} \label{equation:PME-EquivalentPriors}
\begin{split}
\muv_{s_{\ell}}(k)
&= \prod_{a \in N(s_{\ell})} \prod_{s_j \in N(a) \setminus s_{\ell}} \muv_{s_j \to a} (k_j)
= \prod_{a \in N(s_{\ell})} \sum_{k_1 + k_2 \equiv k} \lambdav^{\mathrm{PME}}_{j_1}(k_1)\lambdav^{\mathrm{PME}}_{j_2}(k_2) \\
&= \prod_{a \in N(s_{\ell})}\sum_{k_1 + k_2 \equiv k} \hat{s}_{j_1} \left( q, \hat{\rv}(j_1,k_1), \tau \right) \hat{s}_{j_2} \left( q, \hat{\rv}(j_2,k_2), \tau \right) \\
&= \prod_{a \in N(s_{\ell})}\sum_{k_1 + k_2 \equiv k} \hat{s}_{j_1} \left( q, \Dm \sv(j_1,k_1) + \tau \zetav(j_1,k_1), \tau \right) \hat{s}_{j_2} \left( q, \Dm \sv(j_2,k_2) + \tau \zetav(j_2,k_2), \tau \right),
\end{split}
\end{equation}
where $j_1, j_2$ are implicitly dependent on $a$.
Specifically, they denote the levels that form a triad with $s_{\ell}$ through check node $a$ within the factor graph induced by the outer tree code; that is, $\{ j_1, j_2 \} = N(a) \setminus s_{\ell}$.
Making this relation explicit in \eqref{equation:PME-EquivalentPriors} leads to an overly cumbersome notation, and context should prevent any confusion.

The components that form the left-most product in \eqref{equation:PME-EquivalentPriors} are slightly involved, owing to the mixing phenomenon illustrated in Section \ref{subsection:TreeCodeStructure}.
There are $\Ka^2$ entries\footnote{This characterization assumes that $\Ka^2 \leq m_{\ell}$; while our framework remains valid in spirit when $\Ka^2 > m_{\ell}$, a more careful accounting needs to be performed in this alternate regime.} in vector $\muv_{s_{\ell}}$, including the $\Ka$ locations in $\mathcal{S}_{\ell}^1$, for which the product in \eqref{equation:PME-EquivalentPriors} contains: one dominant term associated with the pair $\sv(j_1, k_1) = \sv(j_2, k_2) = 1$; $2 (\Ka - 1)$ terms for which either $\sv(j_1, k_1) = 1$ or $\sv(j_2, k_2) = 1$, but not both; and the last $m_{\ell} - 2 \Ka + 1$ terms for which both $\sv(j_1, k_1) = \sv(j_2, k_2) = 0$.
Furthermore, for the remaining $m_{\ell}-\Ka^2$ entries in $\muv_{s_{\ell}}$, there are: $2\Ka$ locations where either $\sv(j_1, k_1) = 1$ or $\sv(j_2, k_2) = 1$, but not both; and $m_{\ell} - 2 \Ka$ locations where both $\sv(j_1, k_1) = \sv(j_2, k_2) = 0$.
Unfortunately, these equations do not lend themselves to compact, closed-form expressions.
Moreover, the number of random variables involved in computing these quantities precludes the direct application of numerical integration.
While complex and computationally demanding, it is possible to numerically evaluate the expectations in \eqref{equation:PME-Section-Evolution} through Monte-Carlo simulations using \eqref{equation:PME-EquivalentPriors} and the above characterization.
This requires simulating all the relevant Gaussian random variables $\zetav (j, k)$ and computing the priors $\qv(j,k)$ at once using the fast transform technique.
Altogether, the latter approach provides a blueprint on how to iteratively compute the parameters in \eqref{equation:SigmaEvolution} and \eqref{equation:TauEvolution} with good accuracy for the dynamic PME denoiser.
In Section~\ref{section:SimulationResults}, we demonstrate that the performance of CCS-AMP predicted by state evolution using the aforementioned approach is very close to the empirical performance obtained by simulating the actual system.

\section{Simulation Results and Discussion}
\label{section:SimulationResults}

In this section, we study the empirical performance of the proposed scheme and provide comparisons with the original AMP-based algorithm introduced in \cite{fengler2019sparcs}.
For these simulations, we consider a system with $\Ka \in [10 : 300]$ active users.
The size of the payload corresponding to every active user is $w = 128$~bits.
The total number of channel uses is $n = 38400$.
The length of each block is equal to $16$~bits, i.e., $v_{\ell} = 16$ for all sections $\ell \in [L]$, and the number of sections $L = 16$ when $\Ka < 200$ and $18$ when $\Ka \ge 200$.
The target per-user probability of error is five percent, $\Pe = 0.05$.
It is worth mentioning that these system parameters have become emblematic of recent contributions related to unsourced, uncoordinated random access.

Figure~\ref{figure:architecture} shows the factor graph for the outer tree code employed in simulations when $\Ka < 200$.
We add two additional parity sections to reduce the probability of the decoder producing a list of size greater than $\Ka$ when $\Ka \ge 200$.
These architectures offer very good performance for the parameters of interest, yet we remark that our framework supports many alternate implementations worthy of investigation, possibly, in future studies.
During simulations, sensing matrix $\mathbf{A} \in \mathbb{R}^{n \times m}$ is formed by picking $n$ rows uniformly at random from a Hadamard matrix of dimension $m \times m$.
This Hadamard approach reduces the memory and computational load of the AMP decoder significantly over random Gaussian constructions, owing to the fast transform implementation.
Specifically, the computational complexity corresponding to one iteration of AMP can be reduced from $\mathcal{O}(nm)$ to $\mathcal{O}(m \log m)$ when a sub-sampled Hadamard matrix is utilized as the sensing matrix.
Although this constitutes a small departure from our theoretical analysis, we emphasize that this approach is frequently employed in practice, and it is known to perform well in finite block length regimes~\cite{rush2017capacity,fengler2019sparcs}.

Another common implementation twist we incorporate into our numerical study is running the overall algorithm twice, in the spirit of successive interference cancellation.
That is, we perform one extended round of AMP iterations, whereby the contribution of users decoded with high confidence is removed from the received signal, and then the residual signal is fed back into the AMP decoder for an additional round of composite iterations followed by tree decoding.
To this end, we remove the contribution of top $\Ka-\delta$ decoded messages with the largest likelihoods obtained during the first round, where $\delta$ is chosen empirically.
The second round of decoding then focuses on the recovery of remaining $\delta$ messages from the residual signal. 
Finally, if the decoder produces a list of size greater than $\Ka$, we retain the $\Ka$ messages with largest likelihoods.

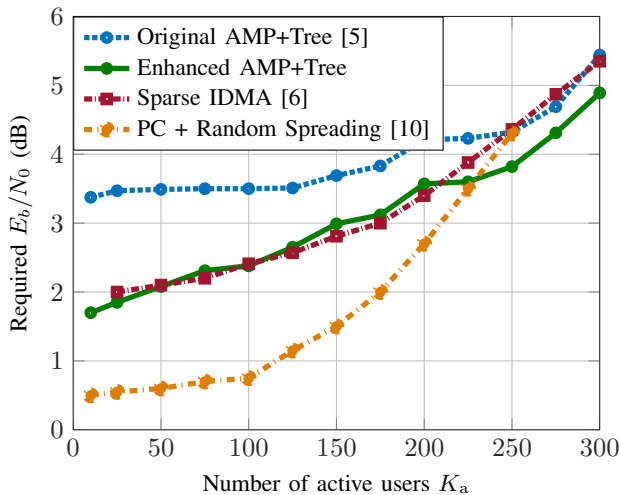
\begin{figure}[h]
\centerline{\input{Figures/results_ISIT.tex}}
  \caption{The figure compares the performance of the proposed scheme with existing schemes.}
  \label{fig:resultsISIT}
\end{figure}

A prime goal of the numerical section is to demonstrate that the enhanced CCS-AMP algorithm described in this article delivers significant improvements over the original algorithm introduced by Fengler et al.~\cite{fengler2019sparcs} without excessively increasing the computational load.
Figure~\ref{fig:resultsISIT} shows the minimum $E_b/N_0$ necessary to achieve the per user probability of error as a function of the number of active devices.
The top curve therein demonstrates the performance of the original AMP-based scheme introduced in~\cite{fengler2019sparcs}, which essentially employs uninformative prior probabilities within every AMP composite iteration.
The performance of the CCS-AMP algorithm that dynamically leverages belief propagation on the factor graph of the tree code is captured by the solid curve.
It can be seen that the enhanced decoder outperforms the original decoder for all values of $\Ka$, and the gain is more pronounced for small values of $\Ka$.
Furthermore, the numerical simulations feature the same parameters, the same underlying factor graphs, sensing matrices, and two-pass decoding.
In this sense, the comparison is straightforward and fair.
The third curve, which is based on sparse-IDMA~\cite{ pradhan2019sparseidma} represents the state-of-the-art for $\Ka \ge 250$.
Our proposed scheme outperforms this benchmark in this same region and exhibits comparable performance in other regions.
The bottom most curve corresponds to a polar coding and random spreading based scheme~\cite{AKPolar} which is the state-of-the-art for $\Ka \le 250$.
This latter scheme outperforms our proposed scheme for small values of $\Ka$, but falls behind when $\Ka > 225$.
It is also worth mentioning that the polar coding scheme is computationally much more demanding, which may prevent its applications in certain scenarios, especially in real-time settings.

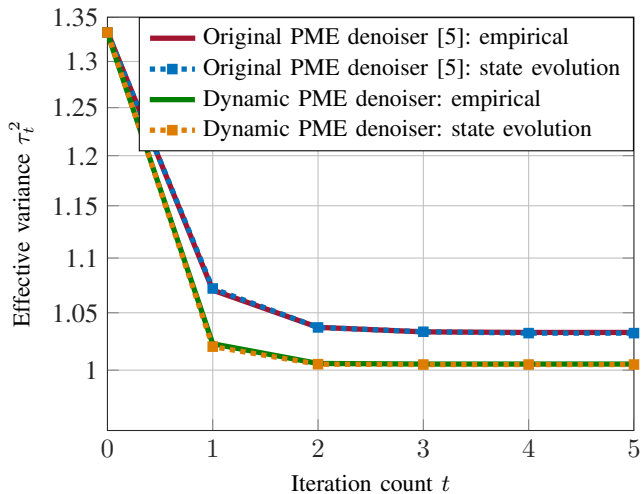
\begin{figure}[h]
\centering
  \input{Figures/State_evolution_results_K25.tex}
  \caption{The figure compares the variance parameter $\tau_t^2$ obtained empirically with that predicted by state evolution for the original PME denoiser in \cite{fengler2019sparcs} and the proposed dynamic PME denoiser. The parameters used to generate these plots are: $\Ka=25, \frac{E_b}{N_0} = 3~dB$.}
  \label{fig:stateEvolution1}
\end{figure}
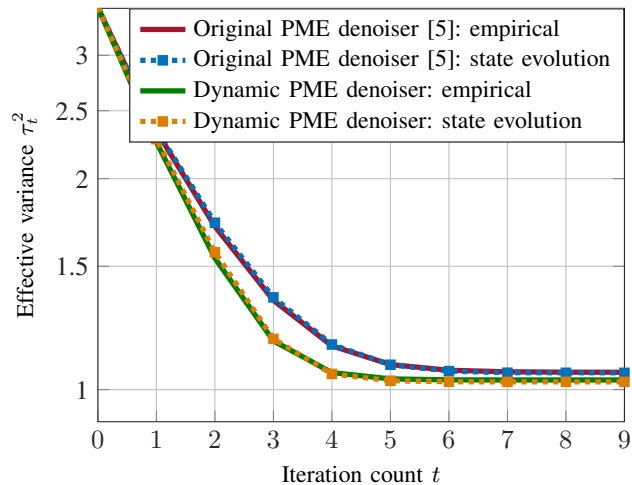
\begin{figure}[h]
\centering
  \input{Figures/State_evolution_results_K150.tex}
  \caption{The figure compares the variance parameter $\tau_t^2$ obtained empirically with that predicted by state evolution for the original PME denoiser in \cite{fengler2019sparcs} and the proposed dynamic PME denoiser. The parameters used to generate these plots are: $\Ka=150, \frac{E_b}{N_0} = 4~dB$.}
  \label{fig:stateEvolution2}
\end{figure}
Another important approach to highlight the benefits of enhanced CCS-AMP over the original AMP-based scheme is through the state evolution discussed in Section~\ref{subSection:SE}.
Recall that parameter $\tau_t^2$ can be interpreted as the variance of the noise in the effective observation at AMP iteration $t$.
A system is performing better when this quantity decreases rapidly as a function of the iteration count, and when its asymptotic value is low.
In Fig.~\ref{fig:stateEvolution1} and Fig.~\ref{fig:stateEvolution2}, we provide a comparison of the two schemes using this viewpoint.
This is accomplished by plotting the variance parameters for the two algorithms obtained empirically through numerical simulations and via the state evolution framework described in Section~\ref{subSection:SE} for $\Ka = 25$ and $\Ka=150$, respectively.
In both cases, the enhanced CCS-AMP algorithm offers noticeable improvements in terms of decay rate and asymptotic value.
We also note that the empirical simulations are very close to the curves afforded by our asymptotic analysis.
This suggests that the state evolution accurately predicts system performance for both the original PME denoiser in~\cite{fengler2019sparcs} and the dynamic PME denoiser put forth in this article, thereby validating the framework developed in Section~\ref{subSection:SE}.
In other words, the system parameters we are interested in are well within the regime where the performance is predicated accurately by the state evolution.

\begin{figure}[h]
\centering
  \input{Figures/runtimeComparison.tex}
  \caption{The figure compares the average run-time of the proposed scheme with the scheme in \cite{fengler2019sparcs}.}
  \label{fig:runtimeComparison}
\end{figure}
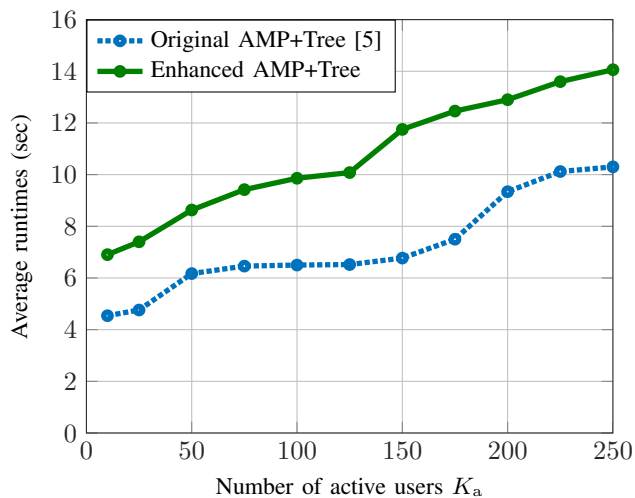
Figure~\ref{fig:runtimeComparison} compares the runtimes of the proposed decoder and the original decoder in~\cite{fengler2019sparcs}.
A recurring theme that underlies much of the research in unsourced, uncoordinated random access is that performance often comes at the cost of computational complexity.
It is therefore pertinent to discuss the computational burden imposed by the dynamic PME denoiser.
We can see from this figure that runtimes are comparable, although then enhanced version does require more resources.
Overall, it seems that for most applications, the performance benefits afforded by the enhanced version would outweigh the computational overhead created by the message passing over the outer factor graph.

\section{Conclusion}
\label{section:Conclusion}

This article presents a novel algorithm for unsourced, uncoordinated random access.
The proposed scheme builds on the connection between coded compressed sensing (CCS) and approximate message passing (AMP), which was first identified by Fengler et al. in~\cite{fengler2019sparcs}.
A main contribution of our work is the realization that the inner and outer codes embedded in this connection can be made to interact dynamically within a unified iterative framework.
Yet, making this interactive framework possible demands several key innovations.
The design of the outer tree code has to be modified in a way that enables belief propagation on a factor graph, while also providing meaningful information to the AMP inner code.
To this end, the class of codes considered herein are based on triadic designs.
Second, parity blocks have to be created in a manner conducive to the application of Fast Fourier techniques or equivalent algorithms.
This is achieved by making sure that the parity precursors have a circular convolution structure.
The dynamic approach yields a non-separable denoiser that is amenable to state evolution, and the overall performance is a significant improvement over the original AMP-based scheme.

The framework introduced in the article also points to several possible avenues moving forward.
The dynamic interaction between the AMP inner code and the BP-based outer code seems natural.
Although the architecture presented in the article is rooted in the lessons learned from previous contributions, it seems that the framework can be extended to a rich class of factor graphs for the outer code.
While a circular structure conducive to FFT techniques is used in our treatment of the algorithm, the proposed paradigm would also extend to other groups or fields.
For instance, it appears that the Walsh–Hadamard transform could be utilized as a means to efficiently compute beliefs on a different factor graph.
This discussion points to possible improvements in performance based on alternate designs for the outer code, possibly exploring hierarchical or cascading structures.
Likewise, there are options to consider with respect to the design of the sensing matrix.
In particular, the design philosophy put forth in the article could be retrofitted to coded compressed sensing as a means to either devise a system with good performance and lower complexity, or as a strategy to handle observation vector that are much larger.
Yet, these and other research avenues are beyond the scope of this article, and they are left as possible future endeavors.

\appendices

\section{Performance of Tree Code Revisited}
\label{Section:PerformanceTreeCodeAppendix}

The performance of the original tree code, as it pertains to coded compressive sensing, is studied at length in~\cite{amalladinne2019coded}.
Herein, we show that, although the parity encoding is fundamentally different for the alternate tree code, the structure of the code and the ensuing statistical properties that underlie overall performance are preserved.
Recall that the difference between the two versions of the tree code lies in the generation of parity bits.
In the original scheme, parities are created using random linear combinations of information bits
\begin{equation} \label{equation:ParityGeneration1}
\pv^{\prime}(\ell) = \sum_{j=1}^{\ell-1} \wv(j) \mathbf{G}_{j,\ell}
\end{equation}
whereas, in the current scheme, we have
\begin{equation} \label{equation:ParityGeneration2}
\pv(\ell) = f_{\mathbb{Z}/2^{p_{\ell}}\mathbb{Z} \rightarrow \mathbb{F}_2^{p_{\ell}}} \left(
\sum_{j=1}^{\ell-1} f_{\mathbb{F}_2^{p_{\ell}} \rightarrow \mathbb{Z}/2^{p_{\ell}}\mathbb{Z}} \big( \wv(j) \mathbf{G}_{j,\ell} \big) \mod 2^{p_{\ell}} \right) .
\end{equation}
Beyond this distinction, the two tree codes are structurally identical.
Below, we show that the statistical properties that serve as a foundation for the analysis of tree decoding in~\cite{amalladinne2019coded} are preserved under the alternate parity generation of \eqref{equation:ParityGeneration2}.

\subsection{Single-Fragment Message}
\label{section:SingleFragment}

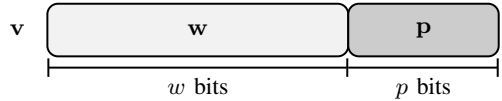
\begin{figure}[htb]
  \centering
  \input{Figures/subvector}
  \caption{Message $\vv$ is obtained by starting with message fragment $\wv$ and appending $p$ parity bits to it.}
  \label{figure:subvector}
\end{figure}

Paralleling the development in~\cite{amalladinne2019coded}, we first examine the case where $\vv$ features a single fragment, as depicted in Fig.~\ref{figure:subvector}.
Specifically, consider a binary message $\wv$ of length $w$.
Parity bits are generated for this message using \eqref{equation:ParityGeneration2} with generator matrix $\mathbf{G}$.
The ensuing codeword, $\vv = \wv \pv$, is then created by taking message $\wv$ and appending parity vector $\pv$ to it, with
\begin{equation*}
\begin{split}
\pv &= f_{\mathbb{Z}/2^{p}\mathbb{Z} \rightarrow \mathbb{F}_2^{p}} \left( f_{\mathbb{F}_2^{p} \rightarrow \mathbb{Z}/2^{p}\mathbb{Z}} \big( \wv \mathbf{G} \big) \mod 2^{p} \right) \\
&= f_{\mathbb{Z}/2^{p}\mathbb{Z} \rightarrow \mathbb{F}_2^{p}} \left( f_{\mathbb{F}_2^{p} \rightarrow \mathbb{Z}/2^{p}\mathbb{Z}} \big( \wv \mathbf{G} \big) \right)
= \wv \mathbf{G} .
\end{split}
\end{equation*}
Since the alternate encoding is equivalent to the original encoding for the one-fragment case, the revised tree encoding trivially inherits the following result.

\begin{lemma}
Fix information vector $\wv$ and parity generating matrix $\mathbf{G}$.
The probability that a randomly selected information vector $\wv_{\mathrm{r}} \in \{ 0, 1 \}^{w}$ produces the same parity sub-component as $\wv$ under $\mathbf{G}$ is given by
\begin{equation*}
\Pr (\pv = \pv_{\mathrm{r}}) = 2^{- \operatorname{rank} (\mathbf{G})} .
\end{equation*}
\end{lemma}

A quantity of interest to our upcoming discussion is the probability distributions of parity patterns for distinct information messages.
Because of the non-linear operations involved in creating parity bits, through multiple representations, the derivation of these distributions is admittedly cumbersome, yet necessary.

\begin{lemma} \label{lemma:RandomGenerator}
Fix erroneous vector $\wv_{\mathrm{e}} \neq \wv$.
Let parity generator matrix $\mathbf{G}$ be a Rademacher matrix of size $w \times p$.
That is, the entries in $\mathbf{G}$ are drawn at random from a uniform Bernoulli distribution, independently of one another.
Under such circumstances,
\begin{equation} \label{equation:DistributionUniform}
f_{\mathbb{F}_2^{p} \rightarrow \mathbb{Z}/2^{p}\mathbb{Z}} \big( \wv \mathbf{G} \big)
- f_{\mathbb{F}_2^{p} \rightarrow \mathbb{Z}/2^{p}\mathbb{Z}} \big( \wv_{\mathrm{e}} \mathbf{G} \big) \mod 2^{p}
\end{equation}
is uniformly distributed over $\mathbb{Z}/2^{p}\mathbb{Z}$ and
the probability of event $\{ \pv = \pv_{\mathrm{e}} \}$ is equal to
\begin{equation*}
\Pr (\pv = \pv_{\mathrm{e}}) = 2^{-p} .
\end{equation*}
\end{lemma}
\begin{IEEEproof}
The event $\{ \pv = \pv_{\mathrm{e}} \}$ is equivalent to $\{ \wv \mathbf{G} = \wv_{\mathrm{e}} \mathbf{G} \}$.
Since $\wv \neq \wv_{\mathrm{e}}$, there exists at least one pair of vector entries, say at location~$i$, such that $(\wv)_i \neq (\wv_{\mathrm{e}})_i$.
In view of the symmetry in the problem, we can assume without loss of generality that $(\wv)_i = 1$ and $(\wv_{\mathrm{e}})_i = 0$.
Then, $\wv_{\mathrm{e}} \mathbf{G} = \sum_{r \neq i} (\wv_{\mathrm{e}})_r \mathbf{G}[r, :]$; we emphasize that the $i$th row of $\mathbf{G}$ does not enter this summation.
For any $k \in \mathbb{Z}/2^{p}\mathbb{Z}$, we can write
\begin{equation*}
\begin{split}
&\Pr \left( f_{\mathbb{F}_2^{p} \rightarrow \mathbb{Z}/2^{p}\mathbb{Z}} \big( \wv \mathbf{G} \big)
- f_{\mathbb{F}_2^{p} \rightarrow \mathbb{Z}/2^{p}\mathbb{Z}} \big( \wv_{\mathrm{e}} \mathbf{G} \big) \equiv k \mod 2^{p} \right) \\
&= \Pr \left( f_{\mathbb{F}_2^{p} \rightarrow \mathbb{Z}/2^{p}\mathbb{Z}} \big( \wv \mathbf{G} \big) \equiv k
+ f_{\mathbb{F}_2^{p} \rightarrow \mathbb{Z}/2^{p}\mathbb{Z}} \left( \sum_{r \neq i} (\wv_{\mathrm{e}})_r \mathbf{G}[r, :] \right) \mod 2^{p} \right) \\
&= \Pr \left( \wv \mathbf{G} = f_{\mathbb{Z}/2^{p}\mathbb{Z} \rightarrow \mathbb{F}_2^{p} } \left( k
+ f_{\mathbb{F}_2^{p} \rightarrow \mathbb{Z}/2^{p}\mathbb{Z}} \left( \sum_{r \neq i} (\wv_{\mathrm{e}})_r \mathbf{G}[r, :] \right) \mod 2^{p} \right) \right) \\
&= \Pr \left( \mathbf{G}[i,:] = f_{\mathbb{Z}/2^{p}\mathbb{Z} \rightarrow \mathbb{F}_2^{p} } \left( k
+ f_{\mathbb{F}_2^{p} \rightarrow \mathbb{Z}/2^{p}\mathbb{Z}} \left( \sum_{r \neq i} (\wv_{\mathrm{e}})_r \mathbf{G}[r, :] \right) \mod 2^{p} \right) - \sum_{r \neq i} (\wv)_r \mathbf{G}[r, :] \right) \\
\end{split}
\end{equation*}
By constructions, $\mathbf{G}[i,:]$ is a sequence of uniform Bernoulli trials, independent of the term on the right-hand-side.
Therefore, the probability that this equality is met is $2^{-p}$, irrespective of $k$.
That is, \eqref{equation:DistributionUniform} is uniformly distributed over $\mathbb{Z}/2^{p}\mathbb{Z}$, as claimed.
The second part of the lemma follows immediately from above and the observation that the event $\{ \pv = \pv_{\mathrm{e}} \}$ is true if and only if \eqref{equation:DistributionUniform} is congruent to zero in $\mathbb{Z}/2^{p}\mathbb{Z}$.
\end{IEEEproof}

\subsection{Multi-Fragment Message}
\label{section:MultiFragment}

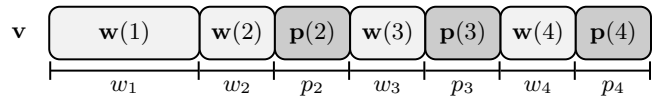
\begin{figure}[htb]
  \centering
  \input{Figures/subvectors}
  \caption{The fragmented nature of tree coding leads to a certain structure in $\vv$, as depicted above.
  This renders the performance analysis of this scheme more challenging due, partly, to the fact that two different  messages can share identical information fragments.
  }
  \label{figure:subvectors}
\end{figure}

The situation becomes slightly more complicated when parity bits are generated over multiple fragments.
This produces a structure akin to the one portrayed in Fig.~\ref{figure:subvectors}.
Every parity pattern $\pv(\ell)$ therein is produced using \eqref{equation:ParityGeneration2}.
We wish to understand how this new way of adding redundancy impacts the probability of erroneous paths staying alive as the tree decoder proceeds forward through the various levels.
To begin, we look at the probability that an erroneous (partial) path produces the same parity pattern as a valid codeword for one block.

\begin{lemma} \label{lemma:ParityDistinctVecotrs}
Suppose information message $\wv$ is fixed.
Consider truncated erroneous vector
\begin{equation} \label{equation:ErroneousCondition}
\wv_{\mathrm{e}}(1) \cdots \wv_{\mathrm{e}}(\ell-1) \neq \wv(0) \cdots \wv(\ell-1) .
\end{equation}
Let $\left\{ \mathbf{G}_{j,\ell} \right\}$ be a collection of independent Rademacher matrices, each of size $w_{j} \times p_{\ell}$.
In other words, the entries in $\mathbf{G}_{j,\ell}$ are drawn at random from a uniform Bernoulli distribution, independently of one another and of other matrices.
Under such circumstances, the probability of event $\{ \pv = \pv_{\mathrm{e}} \}$ is given by
\begin{equation*}
\Pr (\pv(\ell) = \pv_{\mathrm{e}}(\ell)) = 2^{-p_{\ell}} .
\end{equation*}
\end{lemma}
\begin{IEEEproof}
With the revised tree encoding, parity bits at level~$\ell$ are generated using \eqref{equation:ParityGeneration2}.
Accordingly, $\pv(\ell) = \pv_{\mathrm{e}}(\ell)$ if and only if
\begin{equation} \label{equation:ParityCongruent}
\begin{split}
&\sum_{j=1}^{\ell-1} f_{\mathbb{F}_2^{p_{\ell}} \rightarrow \mathbb{Z}/2^{p_{\ell}}\mathbb{Z}} \big( \wv(j) \mathbf{G}_{j,\ell} \big)
\equiv \sum_{j=1}^{\ell-1} f_{\mathbb{F}_2^{p_{\ell}} \rightarrow \mathbb{Z}/2^{p_{\ell}}\mathbb{Z}} \big( \wv_{\mathrm{e}}(j) \mathbf{G}_{j,\ell} \big) \mod 2^{p_{\ell}} .
\end{split}
\end{equation}
This follows because $f_{\mathbb{Z}/2^{p_{\ell}}\mathbb{Z} \rightarrow \mathbb{F}_2^{p_{\ell}}} (\cdot)$ is a bijection.
Since truncated vectors $\wv_{\mathrm{e}}(1) \cdots \wv_{\mathrm{e}}(\ell-1) \neq \wv(1) \cdots \wv(\ell-1)$, there exists a $k$ such that $\wv_{\mathrm{e}}(k) \neq \wv(k)$.
With this $k$, we can rewrite \eqref{equation:ParityCongruent} as
\begin{equation} \label{equation:ParityCongruent1}
\begin{split}
& \left( f_{\mathbb{F}_2^{p_{\ell}} \rightarrow \mathbb{Z}/2^{p_{\ell}}\mathbb{Z}} \big( \wv(k) \mathbf{G}_{k,\ell} \big)
- f_{\mathbb{F}_2^{p_{\ell}} \rightarrow \mathbb{Z}/2^{p_{\ell}}\mathbb{Z}} \big( \wv_{\mathrm{e}}(k) \mathbf{G}_{k,\ell} \big) \right) \\
&+ \sum_{\substack{j=1 \\ j \neq k}}^{\ell-1} \left( f_{\mathbb{F}_2^{p_{\ell}} \rightarrow \mathbb{Z}/2^{p_{\ell}}\mathbb{Z}} \big( \wv(j) \mathbf{G}_{j,\ell} \big) - f_{\mathbb{F}_2^{p_{\ell}} \rightarrow \mathbb{Z}/2^{p_{\ell}}\mathbb{Z}} \big( \wv_{\mathrm{e}}(j) \mathbf{G}_{j,\ell} \big) \right)
\equiv 0 \mod 2^{p_{\ell}} .
\end{split}
\end{equation}
By construction, the isolated difference and random matrix $\mathbf{G}_{k,\ell}$ are independent of the succeeding summation.
Furthermore, Lemma~\ref{lemma:RandomGenerator} asserts that this first term is uniformly distributed over $\mathbb{Z}/2^{p_{\ell}}\mathbb{Z}$.
We then deduce that, under condition \eqref{equation:ErroneousCondition}, $\Pr \left( \pv(\ell) = \pv_{\mathrm{e}}(\ell) \right) = 2^{-p_{\ell}}$, as desired.
\end{IEEEproof}

At this point, we draw a parallel between the properties of the revised tree code and those of the original scheme.
There remains a dichotomy in the impact of parity patterns: 
the parity section $\pv(\ell)$ associated with stage~$\ell$ either acts as a statistically discriminating sequence of independent Bernoulli samples, each with probability half, or the parity conditions are fulfilled trivially when the truncated vectors match.
Beyond this observation, there remain three confounding factors in the analysis of multi-fragment codewords.
First, several fragments within an erroneous candidate codeword may come from a same message;
overlapping fragments reduce the propensity for parity bits to be statistically discriminating.
Second, two different messages may have identical information fragments, as these messages only need to differ one location overall.
Mathematically, when comparing a valid codeword to an erroneous candidate, the two messages are necessarily distinct with $\wv_{\mathrm{e}} \neq \wv$; however, it is possible to have $\wv_{\mathrm{e}}(\ell) = \wv(\ell)$ for a (strict) subset of $\{ 1, \ldots, L \}$.
Finally, the loss of discriminating power from parity constraints may be correlated across fragments in certain cases, which exacerbates the error probability.
Nevertheless, these confounding factors seem intrinsic to the tree code structure.
They remain unchanged irrespective of whether parity bits are constructed using the original scheme or the revised method.
Consequently, these two distinct approaches yield the same expected performance.

\begin{proposition}
The performance of the revised tree code with the parity structure of \eqref{equation:ParityGeneration2} and that of the original tree code introduced in~\cite{amalladinne2018couple,amalladinne2019coded} with random linear combinations as in \eqref{equation:ParityGeneration1} are identical.
\end{proposition}
\begin{IEEEproof}
Consider a situation where the tree decoder seeks to validate codewords that start with root fragment $\wv_{i_1}(1)$.
For a given collection of $\Ka$ transmitted codewords, the list of candidate codewords visited during this phase of the tree decoding process is composed of elements of the form
\begin{equation} \label{equation:CandidateCodewords}
\begin{split}
\vv_{\mathrm{c}} &= \vv_{i_1}(1) \vv_{i_2} (2) \cdots \vv_{i_L}(L) \\
&= \wv_{i_1}(1) \wv_{i_2}(2) \pv_{i_2}(2) \cdots \wv_{i_L}(L) \pv_{i_L}(L) ,
\end{split}
\end{equation}
where $i_{\ell} \in [1:\Ka]$ for all slots $\ell \in [2:L]$.
The ability of the tree decoder to identify an erroneous sequence hinges on the structure of the candidate vector.
In particular, this erroneous codeword will be inconspicuous if the parity patterns are consistent, i.e.,
\begin{equation}
\pv_{i_{\ell}}(\ell) = f_{\mathbb{Z}/2^{p_{\ell}}\mathbb{Z} \rightarrow \mathbb{F}_2^{p_{\ell}}} \left(
\sum_{j=1}^{\ell-1} f_{\mathbb{F}_2^{p_{\ell}} \rightarrow \mathbb{Z}/2^{p_{\ell}}\mathbb{Z}} \big( \wv_{i_j}(j) \mathbf{G}_{j,\ell} \big) \mod 2^{p_{\ell}} \right)
\quad \ell = 2, \ldots, L .
\end{equation}
In view of Lemma~\ref{lemma:ParityDistinctVecotrs} and for a fixed index sequence $i_1, i_2, \ldots, i_L$, the probability of the event above is equal to the probability of the erroneous sequence being undetected under the original tree code, i.e.,
\begin{equation}
\pv_{i_{\ell}}^{\prime}(\ell) = \sum_{j=1}^{\ell-1} \wv_{i_{\ell}}(j) \mathbf{G}_{j,\ell}
= \sum_{j=1}^{\ell-1} \wv_{i_j}(j) \mathbf{G}_{j,\ell}
\quad \ell = 2, \ldots, L .
\end{equation}
That is, the probability that an erroneous sequence survives, conditioned on a given index sequence, is the same in both systems.
Because this statement holds for every possible index sequence, the analysis of performance in~\cite{amalladinne2019coded}, which relies on the groupings of such conditional probabilities, too applies for the revised tree code.
That is, performance in terms of the expected number of surviving paths and overall probability of decoding failure is identical for the two schemes.
\end{IEEEproof}





\section{Message Passing Rules on the Graph}
\label{appendix:MessagePassingRulesGraph}

In this section, we derive the form of the messages passed from check nodes to variable nodes on the factor graph associated with the tree code.
These messages are employed in the context of belief propagation, as discussed in Section~\ref{section:TreeCodeRevisited}.
Paralleling our earlier treatment, we assume that the effective observation available to the tree decoder is of the form $\rv = \mathbf{D} \sv + \tau \zetav$, where $\zetav$ is an i.i.d.\ $\mathcal{N}(0,1)$ random vector and $\tau$ is the standard deviation of the individual noise components.

\subsection{Local Estimates}
\label{subsection:LocalEstimates}

A quantity needed for belief propagation is a local estimate for the probability that a particular device~$i$ is sending a message $\mv_i$ whose $k$th component in section~$\ell$ is a one, conditioned on effective observation $\rv(\ell)$.
As an initial step, we examine a sequence of one-sparse (labeled) candidate blocks $\hat{\mv}_1(\ell), \ldots, \hat{\mv}_{\Ka}(\ell)$ with $\ell \in [L]$.
Under the Gaussian model, the conditional probability of this candidate sequence, given $\rv(\ell)$, is governed by
\begin{equation} \label{equation:MessagePart0}
\begin{split}
&\Pr \left( \hat{\mv}_1(\ell), \ldots, \hat{\mv}_{\Ka}(\ell) | \rv(\ell) \right)
= \frac{\exp\left({- \frac{\left\| \rv(\ell) - d_{\ell} \hat{\sv}(\ell) \right\|^2}{2 \tau^2}}\right)}
{\sum_{\widehat{\mv}_1(\ell), \ldots, \widehat{\mv}_{\Ka}(\ell)} \exp\left({- \frac{\left\| \rv(\ell) - d_{\ell} \widehat{\sv}(\ell) \right\|^2}{2 \tau^2}}\right)}
\end{split}
\end{equation}
where $\hat{\sv}(\ell) = \sum_{i \in [\Ka]} \hat{\mv}_i (\ell)$ and the sum in the denominator ranges over all possible assignments.
We stress that \eqref{equation:MessagePart0} relies on the assumption that message block $\mv_i(\ell)$ is selected uniformly at random from the $m_{\ell}$ one-sparse candidates, independently from other messages.
We further examine the exponent in \eqref{equation:MessagePart0} by writing
\begin{equation} \label{equation:MessageApproximation1}
\begin{split}
&\left\| \rv(\ell) - d_{\ell} \hat{\sv} (\ell) \right\|^2
= \left\| \rv(\ell) \right\|^2 - 2 d_{\ell} \left\langle \rv(\ell) , \hat{\sv} (\ell) \right\rangle + d_{\ell}^2 \left\| \hat{\sv}(\ell) \right\|^2 \\
&= \left\| \rv(\ell) \right\|^2 - 2 d_{\ell} \sum_{i \in [\Ka]} \left\langle \rv(\ell) , \hat{\mv}_i (\ell) \right\rangle + d_{\ell}^2 \Ka + d_{\ell}^2 \left( \left\| \hat{\sv}(\ell) \right\|^2 - \Ka \right) \\
&= \left\| \rv(\ell) \right\|^2
+ \sum_{i \in [\Ka]} \left( \left( \rv \left( \ell,k^{(i)} \right) - d_{\ell} \right)^2 - \rv \left( \ell,k^{(i)} \right)^2 \right) + d_{\ell}^2 \left( \left\| \hat{\sv}(\ell) \right\|^2 - \Ka \right)
\end{split}
\end{equation}
where $k^{(i)}$ is the index of the unique non-zero entry in $\hat{\mv}_i (\ell)$.
We can express \eqref{equation:MessagePart0} using this characterization,
\begin{equation} \label{equation:MessagePart1b}
\begin{split}
\Pr \left( \hat{\mv}_1(\ell) \ldots, \hat{\mv}_{\Ka}(\ell) | \rv(\ell) \right)
&\propto \exp\left({- \frac{d_{\ell}^2 \left( \left\| \hat{\sv}(\ell) \right\|^2 - \Ka \right)}{2 \tau^2}}\right)
\prod_{i \in [\Ka]} \exp\left({- \frac{\left( \rv \left( \ell,k^{(i)} \right) - d_{\ell} \right)^2 - \rv \left( \ell,k^{(i)} \right)^2}{2 \tau^2}}\right) \\
&\propto \exp\left({- \frac{d_{\ell}^2 \left( \left\| \hat{\sv}(\ell) \right\|^2 - \Ka \right)}{2 \tau^2}}\right)
\prod_{i \in [\Ka]} \mathcal{L}_{\ell} \left( k^{(i)} \right)
\end{split}
\end{equation}
where the last line makes use of the likelihood notation $\mathcal{L}_{\ell} (k) = \exp\left({- \frac{\left( \rv(\ell, k) - d_{\ell} \right)^2 - \rv(\ell, k)^2}{2 \tau^2}}\right)$.
Turning to the marginal distribution of a specific device, say device one, we can calculate the probability that its message block~$\ell$ contains a one at location~$k^{(1)}$ based on \eqref{equation:MessagePart1b}.
Specifically, we get the intricate normalized expression
\begin{equation} \label{equation:MessagePart1c}
\begin{split}
\Pr \left( \hat{\mv}_1(\ell) | \rv(\ell) \right)
&= \frac{\sum_{\left( \hat{\mv}_2(\ell) \ldots, \hat{\mv}_{\Ka}(\ell) \right)}
\exp\left({- \frac{d_{\ell}^2 \left( \left\| \hat{\sv}(\ell) \right\|^2 - \Ka \right)}{2 \tau^2}}\right)
\prod_{i \in [\Ka]} \mathcal{L}_{\ell} \left( k^{(i)} \right)}
{\sum_{\left( \widehat{\mv}_1(\ell) \ldots, \widehat{\mv}_{\Ka}(\ell) \right)}
\exp\left({- \frac{d_{\ell}^2 \left( \left\| \widehat{\sv}(\ell) \right\|^2 - \Ka \right)}{2 \tau^2}}\right)
\prod_{i \in [\Ka]} \mathcal{L}_{\ell} \left( k^{(i)} \right)} .
\end{split}
\end{equation}
While this characterization is exact, computing local marginal probabilities based on \eqref{equation:MessagePart1c} is a formidable task for the problem dimensions we are interested in.
This is impractical for the iterative algorithm we wish to develop and, consequently, we seek a suitable low-complexity approximation.

As a first step, we neglect assignments with multiplicities, i.e., $\left\| \hat{\sv}(\ell) \right\|_0 < \Ka$; this seems reasonable because these cases are heavily discounted by the multiplicative factor $\exp \left( {- \frac{d_{\ell}^2 \left( \left\| \hat{\sv}(\ell) \right\|^2 - \Ka \right)}{2 \tau^2}} \right)$ and they are also unlikely when sections are large.
It may be helpful to point out that, in comparison, this factor is equal to one when there are no collisions, i.e., $\left\| \hat{\sv}(\ell) \right\|_0 = \Ka$.
Disregarding collisions, the ratio in \eqref{equation:MessagePart1c} becomes simpler.
Under index notation, marginal estimates can be written as
\begin{equation} \label{equation:MessagePart1d}
\begin{split}
&\Pr \left( \mv_1 (\ell, k) = 1 | \rv(\ell) \right)
\approx \frac{\mathcal{L}_{\ell} (k) \sum_{\kappav : k \in \kappav}
\prod_{\kappa \in \kappav \setminus k} \mathcal{L}_{\ell}(\kappa)}
{\Ka \sum_{\kappav} \prod_{\kappa \in \kappav} \mathcal{L}_{\ell}(\kappa)} \\
&= \frac{\mathcal{L}_{\ell} (k) \sum_{\kappav : k \in \kappav}
\prod_{\kappa \in \kappav \setminus k} \mathcal{L}_{\ell}(\kappa)}
{\Ka \mathcal{L}_{\ell} (k) \sum_{\kappav : k \in \kappav}
\prod_{\kappa \in \kappav \setminus k} \mathcal{L}_{\ell}(\kappa)
+ \sum_{\kappav : k \notin \kappav}
\prod_{\kappa \in \kappav} \mathcal{L}_{\ell}(\kappa)} .
\end{split}
\end{equation}
The set $\kappav$ represents a subset of $\Ka$ distinct indices in $\left\{ 0, \ldots, m_{\ell} - 1 \right\}$.
Every summand accounts for collections of non-colliding assignments $\hat{\mv}_1(\ell), \ldots, \hat{\mv}_{\Ka}(\ell)$ whose equivalent index representations are jointly equal to $\kappav$; we stress that there are $\Ka$ factorial such assignments for a given $\kappav$, one for each index permutation.
The corresponding entries in \eqref{equation:MessagePart1c} all share the same value because the product of likelihoods is permutation invariant.

Computing \eqref{equation:MessagePart1d} remains a sizeable task for parameters of interest, and it may be too costly to embed in an AMP iterate.
Accordingly, we simplify our marginal estimator by replacing the likelihoods by a scaled version of the marginal posterior mean estimate (PME) introduced by Fengler et al.~\cite{fengler2019sparcs}.
This estimate is discussed at length in Section~\ref{subsection:TestStatistics}.
For the time being, it suffices to state that
\begin{equation} \label{equation:MessagePart1d-PME}
\Pr \left( \mv_1 (\ell, k) = 1 | \rv(\ell) \right)
\approx \frac{1}{\Ka} \frac{q \exp \left( - \frac{ \left( \rv(\ell, k) - d_{\ell} \right)^2}{2 \tau^2} \right)}
{(1-q) \exp \left( -\frac{\rv(\ell, k)^2}{2 \tau^2} \right)
+ q \exp \left( - \frac{ \left( \rv(\ell, k) - d_{\ell} \right)^2}{2 \tau^2} \right)}
\end{equation}
where $q$ and $\tau$ are a judiciously selected constants.
While \eqref{equation:MessagePart1d-PME} is guaranteed to produce bounded, non-negative elements; the ensuing vector may not be normalized.
We can rectify this situation by adding a normalizing factor to the estimates.
Altogether, after accounting for the reshuffling, we propose to use block estimate $\lambdav_{\ell}(k)$ as a proxy for the marginal probability $\Pr \left( \mv_1 (\ell, k) = 1 | \rv(\ell) \right)$, where we define
\begin{equation} \label{equation:MessageEstimate}
\begin{split}
\lambdav_{\ell} (k)
&\propto \frac{q \exp \left( - \frac{ \left( \rv(\ell, k) - d_{\ell} \right)^2}{2 \tau^2} \right)}
{(1-q) \exp \left( -\frac{\rv(\ell, k)^2}{2 \tau^2} \right)
+ q \exp \left( - \frac{ \left( \rv(\ell, k) - d_{\ell} \right)^2}{2 \tau^2} \right)} .
\end{split}
\end{equation}
The `$\propto$' symbol accounts for the normalization constant that makes $\left\| \lambdav_{\ell} \right\|_1 = 1$.
At low signal-to-noise ratios, this estimator remains very uninformative and outputs near uniform probabilities.
However, as the signal-to-noise ratio increases, the mass of the estimated vector starts to concentrate on fewer entries.
This is the type of behavior needed for tree decoding to help improve AMP performance.

\subsection{Global Estimates}

Proceeding forward, we turn our attention to the \emph{a posteriori} probability of message sequence $\hat{\mv}_1, \ldots, \hat{\mv}_{\Ka}$ conditioned on $\rv$.
Taking into consideration the parity requirements for valid messages through indicator function $\mathcal{G} (\cdot)$, we can write
\begin{equation} \label{equation:MessagePart1}
\Pr \left( \hat{\mv}_1 \ldots, \hat{\mv}_{\Ka} | \rv \right)
\propto \prod_{i \in [\Ka]} \mathcal{G} \left( \hat{\mv}_{i} \right)
\prod_{j \in [L]} \Pr \left( \hat{\mv}_1(j) \ldots, \hat{\mv}_{\Ka}(j) | \rv(j) \right) .
\end{equation}
The first component in the product enforces parity consistency on every message individually.
The second term comes from the Gaussian model for the effective observation.
Incorporating the approximate expressions derived in the previous section, this equation evolves into
\begin{equation} \label{equation:MessagePart2}
\begin{split}
\Pr \left( \hat{\mv}_1 \ldots, \hat{\mv}_{\Ka} | \rv \right)
&\approx \prod_{i \in [\Ka]} \mathcal{G} \left( \hat{\mv}_{i} \right)
\prod_{j \in [L]}
\prod_{i \in [\Ka]} \Pr \left( \hat{\mv}_i(j) | \rv(j) \right) \\
&\approx \prod_{i \in [\Ka]}
\left( \mathcal{G} \left( \hat{\mv}_{i} \right) \Pr \left( \hat{\mv}_i(j) | \rv(j) \right) \right) .
\end{split}
\end{equation}
The terms in \eqref{equation:MessagePart2} decompose into a product form, one probability element for every active device.
We can further simplify this expression using the marginal estimates from \eqref{equation:MessageEstimate}.
The probability that a fixed device has message $\hat{\mv}$ is approximated by
\begin{equation} \label{equation:MessagePart4}
\begin{split}
\Pr \left( \hat{\mv} | \rv \right)
&\approx \mathcal{G} \left( \hat{\mv} \right) \prod_{\ell \in [L]} \Pr \left( \hat{\mv}(\ell) | \rv(\ell) \right) \\
&\approx \mathcal{G} \left( \hat{\mv} \right) \prod_{\ell \in [L]} \lambdav_{\ell} (k_{\ell})
\end{split}
\end{equation}
where $k_{\ell}$ is the location of the unique non-zero entry in $\hat{\mv} (\ell)$.
We see that the information afforded by the effective observation enters this equation through the local measure $\lambdav_{\ell}$.
It then suffices to collect summary vectors $\left( \lambdav_{\ell} : \ell \in [L] \right)$, where of $\lambdav_{\ell}$ has length $m_{\ell}$, to run belief propagation on the factor graph.

Inspecting \eqref{equation:MessagePart4} reveals that the graphical structure of the tree code is linked to the probability that a fixed device picks a certain index for a particular block.
We connect this quantity explicitly to the variable nodes on the factor graph by interpreting $p_{s_{\ell}}(k)$ as an estimate of the probability that this fixed device has a message $\mv$ containing block~$k$ in section~$\ell$.
Accordingly, we design message passing rules to deliver estimates for $p_{s_{\ell}}(k)$ upon completion of the process.
We can write the block estimates in a format that highlights the underlying factors,
\begin{equation} \label{equation:MessagePart6}
\begin{split}
p_{s_{\ell}} (k)
&\propto \sum_{\kv: k_{\ell} = k} \mathcal{G} \left( \kv \right) \prod_{\ell \in [L]} \lambdav_{\ell} (k_{\ell}) \\
&= \sum_{\kv: k_{\ell} = k} \prod_{a \in \mathcal{P}} \mathcal{G}_a \left( \kv_a \right)
\prod_{j \in [L]} \lambdav_j (k_j) ,
\end{split}
\end{equation}
where $\kv = ( k_1, \ldots, k_{\Ka} )$ is the compact index notation for $\hat{\mv}$ obtained through the relation $k_j = \left[ \hat{\vv}(j) \right]_2$.
Also, we use the localized compact notation applied to neighborhoods, $\kv_{a} = \left( k_j : j \in N(a) \right)$.
Function $\mathcal{G} (\cdot)$ verifies the parity consistency of its vector argument under the tree code structure, as defined in \eqref{equation:ParityGeneration}.
For the tree code, the validity characteristic function $\mathcal{G}(\cdot)$ becomes the product of local factor functions $\left\{ \mathcal{G}_a(\cdot) : a \in \mathcal{P} \right\}$.

Before deriving message passing rules, we must account for the contribution of the local observations through $\left( \lambdav_{\ell} : \ell \in [L] \right)$.
Following established techniques~\cite{kschischang2001factor}, this is achieved by augmenting the graph with the (trivial) factor nodes afforded by the sections of $\rv$.
Note that these latter factors have only one connection each and, as such, their messages are static; they do not evolved with iterations.
With this augmentation, we obtain standard expressions for the message passing rules.
Note that we simply fold the static messages in the expressions without explicitly defining the augmented graph, as this step is common to the treatment of codes.
The message passing rules compute estimates on the marginal distributions $\left( p_{s_{\ell}} : \ell \in [L] \right)$.

\subsection{Message Passing Rules}

A message passed from check node $a_p$ to variable node $s \in N(a_p)$ subscribe to the format
\begin{equation} \label{equation:MessageCheck2Variable}
\muv_{a_p \to s} (k)
= \sum_{\kv_{a_p}: k_p = k} \mathcal{G}_{a_p} \left( \kv_{a_p} \right)
\prod_{s_j \in N(a_p) \setminus s} \muv_{s_j \to a_p} (k_j) .
\end{equation}
Similarly, a message going from variable node $s_{\ell}$ to check node $a \in N(s_{\ell})$ assumes the form
\begin{equation} \label{equation:MessageVariable2Check}
\muv_{s_{\ell} \rightarrow a} (k)
\propto \lambdav_{\ell} (k) \prod_{a_p \in N(s_{\ell}) \setminus a} \muv_{a_p \to s_{\ell}} (k) .
\end{equation}
The `$\propto$' symbol indicates that the measure is renormalized before being sent as a message.
All the dynamic messages are initialized with $\muv_{s \to a} = \onev$ and $\muv_{a \to s} = \onev$.
The parallel sum-product algorithm then iterates between \eqref{equation:MessageCheck2Variable} and \eqref{equation:MessageVariable2Check}.
At any stage of this iterative process, the estimated marginal distribution of a specific device having transmitted block~$k$ at variable node~$s_{\ell}$ is proportional to the product of the current messages from all adjoining factors,
\begin{equation} \label{equation:MessageMarginals}
p_{s_{\ell}} (k) \propto \lambdav_{\ell} (k) \prod_{a \in N(s)} \muv_{a \to s_{\ell}} (k) .
\end{equation}
As usual, this procedure is guaranteed to converge for acyclic graphical models, but not for arbitrary graphs.
Still, it is known to perform well in many cases where the factor graph features cycles.
In our proposed algorithm, the number of belief propagation steps within one composite AMP algorithm is envisioned to be small.

\section{Lipschitz Dynamic PME Denoiser}
\label{appendix:LipschitzPMEDenoiser}

In this section, we show that the dynamic PME denoiser, with one round of message passing on the factor graph of the tree code, is Lipschitz continuous.
The reader may notice that even though the treatment of this proof is restricted to triadic designs for the outer code, it can be extended to accommodate cases beyond those discussed in this paper.
Our proof technique may serve as a blueprint to analyze generic architectures which use a combination of inner AMP and an outer decoder with dynamic interactions between the two.
Our strategy is to show that the magnitudes of the entries in the Jacobian matrix of $\etav_t^{\mathrm{PME}}(\rv)$ with respect to $\rv$ are uniformly bounded.
Recall that
\begin{align*}
\etav_t^{\mathrm{PME}}(\rv)
&= \hat{\sv}_{1}^{\mathrm{PME}} \left( \rv, \tau_t \right) \cdots \hat{\sv}_{L}^{\mathrm{PME}} \left( \rv, \tau_t \right) \\
\hat{\sv}_{\ell}^{\mathrm{PME}} \left( \rv, \tau_t \right)
&= \big( \hat{s}_{\ell} \left( \qv(\ell, k), \rv(\ell, k), \tau_t \right) : k \in 0, \ldots, m_{\ell} - 1 \big) .
\end{align*}
Consequently, we are interested in objects of the form
\begin{equation*}
\frac{\partial \hat{s}_{\ell} \left( \qv(\ell, k), \rv(\ell, k), \tau_t \right)}
{\partial \rv(j,k')} .
\end{equation*}
There are three categories to consider: $j = \ell$, $j \in N(s_{\ell})$, and $j \notin \{ \ell \} \cup N(s_{\ell})$.
The third group is the easiest to treat because the partial derivatives are each equal to zero when only one round of belief propagation is performed on the underlying factor graph.
Next, we turn our attention to cases where $j = \ell$.
Applying Lemma~\ref{lemma:PartialSestimate}, we immediately get
\begin{equation}
\left| \frac{\partial \hat{s}_{\ell} \left( \qv(\ell, k), \rv(\ell, k), \tau_t \right)}
{\partial \rv(\ell,k)} \right|
= \frac{d_{\ell}}{\tau^2} \hat{s}_{\ell} \left( \qv(\ell, k), \rv(\ell, k), \tau_t \right)
\left( 1 - \hat{s}_{\ell} \left( \qv(\ell, k), \rv(\ell, k), \tau_t \right) \right)
\leq \frac{d_{\max}}{2 \tau^2} .
\end{equation}
Furthermore, when $k' \neq k$, we have
\begin{equation}
\frac{\partial \hat{s}_{\ell} \left( \qv(\ell, k), \rv(\ell, k), \tau_t \right)}
{\partial \rv(\ell,k')}
= 0 .
\end{equation}
Therefore, entries in the Jacobian matrix corresponding to this category are uniformly bounded.
It remains to derive a similar bound for scenarios where $j \in N(s_{\ell})$.
Establishing the desired property for this category requires a few steps.
We begin with an analog to Lemma~\ref{lemma:PartialSestimate}.

\begin{lemma} \label{lemma:PartialSestimateQ}
The partial derivative of the posterior mean estimator (PME) defined in Lemma~\ref{lemma:PME} with respect to $q$ is
\begin{equation}
\frac{\partial\hat{s}_{\ell}(q, r, \tau)}{\partial q}
= \frac{\hat{s}_{\ell}(q, r, \tau) \left( 1 - \hat{s}_{\ell}(q, r, \tau) \right)}{q(1-q)} .
\end{equation}
\end{lemma}
\begin{IEEEproof}
Recall that the PME can be rewritten as
\begin{equation}
\begin{split}
\hat{s}_{\ell}(q, r, \tau)
&= \frac{q}{q + (1-q) \exp \left( \frac{d_{\ell}^2 - 2 r d_{\ell}}{2 \tau^2} \right)} .
\end{split}
\end{equation}
By the chain rule of differentiation, we get
\begin{equation*}
\begin{split}
\frac{\partial \hat{s}_{\ell}(q, r, \tau)}{\partial q}
&= \frac{\exp \left( \frac{d_{\ell}^2 - 2 r d_{\ell}}{2 \tau^2} \right)}
{\left( q + (1-q) \exp \left( \frac{d_{\ell}^2 - 2 r d_{\ell}}{2 \tau^2} \right) \right)^2} \\
&= \frac{\hat{s}_{\ell}(q, r, \tau) \left( 1 - \hat{s}_{\ell}(q, r, \tau) \right)}{q(1-q)} ,
\end{split}
\end{equation*}
as stated.
\end{IEEEproof}

Our next step is to link $\qv(\ell,k)$ to quantities defined on the factor graph of the tree code.
Recall that we introduced $\qv(\ell,k)$ in \eqref{equation:DynamicDenoiserPriors} with
\begin{equation} 
\qv(\ell,k) = 1 - \left( 1-\frac{\muv_{s_{\ell}}(k)}{\| \muv_{s_{\ell}} \|_1} \right)^{\Ka}
\end{equation}
As an immediate consequence of this definition, we obtain the inequality
\begin{equation} \label{equation:qmu}
\qv(\ell, k)
= 1 - \left( 1 - \frac{\muv_{s_{\ell}}(k)}{\left\| \muv_{s_{\ell}} \right\|_1} \right)^{\Ka}
\geq 1 - \left( 1 - \frac{\muv_{s_{\ell}}(k)}{\left\| \muv_{s_{\ell}} \right\|_1} \right)
= \frac{\muv_{s_{\ell}}(k)}{\left\| \muv_{s_{\ell}} \right\|_1} .
\end{equation}
Another straightforward, yet useful result related to $\qv(\ell,k)$ is
\begin{equation} \label{equation:qmu-derivative}
\begin{split}
\frac{\partial \qv(\ell, k)}{\partial \rv(j,k')}
&= \Ka \left( 1-\frac{\muv_{s_{\ell}}(k)}{\| \muv_{s_{\ell}} \|_1} \right)^{\Ka-1}
\frac{\partial}{\partial \rv(j,k')} \frac{\muv_{s_{\ell}}(k)}{\left\| \muv_{s_{\ell}} \right\|_1} \\
&= \Ka \frac{1 - \qv(\ell, k)}{1 - \frac{\muv_{s_{\ell}}(k)}{\| \muv_{s_{\ell}} \|_1}}
\frac{\partial}{\partial \rv(j,k')} \frac{\muv_{s_{\ell}}(k)}{\left\| \muv_{s_{\ell}} \right\|_1} .
\end{split}
\end{equation}
Then, by Lemma~\ref{lemma:PartialSestimateQ} and \eqref{equation:qmu-derivative}, we have
\begin{equation}
\begin{split}
&\frac{\partial \hat{s}_{\ell} \left( \qv(\ell, k), \rv(\ell,k), \tau \right)}{\partial \rv(j,k')}
= \frac{\hat{s}_{\ell}(\qv(\ell, k), \rv(\ell,k), \tau) \left( 1 - \hat{s}_{\ell}(\qv(\ell, k), \rv(\ell,k), \tau) \right)}
{\qv(\ell, k) \left( 1 - \qv(\ell, k) \right)} \frac{\partial \qv(\ell, k)}{\partial \rv(j,k')} \\
&= \Ka \frac{\hat{s}_{\ell}(\qv(\ell, k), \rv(\ell,k), \tau) \left( 1 - \hat{s}_{\ell}(\qv(\ell, k), \rv(\ell,k), \tau) \right)}
{\qv(\ell, k) \left( 1-\frac{\muv_{s_{\ell}}(k)}{\| \muv_{s_{\ell}} \|_1} \right)}
\frac{\partial}{\partial \rv(j,k')} \frac{\muv_{s_{\ell}}(k)}{\left\| \muv_{s_{\ell}} \right\|_1} .
\end{split}
\end{equation}
Taking the absolute value of the gradient and incorporating \eqref{equation:qmu}, we get
\begin{equation} \label{equation:partials}
\begin{split}
\left| \frac{\partial \hat{s}_{\ell} \left( \qv(\ell, k), \rv(\ell,k), \tau \right)}{\partial \rv(j,k')} \right|
&= \Ka \frac{\hat{s}_{\ell}(\qv(\ell, k), \rv(\ell,k), \tau) \left( 1 - \hat{s}_{\ell}(\qv(\ell, k), \rv(\ell,k), \tau) \right)}
{\qv(\ell, k) \left( 1 - \frac{\muv_{s_{\ell}}(k)}{\| \muv_{s_{\ell}} \|_1} \right)}
\left| \frac{\partial}{\partial \rv(j,k')} \frac{\muv_{s_{\ell}}(k)}{\left\| \muv_{s_{\ell}} \right\|_1} \right| \\
&\leq \frac{\Ka}{2} \frac{1}
{\frac{\muv_{s_{\ell}}(k)}{\| \muv_{s_{\ell}} \|_1} \left( 1 - \frac{\muv_{s_{\ell}}(k)}{\| \muv_{s_{\ell}} \|_1} \right)}
\left| \frac{\partial}{\partial \rv(j,k')} \frac{\muv_{s_{\ell}}(k)}{\left\| \muv_{s_{\ell}} \right\|_1} \right| .
\end{split}
\end{equation}
At this point, we shift our focus to the partial derivative of $\muv_{s_{\ell}}(k) / \left\| \muv_{s_{\ell}} \right\|_1$.
For the purpose of this derivation, we introduce the compact notation
\begin{equation*}
\muv_{s_{\ell} \sim a}(k) = \prod_{a_p \in N(s_{\ell}) \setminus a} \muv_{a_p \to s_{\ell}} (k) ,
\end{equation*}
where $a \in N(s_{\ell})$.
It is pertinent to note that, in triadic designs, $N(s_{\ell}) \setminus a$ is the empty set whenever $s_{\ell}$ is a parity section.
This is because, in such cases, the cardinality of $N(s_{\ell})$ is one, as illustrated in Fig.~\ref{figure:LocalTreeBP1}.
For such nodes, we take $\muv_{s_{\ell} \sim a}(k) = 1$ for all values of $k$.

Getting back to the proof of the last category, let $a$ be the sole check in $N(s_{\ell})$ such that $a \in N(s_j)$.
In other words, $a \in N(s_{\ell}) \cap N(s_{j})$ denotes the unique check node shared by $s_{\ell}$ and $s_j$ in the triadic design.
Then, we can write
\begin{equation} \label{equation:dmu}
\begin{split}\frac{\partial}{\partial \rv(j,k')}
\muv_{s_{\ell}}(k)
&= \frac{\partial}{\partial \rv(j,k')}
\prod_{a_p \in N(s_{\ell})} \muv_{a_p \to s_{\ell}} (k) \\
&= \muv_{s_{\ell} \sim a}(k)
\frac{\partial}{\partial \rv(j,k')} \sum_{k_1 + k_2 \equiv k} \lambdav_{j}^{\mathrm{PME}}(k_1) \lambdav_{j^{*}}^{\mathrm{PME}}(k_2) \\
&= \muv_{s_{\ell} \sim a}(k)
\frac{\partial}{\partial \rv(j,k')} \lambdav_{j}^{\mathrm{PME}}(k') \lambdav_{j^{*}}^{\mathrm{PME}}(k-k') \\
&= \frac{d_{j}}{\tau^2} \muv_{s_{\ell} \sim a}(k)
\left( 1 - \lambdav_{j}^{\mathrm{PME}}(k') \right) \lambdav_{j}^{\mathrm{PME}}(k') \lambdav_{j^{*}}^{\mathrm{PME}}(k-k')
\end{split}
\end{equation}
where $s_{j^{*}}$ is the unique section that forms a triad with $s_\ell$ and $s_{j}$ through check node $a$, i.e., $N(a) = \left\{ s_{\ell}, s_j, s_{j^{*}} \right\}$.
Likewise, we can compute the partial derivative of $\muv_{s_{\ell}}(k) / \left\| \muv_{s_{\ell}} \right\|_1$,
\begin{equation} \label{equation:partialmu}
\begin{split}
&\frac{\partial}{\partial \rv(j,k')} \frac{\muv_{s_{\ell}}(k)}{\left\| \muv_{s_{\ell}} \right\|_1}
= \frac{1}{\left\| \muv_{s_{\ell}} \right\|_1}
\frac{\partial \muv_{s_{\ell}}(k)}{\partial \rv(j,k')}
- \frac{\muv_{s_{\ell}}(k)}{\left\| \muv_{s_{\ell}} \right\|_1^2}
\sum_{k_0} \frac{\partial \muv_{s_{\ell}}(k_0)}{\partial \rv(j,k')} \\
&= \frac{d_{j}}{\tau^2} \left( 1 - \lambdav_{j}^{\mathrm{PME}}(k') \right) \frac{\lambdav_{j}^{\mathrm{PME}}(k')}{\left\| \muv_{s_{\ell}} \right\|_1} \left(\muv_{s_{\ell} \sim a}(k)\lambdav_{j^{*}}^{\mathrm{PME}}(k-k') - \frac{\muv_{s_{\ell}}(k)}{\left\| \muv_{s_{\ell}} \right\|_1}\sum_{k_0}\muv_{s_{\ell} \sim a}(k_0)\lambdav_{j^{*}}^{\mathrm{PME}}(k_0-k')\right)
\end{split}
\end{equation}
where we have leveraged \eqref{equation:dmu} in simplifying the expression above.
From \eqref{equation:partials} and \eqref{equation:partialmu}, we get
\begin{equation} \label{equation:partials2}
\begin{split}
\left| \frac{\partial \hat{s}_{\ell} \left( \qv(\ell, k), \rv(\ell,k), \tau \right)}{\partial \rv(j,k')} \right| \le \frac{\Ka}{2}&\frac{d_{j}}{\tau^2}\frac{\left( 1 - \lambdav_{j}^{\mathrm{PME}}(k') \right)}{\frac{\muv_{s_{\ell}}(k)}{\| \muv_{s_{\ell}} \|_1} \left( 1 - \frac{\muv_{s_{\ell}}(k)}{\| \muv_{s_{\ell}} \|_1} \right)}\Bigg| \frac{\lambdav_{j}^{\mathrm{PME}}(k') \lambdav_{j^{*}}^{\mathrm{PME}}(k - k')
\muv_{s_{\ell} \sim a}(k)}{\left\| \muv_{s_{\ell}} \right\|_1} \\
&- \frac{\muv_{s_{\ell}} (k)}{\left\| \muv_{s_{\ell}} \right\|_1}
\sum_{k_0} \frac{\lambdav_{j}^{\mathrm{PME}}(k') \lambdav_{j^{*}}^{\mathrm{PME}}(k_0 - k')
\muv_{s_{\ell} \sim a}(k_0)} {\left\| \muv_{s_{\ell}} \right\|_1} \Bigg| .
\end{split}
\end{equation}
We upper bound for the last term in inequality \eqref{equation:partials2} as follows,
\begin{equation} \label{equation:ub}
\begin{split}
&\left| \frac{\lambdav_{j}^{\mathrm{PME}}(k') \lambdav_{j^{*}}^{\mathrm{PME}}(k - k')
\muv_{s_{\ell} \sim a}(k)}{\left\| \muv_{s_{\ell}} \right\|_1}
- \frac{\muv_{s_{\ell}} (k)}{\left\| \muv_{s_{\ell}} \right\|_1}
\sum_{k_0} \frac{\lambdav_{j}^{\mathrm{PME}}(k') \lambdav_{j^{*}}^{\mathrm{PME}}(k_0 - k')
\muv_{s_{\ell} \sim a}(k_0)} {\left\| \muv_{s_{\ell}} \right\|_1} \right| \\
&= \left| \left( 1 - \frac{\muv_{s_{\ell}} (k)}{\left\| \muv_{s_{\ell}} \right\|_1} \right) \frac{\lambdav_{j}^{\mathrm{PME}}(k') \lambdav_{j^{*}}^{\mathrm{PME}}(k - k')
\muv_{s_{\ell} \sim a}(k)}{\left\| \muv_{s_{\ell}} \right\|_1}
- \frac{\muv_{s_{\ell}} (k)}{\left\| \muv_{s_{\ell}} \right\|_1}
\sum_{k_0 \neq k} \frac{\lambdav_{j}^{\mathrm{PME}}(k') \lambdav_{j^{*}}^{\mathrm{PME}}(k_0 - k')
\muv_{s_{\ell} \sim a}(k_0)} {\left\| \muv_{s_{\ell}} \right\|_1} \right| \\
&\leq \left| \left( 1 - \frac{\muv_{s_{\ell}} (k)}{\left\| \muv_{s_{\ell}} \right\|_1} \right) \frac{\lambdav_{j}^{\mathrm{PME}}(k') \lambdav_{j^{*}}^{\mathrm{PME}}(k - k')
\muv_{s_{\ell} \sim a}(k)}{\left\| \muv_{s_{\ell}} \right\|_1} \right|
+ \left| \frac{\muv_{s_{\ell}} (k)}{\left\| \muv_{s_{\ell}} \right\|_1}
\sum_{k_0 \neq k} \frac{\lambdav_{j}^{\mathrm{PME}}(k') \lambdav_{j^{*}}^{\mathrm{PME}}(k_0 - k')
\muv_{s_{\ell} \sim a}(k_0)} {\left\| \muv_{s_{\ell}} \right\|_1} \right| \\
&\leq \left| \left( 1 - \frac{\muv_{s_{\ell}} (k)}{\left\| \muv_{s_{\ell}} \right\|_1} \right) \frac{\muv_{s_{\ell}} (k)}{\left\| \muv_{s_{\ell}} \right\|_1} \right|
+ \left| \frac{\muv_{s_{\ell}} (k)}{\left\| \muv_{s_{\ell}} \right\|_1}
\left( 1 - \frac{\muv_{s_{\ell}} (k)}{\left\| \muv_{s_{\ell}} \right\|_1} \right) \right|
= 2 \frac{\muv_{s_{\ell}} (k)}{\left\| \muv_{s_{\ell}} \right\|_1}
\left( 1 - \frac{\muv_{s_{\ell}} (k)}{\left\| \muv_{s_{\ell}} \right\|_1} \right) .
\end{split}
\end{equation}
Combining \eqref{equation:partials2}, \eqref{equation:ub}, and the fact that $1 - \lambdav_{j_1}^{\mathrm{PME}}(k') \le 1$, we arrive at a uniform bound for the last category,
\begin{equation}
\left| \frac{\partial \hat{s}_{\ell} \left( \qv(\ell, k), \rv(\ell,k), \tau \right)}{\partial \rv(j,k')} \right| \le \frac{\Ka d_{\max}}{\tau^2} .
\end{equation}
Altogether, we gather that the components of the Jacobian matrix are uniformly bounded and, consequently, we conclude that the denoiser is Lipschitz continuous.

\bibliographystyle{IEEEbib}
\bibliography{IEEEabrv,MACcollison}

\end{document}

%% file: preamble.tex
\usepackage{amsmath,amssymb,amsthm}
\usepackage{enumerate}
\usepackage{stfloats}
\usepackage{comment}
\usepackage{subcaption}
\usepackage{siunitx}
\usepackage{mathtools}
\usepackage{accents,latexsym,cancel}
\usepackage{cite}

\usepackage[dvipsnames]{xcolor}
\definecolor{myPink}{RGB}{255,105,183}

\usepackage[T1]{fontenc}
\usepackage{graphics} 
\usepackage{epsfig} 
\usepackage[mathscr]{euscript}
\usepackage{algorithm}
\usepackage[noend]{algpseudocode}
\usepackage{bbm}
\makeatletter
\def\BState{\State\hskip-\ALG@thistlm}
\makeatother

\usepackage{tikz}
\usetikzlibrary{arrows,shapes,chains,matrix,positioning,scopes,patterns,calc,
decorations.markings,
decorations.pathmorphing,
}

\usepackage{pgfplots}
\pgfplotsset{compat=1.3}
\usepgflibrary{shapes}

\newtheorem{theorem}{Theorem}
\newtheorem{lemma}[theorem]{Lemma}
\newtheorem{proposition}[theorem]{Proposition}
\newtheorem{definition}[theorem]{Definition}
\newtheorem{example}[theorem]{Example}
\newtheorem{remark}[theorem]{Remark}

\renewcommand{\epsilon}{\varepsilon}

\newcommand{\Pe}{\ensuremath{P_{\mathrm{e}}}}

\newcommand{\RNum}[1]{\uppercase\expandafter{\romannumeral #1\relax}}

\newcommand{\gv}{\ensuremath{\mathbf{g}}}
\newcommand{\kv}{\ensuremath{\mathbf{k}}}
\newcommand{\Lv}{\ensuremath{\mathbf{L}}}
\newcommand{\mv}{\ensuremath{\mathbf{m}}}

\newcommand{\pv}{\ensuremath{\mathbf{p}}}
\newcommand{\qv}{\ensuremath{\mathbf{q}}}
\newcommand{\rv}{\ensuremath{\mathbf{r}}}
\newcommand{\sv}{\ensuremath{\mathbf{s}}}

\newcommand{\vv}{\ensuremath{\mathbf{v}}}

\newcommand{\wv}{\ensuremath{\mathbf{w}}}
\newcommand{\xv}{\ensuremath{\mathbf{x}}}
\newcommand{\yv}{\ensuremath{\mathbf{y}}}
\newcommand{\zv}{\ensuremath{\mathbf{z}}}

\newcommand{\zerov}{\ensuremath{\mathbf{0}}}
\newcommand{\onev}{\ensuremath{\mathbf{1}}}
\newcommand{\etav}{\ensuremath{\boldsymbol{\eta}}}
\newcommand{\kappav}{\ensuremath{\boldsymbol{\kappa}}}
\newcommand{\lambdav}{\ensuremath{\boldsymbol{\lambda}}}

\newcommand{\muv}{\ensuremath{\boldsymbol{\mu}}}

\newcommand{\Xiv}{\ensuremath{\boldsymbol{\xi}}}

\newcommand{\zetav}{\ensuremath{\boldsymbol{\zeta}}}

\newcommand{\Am}{\ensuremath{\mathbf{A}}}
\newcommand{\Dm}{\ensuremath{\mathbf{D}}}

\newcommand{\Ktot}{\ensuremath{K_{\mathrm{tot}}}}
\newcommand{\Ka}{\ensuremath{K_{\mathrm{a}}}}

\def\Pr{\mathrm{Pr}}

\DeclareMathAlphabet{\mcl}{OMS}{cmsy}{m}{n}

\newlength\tikzwidth
\newlength\tikzheight

\definecolor{mycolor1}{rgb}{0.63529,0.07843,0.18431}%
\definecolor{mycolor2}{rgb}{0.00000,0.44706,0.74118}%
\definecolor{mycolor3}{rgb}{0.00000,0.49804,0.00000}%
\definecolor{mycolor4}{rgb}{0.87059,0.49020,0.00000}%
\definecolor{mycolor5}{rgb}{0.00000,0.44700,0.74100}%
\definecolor{mycolor6}{rgb}{0.74902,0.00000,0.74902}%


%% file: Figures/graph1.tex
\begin{tikzpicture}
  [
  font=\small, line width=1pt, draw=black,
  check/.style={rectangle, minimum height=6mm, minimum width=6mm, draw=black, fill=gray!20},
  section/.style={circle, minimum size=7mm, draw=black}
  ]

\foreach \m in {1,2,3,4,5} {
  \node[section] (s\m) at (0,2.7-0.9*\m) {$s_{\m}$};
}

\node[check] (a3) at (3,0.9) {$a_3$};
\node[check] (a5) at (3,-0.9) {$a_5$};

\node[rotate=90] (variable) at (-0.7,0) {Variable nodes};
\node[rotate=-90] (check) at (3.7,0) {Factor nodes};

\draw (s1) -- (a3.west);
\draw (s2) -- (a3.west);
\draw (s3) -- (a3.west);
\draw (s1) -- (a5.west);
\draw (s2) -- (a5.west);
\draw (s4) -- (a5.west);
\draw (s5) -- (a5.west);
\end{tikzpicture}

%% file: Figures/graph2.tex
\begin{tikzpicture}
  [
  font=\small, line width=1pt, draw=black,
  check/.style={rectangle, minimum height=6mm, minimum width=6mm, draw=black, fill=gray!20},
  trivialcheck/.style={rectangle, minimum height=4mm, minimum width=4mm, draw=black, fill=gray!20},
  section/.style={circle, minimum size=7mm, draw=black}
  ]

\foreach \m in {1,2,3,4,5} {
  \node[section] (s\m) at (0,2.7-0.9*\m) {$s_{\m}$};
}

\foreach \t in {1,2,3,4,5} {
  \node[trivialcheck] (t\t) at (-1.5,2.7-0.9*\t) {}
    edge (s\t);
}

\node[check] (a3) at (3,0.9) {$a_3$};
\node[check] (a5) at (3,-0.9) {$a_5$};

\node[rotate=90] (variable) at (-2.2,0) {Local observations};
\node[rotate=-90] (check) at (3.7,0) {FFT-based factors};

\draw (s1) -- (a3.west);
\draw (s2) -- (a3.west);
\draw (s3) -- (a3.west);
\draw (s1) -- (a5.west);
\draw (s2) -- (a5.west);
\draw (s4) -- (a5.west);
\draw (s5) -- (a5.west);

\draw[shorten <=0.8cm,shorten >=0.65cm,->] (s1)++(0,0.2cm) -- node[above,rotate=-19] {$\muv_{s_1 \to a_3}$} ([yshift=0.2cm]a3.west);
\draw[shorten <=0.7cm,shorten >=0.75cm,<-] (s5)++(0,-0.2cm) -- node[below,rotate=19] {$\muv_{a_5 \to s_5}$} ([yshift=-0.2cm]a5.west);
\draw[shorten <=0.4cm,shorten >=0.4cm,->] (t3)++(0,0.2cm) -- node[above] {$\lambdav_3$} (-0.2,0.2cm);

\end{tikzpicture}

%% file: Figures/subvectors2_new.tex
\begin{tikzpicture}[
  font=\small, >=stealth', line width=0.75pt,
  infobits/.style={rectangle, minimum height=7mm, minimum width=20mm, draw=black, fill=gray!10, rounded corners},
  paritybits/.style={rectangle, minimum height=7mm, minimum width=20mm, draw=black, fill=gray!40, rounded corners}
]

\node[infobits] (vb1) at (1,0) {$\vv(1)$};
\node[infobits] (vb2) at (3,0) {$\vv(2)$};
\node[paritybits] (vp3) at (5,0) {$\vv(3)$};
\draw[|-|] (0,-0.5) to node[midway,below] {$v_1$} (2,-0.5);
\draw[-|] (2,-0.5) to node[midway,below] {$v_2$} (4,-0.5);
\draw[-|] (4,-0.5) to node[midway,below] {$v_3$} (6,-0.5);

\end{tikzpicture}

%% file: Figures/subvectors4_new.tex
\begin{tikzpicture}[
  font=\small, >=stealth', line width=0.75pt,
  infobits/.style={rectangle, minimum height=6.5mm, minimum width=10mm, draw=black, fill=gray!20, rounded corners},
  paritybits/.style={rectangle, minimum height=6.5mm, minimum width=10mm, draw=black, fill=gray!20, rounded corners},
  impossibleblocks/.style={rectangle, minimum height=6.5mm, minimum width=10mm, draw=black, fill=none, rounded corners}
]

\node[infobits] (v10) at (0,0) {0};
\node[impossibleblocks] (v11) at (0,1) {\xcancel{1}};
\node[infobits] (v12) at (0,2) {2};
\node[impossibleblocks] (v13) at (0,3) {\xcancel{3}};

\node[impossibleblocks] (v20) at (3,0) {\xcancel{0}};
\node[infobits] (v21) at (3,1) {1};
\node[infobits] (v22) at (3,2) {2};
\node[impossibleblocks] (v23) at (3,3) {\xcancel{3}};

\node[impossibleblocks] (v30) at (6,0) {\xcancel{0}};
\node[impossibleblocks] (v31) at (6,1) {\xcancel{1}};
\node[paritybits] (v32) at (6,2) {2};
\node[paritybits] (v33) at (6,3) {3};

\foreach \y/\ya in {0/-0.15,1/-0.05,2/0.05,3/0.15}{
    \foreach \yb in {0,1,2,3}{
        \draw[dashed] (0.5,\y+\ya) -- (2.5,\yb+\ya);
    }
}

\foreach \yexit/\yentry in {0/0,1/1,2/2,3/3}{
    \draw[dashed] (3.5,\yexit-0.15) -- (5.5,\yentry-0.15);
}
\foreach \yexit/\yentry in {0/1,1/2,2/3,3/0}{
    \draw[dashed] (3.5,\yexit-0.05) -- (5.5,\yentry-0.05);
}
\foreach \yexit/\yentry in {0/2,1/3,2/0,3/1}{
    \draw[dashed] (3.5,\yexit+0.05) -- (5.5,\yentry+0.05);
}
\foreach \yexit/\yentry in {0/3,1/0,2/1,3/2}{
    \draw[dashed] (3.5,\yexit+0.15) -- (5.5,\yentry+0.15);
}

\draw[line width=1.25pt] (0.5,0-0.15) -- (2.5,2-0.15);
\draw[line width=1.25pt] (3.5,2-0.15) -- (5.5,2-0.15);

\draw[line width=1.25pt] (0.5,2+0.05) -- (2.5,1+0.05);
\draw[line width=1.25pt] (3.5,1+0.05) -- (5.5,3+0.05);

\node (w1) at (0,-1) {$\left[ \hat{\vv}(1) \mathbf{G}_{1,3} \right]$};
\node (w2) at (3,-1) {$\left[ \hat{\vv}(2) \mathbf{G}_{2,3} \right]$};
\node (p3) at (6,-1) {$\vv(3)$};

\end{tikzpicture}

%% file: Figures/architecture.tex
\begin{tikzpicture}[
  font=\small, >=stealth', line width=0.75pt,
  infobits/.style={rectangle, minimum height=3mm, minimum width=10mm, draw=black, fill=gray!10, rounded corners},
  paritybits/.style={rectangle, minimum height=3mm, minimum width=10mm, draw=black, fill=gray!40, rounded corners}
]

\node[infobits] (vb1) at (1,0) {$\vv(1)$};
\node[infobits] (vb2) at (2,0) {$\vv(2)$};
\node[infobits] (vb4) at (3,0) {$\vv(4)$};
\node[infobits] (vb5) at (4,0) {$\vv(5)$};
\node[infobits] (vb7) at (5,0) {$\vv(7)$};
\node[infobits] (vb8) at (6,0) {$\vv(8)$};
\node[infobits] (vb10) at (7,0) {$\vv(10)$};
\node[infobits] (vb11) at (8,0) {$\vv(11)$};

\node[paritybits] (vp3) at (1.5,-1) {$\vv(3)$};
\node[paritybits] (vp6) at (3.5,-1) {$\vv(6)$};
\node[paritybits] (vp9) at (5.5,-1) {$\vv(9)$};
\node[paritybits] (vp12) at (7.5,-1) {$\vv(12)$};
\node[paritybits] (vp13) at (1.5,1) {$\vv(13)$};
\node[paritybits] (vp14) at (3.5,1) {$\vv(14)$};
\node[paritybits] (vp15) at (5.5,1) {$\vv(15)$};

\node[paritybits] (vp16) at (7.5,1) {$\vv(16)$};

\draw  (vb1.south) edge (vp3);
\draw  (vb2.south) edge (vp3);
\draw  (vb4.south) edge (vp6);
\draw  (vb5.south) edge (vp6);
\draw  (vb7.south) edge (vp9);
\draw  (vb8.south) edge (vp9);
\draw  (vb10.south) edge (vp12);
\draw  (vb11.south) edge (vp12);

\draw  (vb1.north) edge (vp13);
\draw  (vb7.north) edge (vp13);
\draw  (vb2.north) edge (vp14);
\draw  (vb10.north) edge (vp14);
\draw  (vb4.north) edge (vp15);
\draw  (vb8.north) edge (vp15);
\draw  (vb5.north) edge (vp16);
\draw  (vb11.north) edge (vp16);

\end{tikzpicture}

%% file: Figures/graphBP0.tex
\begin{tikzpicture}
  [
  font=\small, line width=1pt, draw=black,
  check/.style={rectangle, minimum height=6mm, minimum width=6mm, draw=black, fill=gray!20},
  trivialcheck/.style={rectangle, minimum height=4mm, minimum width=4mm, draw=black, fill=gray!20},
  section/.style={circle, minimum size=7mm, draw=black}
  ]

\node[section] (s1) at (0,0) {$s_{\ell}$};
\node[trivialcheck] (t1) at (-1.5,0) {}
    edge (s1);

\draw[shorten <=0.4cm,shorten >=0.4cm,->] (t1)++(0,0.2cm) -- node[above] {$\lambdav_{\ell}$} (-0.2,0.2cm);

\end{tikzpicture}

%% file: Figures/graphBP1.tex
\begin{tikzpicture}
  [
  font=\small, line width=1pt, draw=black,
  check/.style={rectangle, minimum height=4mm, minimum width=4mm, draw=black, fill=gray!20},
  trivialcheck/.style={rectangle, minimum height=3mm, minimum width=3mm, draw=black, fill=gray!20},
  section/.style={circle, minimum size=7mm, draw=black},
  emptysection/.style={circle, minimum size=5mm, draw=black}
  ]

\node[section] (sp) at (0,0) {$s_{j}$};
\node[trivialcheck] (tp) at (-0.66,-1) {};
\node[check] (ap) at (0.66,-1) {};

\node[emptysection] (sp1) at (0.66-0.35,-0.9-1.2) {};
\node[trivialcheck] (tp1) at (0.66-0.35,-1.5-1.2) {}
    edge[-] (sp1);
\node[emptysection] (sp2) at (0.66+0.35,-0.9-1.2) {};
\node[trivialcheck] (tp2) at (0.66+0.35,-1.5-1.2) {}
    edge[-] (sp2);

\draw (sp) -- (tp.north);
\draw (sp) -- (ap.north);
\draw (sp1) -- (ap.south);
\draw (sp2) -- (ap.south);

\node[section] (sl) at (4,0) {$s_{\ell}$};
\node[trivialcheck] (tl) at (4,-1.3) {};
\node[check] (al1) at (4-0.9,-0.9) {};
\node[check] (al2) at (4+0.9,-0.9) {};

\node[emptysection] (sl11) at (3.1-0.35,-0.9-1.2) {};
\node[trivialcheck] (tl11) at (3.1-0.35,-1.5-1.2) {}
    edge[-] (sl11);
\node[emptysection] (sl12) at (3.1+0.35,-0.9-1.2) {};
\node[trivialcheck] (tl12) at (3.1+0.35,-1.5-1.2) {}
    edge[-] (sl12);

\node[emptysection] (sl21) at (4.9-0.35,-0.9-1.2) {};
\node[trivialcheck] (tl21) at (4.9-0.35,-1.5-1.2) {}
    edge[-] (sl21);
\node[emptysection] (sl22) at (4.9+0.35,-0.9-1.2) {};
\node[trivialcheck] (tl22) at (4.9+0.35,-1.5-1.2) {}
    edge[-] (sl22);

\draw (sl) -- (tl.north);
\draw (sl) -- (al1.north);
\draw (sl) -- (al2.north);
\draw (sl11) -- (al1.south);
\draw (sl12) -- (al1.south);
\draw (sl21) -- (al2.south);
\draw (sl22) -- (al2.south);

\draw[shorten <=0.22cm,<-] (sp.south west)++(0,0.2cm) -- node[above,rotate=58,xshift=-0.15cm] {$\lambdav_j$} ([yshift=0.2cm]tp.north);
\draw[shorten <=0.12cm,<-] (al1.south)++(-0.15cm,0) -- node[above,rotate=68,xshift=-0.1cm] {$\lambdav$} ([yshift=0.2cm]sl11.north);
\draw[shorten <=0.22cm,<-] (sl.south east)++(0,0.25cm) -- node[above,rotate=-45,xshift=0.15cm] {${\muv}$} ([yshift=0.2cm]al2.north);

\end{tikzpicture}

%% file: Figures/results_ISIT.tex
\begin{tikzpicture}
\definecolor{mycolor1}{rgb}{0.63529,0.07843,0.18431}%
\definecolor{mycolor2}{rgb}{0.00000,0.44706,0.74118}%
\definecolor{mycolor3}{rgb}{0.00000,0.49804,0.00000}%
\definecolor{mycolor4}{rgb}{0.87059,0.49020,0.00000}%
\definecolor{mycolor5}{rgb}{0.00000,0.44700,0.74100}%
\definecolor{mycolor6}{rgb}{0.74902,0.00000,0.74902}%

\begin{axis}[%
font=\small,
width=7cm,
height=5.5cm,
scale only axis,
every outer x axis line/.append style={white!15!black},
every x tick label/.append style={font=\color{white!15!black}},
xmin=0,
xmax=300,
xtick = {0,50,100,...,300},
xlabel={Number of active users $\Ka$},
xmajorgrids,
every outer y axis line/.append style={white!15!black},
every y tick label/.append style={font=\color{white!15!black}},
ymin=0,
ymax=6,
ytick = {0,...,6},
ylabel={Required $E_b/N_0$ (dB)},
ymajorgrids,
legend style={at={(0,1)},anchor=north west, draw=black,fill=white,legend cell align=left}
]

\addplot [color=mycolor2,densely dotted,line width=2.0pt,mark size=1.4pt,mark=o, mark options={solid}]
  table[row sep=crcr]{10 3.375 \\
25  3.47 \\
50 3.49\\
75 3.5 \\
100 3.5 \\
125 3.51 \\
150 3.69 \\
175 3.83 \\
200 4.21 \\
225 4.23 \\
250 4.32 \\
275 4.69 \\
300 5.44 \\
};
\addlegendentry{Original AMP+Tree \cite{fengler2019sparcs}};

\addplot [color=mycolor3,solid,line width=2.0pt,mark size=1.4pt,mark=o,mark options={solid}]
  table[row sep=crcr]{10 1.7 \\
25 1.85 \\
50 2.08 \\
75 2.31 \\
100 2.38 \\
125 2.65\\
150 2.99 \\
175 3.12 \\
200 3.57\\
225 3.6 \\
250 3.82 \\
275 4.31 \\
300 4.89 \\
};
\addlegendentry{Enhanced AMP+Tree};

\addplot [color=mycolor1,densely dashdotted,line width=2.0pt,mark size=1.4pt,mark=square,mark options={solid}]
  table[row sep=crcr]{
  25  2\\
50	2.1\\
75	2.2\\
100	2.41\\
125	2.57\\
150	2.81\\
175	3\\
200 3.4\\
225 3.88\\
250 4.36\\
275 4.87\\
300 5.35\\
};
\addlegendentry{Sparse IDMA \cite{pradhan2019sparseidma}};

\addplot [color=mycolor4,dashdotted,mark=*,line width=2.0pt]
  table[row sep=crcr]{
  10    0.5\\
 25	0.55\\
50	0.6\\
75 0.7\\
100 0.75\\
125 1.15\\
150 1.5\\
175 2\\
200 2.7\\
225 3.5\\
250 4.3\\
};
\addlegendentry{PC + Random Spreading \cite{AKPolar}};
\end{axis}

\end{tikzpicture}%

%% file: Figures/State_evolution_results_K25.tex
\begin{tikzpicture}
\definecolor{mycolor1}{rgb}{0.63529,0.07843,0.18431}%
\definecolor{mycolor2}{rgb}{0.00000,0.44706,0.74118}%
\definecolor{mycolor3}{rgb}{0.00000,0.49804,0.00000}%
\definecolor{mycolor4}{rgb}{0.87059,0.49020,0.00000}%
\definecolor{mycolor5}{rgb}{0.00000,0.44700,0.74100}%
\definecolor{mycolor6}{rgb}{0.74902,0.00000,0.74902}%

\begin{semilogyaxis}[%
font=\small,
width=7cm,
height=5.5cm,
scale only axis,
every outer x axis line/.append style={white!15!black},
every x tick label/.append style={font=\color{white!15!black}},
xmin=0,
xmax=5,
xtick = {0,1,2,...,5},
xlabel={Iteration count $t$},
xmajorgrids,
every outer y axis line/.append style={white!15!black},
every y tick label/.append style={font=\color{white!15!black}},
ymin=0.95,
ymax=1.35,
ytick={1,1.05,1.1,1.15,1.2,1.25,1.3,1.35},  
yticklabels={1,1.05,1.1,1.15,1.2,1.25,1.3,1.35},
ylabel={Effective variance $\tau_t^2$},
ymajorgrids,
legend style={at={(1,1)},anchor=north east, draw=black,fill=white,legend cell align=left}
]

\addplot [color=mycolor1,solid,line width=2.0pt]
  table[row sep=crcr]{0 1.3334   \\
1  1.0704    \\
2  1.0371   \\
3 1.0331   \\
4   1.0326    \\
5  1.0326 \\
};
\addlegendentry{Original PME denoiser \cite{fengler2019sparcs}: empirical};

\addplot [color=mycolor2, dotted,line width=2.0pt,mark size=1pt,mark=square, mark options={solid}]
  table[row sep=crcr]{0 1.3325 \\
1   1.0719    \\
2   1.0370    \\
3 1.0332    \\
4 1.0320    \\
5 1.0320 \\
};
\addlegendentry{Original PME denoiser \cite{fengler2019sparcs}: state evolution};

\addplot [color=mycolor3,solid,line width=2.0pt]
  table[row sep=crcr]{0 1.3316 \\
1 1.0227 \\
2 1.0057 \\
3 1.0052 \\
4  1.0052\\
5 1.0052\\
};
\addlegendentry{Dynamic PME denoiser: empirical};

\addplot [color=mycolor4, dotted,line width=2.0pt,mark size=1pt,mark=square, mark options={solid}]
  table[row sep=crcr]{0 1.3325  \\
1  1.0202   \\
2   1.0051   \\
3 1.0048      \\
4 1.0048     \\
5  1.0048  \\
};
\addlegendentry{Dynamic PME denoiser: state evolution};

\end{semilogyaxis}

\end{tikzpicture}%

%% file: Figures/State_evolution_results_K150.tex
\begin{tikzpicture}
\definecolor{mycolor1}{rgb}{0.63529,0.07843,0.18431}%
\definecolor{mycolor2}{rgb}{0.00000,0.44706,0.74118}%
\definecolor{mycolor3}{rgb}{0.00000,0.49804,0.00000}%
\definecolor{mycolor4}{rgb}{0.87059,0.49020,0.00000}%
\definecolor{mycolor5}{rgb}{0.00000,0.44700,0.74100}%
\definecolor{mycolor6}{rgb}{0.74902,0.00000,0.74902}%

\begin{semilogyaxis}[%
font=\small,
width=7cm,
height=5.5cm,
scale only axis,
every outer x axis line/.append style={white!15!black},
every x tick label/.append style={font=\color{white!15!black}},
xmin=0,
xmax=9,
xtick = {0,1,2,...,9},
xlabel={Iteration count $t$},
xmajorgrids,
every outer y axis line/.append style={white!15!black},
every y tick label/.append style={font=\color{white!15!black}},
ymin=0.9,
ymax=3.5,
ytick = {1,1.5,2,2.5,3},
yticklabels = {1,1.5,2,2.5,3},
ylabel={Effective variance $\tau_t^2$},
ymajorgrids,
yminorgrids,
legend style={at={(1,1)},anchor=north east, draw=black,fill=white,legend cell align=left}
]

\addplot [color=mycolor1,solid,line width=2.0pt]
  table[row sep=crcr]{0 3.5092 \\
  1 2.3399 \\
  2 1.7099 \\
  3 1.3412 \\
  4 1.1547 \\
  5 1.0860 \\
  6 1.0658 \\
  7 1.0604 \\
  8 1.0591 \\
  9 1.0588 \\
};
\addlegendentry{Original PME denoiser \cite{fengler2019sparcs}: empirical};

\addplot [color=mycolor2, dotted,line width=2.0pt,mark size=1pt,mark=square, mark options={solid}]
  table[row sep=crcr]{0 3.5119 \\
  1 2.3499 \\
  2 1.7303 \\
  3 1.3540 \\
  4 1.1599 \\
  5 1.0853 \\
  6 1.0628 \\
  7 1.0572 \\
  8 1.0570 \\
  9 1.0569 \\
};
\addlegendentry{Original PME denoiser \cite{fengler2019sparcs}: state evolution};

\addplot [color=mycolor3,solid,line width=2.0pt]
  table[row sep=crcr]{0 3.5129 \\
  1 2.2551 \\
  2 1.5402 \\
  3 1.1738 \\
  4 1.0576 \\
  5 1.0360 \\
  6 1.0327 \\
  7 1.0323 \\
  8 1.0323 \\
  9 1.0323 \\
};
\addlegendentry{Dynamic PME denoiser: empirical};

\addplot [color=mycolor4, dotted,line width=2.0pt,mark size=1pt,mark=square, mark options={solid}]
  table[row sep=crcr]{0 3.5119 \\
  1 2.2770 \\
  2 1.5706 \\
  3 1.1816 \\
  4 1.0530 \\
  5 1.0298 \\
  6 1.0264 \\
  7 1.0260 \\
  8 1.0261 \\
  9 1.0259 \\
};
\addlegendentry{Dynamic PME denoiser: state evolution};

\end{semilogyaxis}

\end{tikzpicture}%

%% file: Figures/runtimeComparison.tex
\begin{tikzpicture}
\definecolor{mycolor1}{rgb}{0.63529,0.07843,0.18431}%
\definecolor{mycolor2}{rgb}{0.00000,0.44706,0.74118}%
\definecolor{mycolor3}{rgb}{0.00000,0.49804,0.00000}%
\definecolor{mycolor4}{rgb}{0.87059,0.49020,0.00000}%
\definecolor{mycolor5}{rgb}{0.00000,0.44700,0.74100}%
\definecolor{mycolor6}{rgb}{0.74902,0.00000,0.74902}%

\begin{axis}[%
font=\small,
width=7cm,
height=5.5cm,
scale only axis,
every outer x axis line/.append style={white!15!black},
every x tick label/.append style={font=\color{white!15!black}},
xmin=0,
xmax=250,
xtick = {0,50,100,...,250},
xlabel={Number of active users $\Ka$},
xmajorgrids,
every outer y axis line/.append style={white!15!black},
every y tick label/.append style={font=\color{white!15!black}},
ymin=0,
ymax=16,
ytick = {0,2,4,6,...,18},
ylabel={Average runtimes (sec)},
ymajorgrids,
legend style={at={(0,1)},anchor=north west, draw=black,fill=white,legend cell align=left}
]

\addplot [color=mycolor2,densely dotted,line width=2.0pt,mark size=1.4pt,mark=o, mark options={solid}]
  table[row sep=crcr]{10 4.54\\
25	4.76\\
50	6.17\\
75	6.46\\
100	6.5\\
125	6.52\\
150	6.77\\
175 7.5\\
200 9.34\\
225 10.12\\
250 10.3\\
};
\addlegendentry{Original AMP+Tree \cite{fengler2019sparcs}};

\addplot [color=mycolor3,solid,line width=2.0pt,mark size=1.4pt,mark=o,mark options={solid}]
  table[row sep=crcr]{10 6.903\\
  25  7.4\\
50	8.63\\
75	9.42\\
100	9.86\\
125	10.08\\
150 11.75\\
175 12.46\\
200 12.9\\
225 13.6\\
250 14.06\\
};
\addlegendentry{Enhanced AMP+Tree};

\end{axis}

\end{tikzpicture}%

%% file: Figures/subvector.tex
\begin{tikzpicture}[
  font=\small, >=stealth', line width=0.75pt,
  infobits/.style={rectangle, minimum height=7mm, minimum width=40mm, draw=black, fill=gray!10, rounded corners},
  paritybits/.style={rectangle, minimum height=7mm, minimum width=20mm, draw=black, fill=gray!40, rounded corners}
]

\node[infobits] (vb) at (2,0) {$\wv$};
\node[paritybits] (vp) at (5,0) {$\pv$};
\node (bits) at (-0.4,0) {$\vv$};
\draw[|-|] (0,-0.5) to node[midway,below] {$w$ bits} (4,-0.5);
\draw[-|] (4,-0.5) to node[midway,below] {$p$ bits} (6,-0.5);

\end{tikzpicture}

%% file: Figures/subvectors.tex
\begin{tikzpicture}[
  font=\small, >=stealth', line width=0.75pt,
  infobits/.style={rectangle, minimum height=7mm, minimum width=10mm, draw=black, fill=gray!10, rounded corners},
  paritybits/.style={rectangle, minimum height=7mm, minimum width=10mm, draw=black, fill=gray!40, rounded corners}
]

\node[infobits, minimum width=20mm] (vb0) at (1,0) {$\wv(1)$};
\node[infobits] (vb1) at (2.5,0) {$\wv(2)$};
\node[paritybits] (vp1) at (3.5,0) {$\pv(2)$};
\node[infobits] (vb2) at (4.5,0) {$\wv(3)$};
\node[paritybits] (vp2) at (5.5,0) {$\pv(3)$};
\node[infobits] (vb3) at (6.5,0) {$\wv(4)$};
\node[paritybits] (vp3) at (7.5,0) {$\pv(4)$};
\node (bits) at (-0.4,0) {$\vv$};
\draw[|-|] (0,-0.5) to node[midway,below] {$w_1$} (2,-0.5);
\draw[-|] (2,-0.5) to node[midway,below] {$w_2$} (3,-0.5);
\draw[-|] (3,-0.5) to node[midway,below] {$p_2$} (4,-0.5);
\draw[-|] (4,-0.5) to node[midway,below] {$w_3$} (5,-0.5);
\draw[-|] (5,-0.5) to node[midway,below] {$p_3$} (6,-0.5);
\draw[-|] (6,-0.5) to node[midway,below] {$w_4$} (7,-0.5);
\draw[-|] (7,-0.5) to node[midway,below] {$p_4$} (8,-0.5);

\end{tikzpicture}